\documentclass[aps,prl,twocolumn,superscriptaddress,showpacs,amsmath,amssymb,longbibliography]{revtex4-1}
\usepackage[T1]{fontenc}
\usepackage[latin9]{inputenc}
\usepackage{color}
\usepackage{babel}
\usepackage{amstext}
\usepackage{graphicx}
\usepackage{esint}
\usepackage[unicode=true,pdfusetitle,
 bookmarks=true,bookmarksnumbered=false,bookmarksopen=false,
 breaklinks=false,pdfborder={0 0 1},backref=false,colorlinks=false]
 {hyperref}

\makeatletter
\@ifundefined{textcolor}{}
{%
 \definecolor{BLACK}{gray}{0}
 \definecolor{WHITE}{gray}{1}
 \definecolor{RED}{rgb}{1,0,0}
 \definecolor{GREEN}{rgb}{0,1,0}
 \definecolor{BLUE}{rgb}{0,0,1}
 \definecolor{CYAN}{cmyk}{1,0,0,0}
 \definecolor{MAGENTA}{cmyk}{0,1,0,0}
 \definecolor{YELLOW}{cmyk}{0,0,1,0}
}

\usepackage{multirow}

\makeatother

\begin{document}
%\title{Semiclassical dynamics as an upper bound for relaxation rates \\in a disordered spinless-fermion model}
\title{Semiclassical bounds on dynamics of two-dimensional interacting disordered fermions}
\author{\L ukasz Iwanek}
\affiliation{Department of Theoretical Physics, Faculty of Fundamental Problems of Technology, Wroc\l aw University of Science and Technology, 50-370 Wroc\l aw, Poland}
\author{Marcin Mierzejewski}
\affiliation{Department of Theoretical Physics, Faculty of Fundamental Problems of Technology, Wroc\l aw University of Science and Technology, 50-370 Wroc\l aw, Poland}
\author{Anatoli Polkovnikov}
\affiliation{Department of Physics, Boston University, 590 Commonwealth Avenue, Boston, Massachusetts 02215, USA}
\author{Dries Sels}
\affiliation{Department of Physics, New York University, New York, NY, USA}
\affiliation{Center for Computational Quantum Physics, Flatiron Institute, New York, NY, USA}
\author{Adam S. Sajna}
\affiliation{Department of Theoretical Physics, Faculty of Fundamental Problems of Technology, Wroc\l aw University of Science and Technology, 50-370 Wroc\l aw, Poland}

\begin{abstract}
Using the truncated Wigner approximation (TWA) we study quench dynamics
of two-dimensional lattice systems consisting of interacting spinless fermions with potential disorder. 
First, we demonstrate that the semiclassical dynamics generally relaxes faster than the full quantum dynamics.  We obtain this result by comparing the semiclassical dynamics with exact
diagonalization and Lanczos propagation of one-dimensional chains.  Next, exploiting the 
TWA capabilities of simulating  large lattices,  we investigate
how the relaxation rates depend on the dimensionality of the studied system.
We show that strongly disordered one-dimensional and two-dimensional systems exhibit a transient, logarithmic-in-time relaxation, which was recently established for one-dimensional chains.  Such relaxation corresponds to the infamous $1/f$-noise at strong disorder.
% It is shown that already quasi-two-dimensional
%lattice structure of ladder type exhibit two-dimensional-like dynamics.
%Interestingly in the stronger disorder regime, the spectral properties
%of the imbalance goes into the robust $1/\omega$ behavior at low frequencies which is also
 %visible in the one dimensional exact simulations. However,
%in order to obtain such logarithmic in-time relaxation processes in
%wo dimension, larger disorder strength are needed in comparison to
%the one dimensional case.
\end{abstract}
\maketitle

\section{Introduction}

Anomalous dynamics of strongly disordered systems with many-body interactions has recently attracted significant interest leading to numerous experimental and theoretical studies. 
The problem emerged from studying role of electron-electron interactions on the fate of the Anderson localization of noninteracting particles~\cite{basko06,oganesyan07}.  Numerical studies of one-dimensional (1D) chains indicated
that at sufficiently strong disorder interacting finite size systems remain localized or nearly localized even in the presence of local two-body interactions~\cite{monthus10,luitz15,ZZZ5_4,Ponte2015,lazarides15,vasseur15a,serbyn2014a,
pekker2014,torres15,torres16,laumann2015,huse14,gopal17,Hauschild_2016,herbrych13,imbrie16,
steinigeweg16,Herbrych17}. The ultimate stability of the many-body localization (MBL) in macroscopic systems is under debate~\cite{Panda2020,Sierant2020,sierant_lewenstein_20,Morningstar2022,abanin_bardarson_21,Morningstar2022} and it has been questioned in a series of recent works~\cite{suntajs_bonca_20a,suntajs_bonca_20,sels2020,Sels_2022,Sels_dilute_2021}.
However, it is well established that at strong disorder interacting chains exhibit very slow logarithmic in time relaxation~\cite{znidaric08,bardarson12,kjall14,serbyn15,luitz16,serbyn13_1,bera15,altman15,agarwal15,gopal15,
znidaric16,mierzejewski2016,lev14,lev15,barisic16,bonca17,bordia2017_1,zakrzewski16,protopopov2018,
sankar2018,zakrzewski2018, Chandran2014,potter16,prelovsek16,proto2017,friedman2017}. Such slow dynamics in one dimension was also found in systems where the noninteracting limit does not correspond to the localized phase~\cite{zakrzewski16,barlev2016,lisarma17,mierzejewski2016}.
The  finite-time dynamics of strongly disordered systems is typically subdiffusive \cite{luitz2016prl,luitz116,znidaric16,gopal17,kozarzewski18,prelovsek217,new_karrasch,prelovsek2018a}. 
Such slow dynamics was frequently considered as a precursor to localization~\cite{luitz2016prl,luitz116,znidaric16,gopal17,kozarzewski18,prelovsek217,new_karrasch,prelovsek2018a} 
and was attributed to the Griffiths effects due to the presence of weak links responsible for the existence of rare localized regions \cite{agarwal15,bordia2017_1,agarwal16,luschen17}.

Despite that the localized phase in thermodynamic limit is likely unstable to interactions, there is a key open question about long-time dynamics in such systems.  Existing computational methods have severe limitations on accessible system sizes and/or accessible time scales. Due to these limitations, previous numerical studies focused mainly on the dynamics of 1D finite-size systems.  At the same time, several recent experiments show signatures that drastic slowing down of dynamics at large disorder also exists in two-dimensional (2D) systems \cite{Choi2016,bordia2017_1,1910.06024} and three-dimensional systems \cite{kondov15}.  Theoretically  dynamics of strongly disordered  systems beyond 1D  remains largely an open problem \cite{mierzejewski2020,strkalj2022}.

In this paper, we demonstrate that the semiclassical description in terms of fermionic truncated Wigner approximation (fTWA) \cite{Davidson2017, S.M.Davidson.thesis, PhysRevB.99.134301,PhysRevA.102.033338, 2007.05063,2205.06620} allows one to partially overcome the limitations of other numerical methods and analyze long-time dynamics both in 1D and in 2D systems. While the semiclassical approach is not expected to be quantitatively reliable at long times,  namely it leads to faster relaxation dynamics than seen within exact numerical methods, it shows qualitative agreement with exact dynamics in 1D systems. At the same time, fTWA allows one to overcome small size, short time and dimensionality limitations intrinsic to other methods because 
the complexity of the fTWA-calculations scales only polynomially with the system size.  Utilizing this approach, we show for strongly disordered 2D systems that the  imbalance decays logarithmically in time characteristic of glassy behavior.  Because fTWA gives a faster decay than in actual systems this result implies that the decay should also be at most logarithmic in time. Such logarithmic  time-dependence is reflected in the spectral functions showing approximate $1/\omega$ dependence also established in 1D disordered systems~\cite{mierzejewski2016, serbyn2017, sels2020,vidmar2021}. The emergence of such inverse frequency spectral functions form, at least within fTWA, is thus not special to 1D systems.  We note that this form of the spectral function is also known as $1/f$ noise,  which was observed experimentally in a broad range of physical systems~\cite{Ward:2007}.

The remainder of this paper is organized as follows: in Sec. \ref{sec-ftwa} we introduce the disordered spinless fermionic model and the implementation of the fTWA method.  In Sec. \ref{bound} we show how fTWA bounds the actual decay of correlations in quantum systems.  In Sec.~\ref{spectral} we present an analysis of the relaxation dynamics and the spectral function focusing on 2D systems.  Finally we summarize our results.  In the Appendix we show the analysis of finite size effects on the system dynamics.

\section{Dynamics of spinless fermions  within fTWA} \label{sec-ftwa}

In this work we consider spinless fermions whose dynamics is given
by the following Hamiltonian
\begin{align}
\hat{H}=&-J\sum_{\left\langle ij\right\rangle }\left(\hat{c}_{i}^{\dagger}\hat{c}_{j}+h.c.\right)+V\sum_{\left\langle ij\right\rangle }\left(\hat{n}_{i}-\frac{1}{2}\right)\left(\hat{n}_{j}-\frac{1}{2}\right)\nonumber\\
&+\sum_{i}\Delta_{i}\left(\hat{n}_{i}-\frac{1}{2}\right),\label{eq: hamiltonian}
\end{align}
where $\langle ij\rangle$ stands for nearest neighbor sites,
$\hat{c}_{i}$ ($\hat{c}_{i}^{\dagger}$) is the fermionic annihilation (creation)
operator on the site $i$,  $\hat{n}_{i}=\hat{c}_{i}^{\dagger}\hat{c}_{i}$
is the corresponding number operator,  $J$ is the hoping amplitude,  $V$ is the nearest
neighbor interaction coupling,  and $\Delta_{i}$ is a local random potential
drawn from a uniform distribution in the range $[-W,\,W]$.  

\begin{figure*}[t]
\includegraphics[scale=0.7]{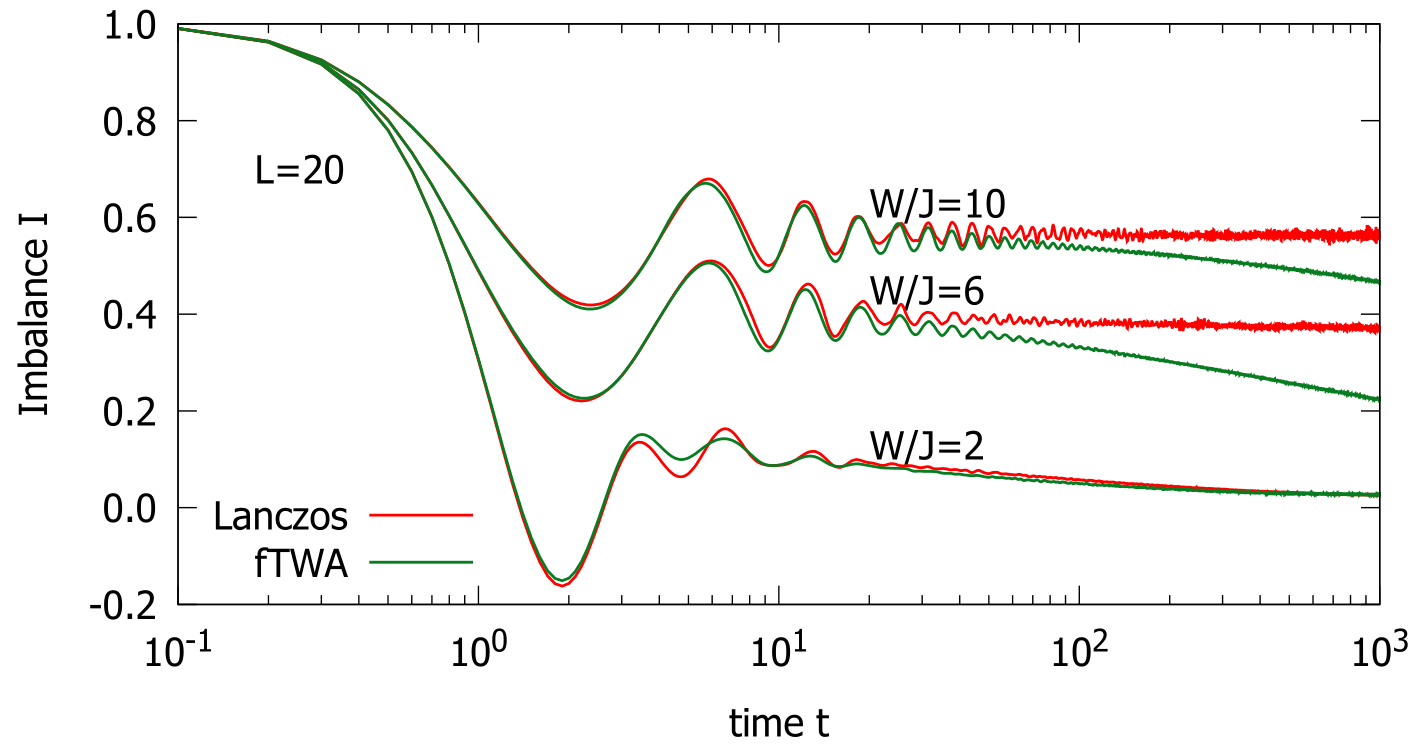}
\caption{Time dependence of the imbalance for different disorder strengths $W$.  Numerical
calculations are performed for 1D lattice with $L=20$ sites.
The fTWA and Lanczos data are represented by a green and red line, respectively. Data from top to bottom are obtained for disorders
with strength $W/J=10,\,6,\,2$, respectively, and are averaged over 200 disorders realizations.
The rest of the parameters are $V/J=1,$ $J=0.5$. In fTWA we use 500 trajectories
for each disorder realization.\label{fig: time fTWA vs ED}}
\end{figure*}

We describe the dynamics of the electrons within the semiclassical fTWA method \cite{Davidson2017}.  Within
this framework,  fermionic bilinears $\hat{E}^{i}_{j}=(\hat{c}_{i}^{\dagger}\hat{c}_{j}-\hat{c}_{j}\hat{c}_{i}^{\dagger})/2$
are mapped to the complex phase space variables $\rho_{ij}$ (e.g. $\hat{n}_i$ operator maps to $\rho_{ii}+1/2$). The phase space variables satisfy the canonical Poisson bracket relations  \footnote{In comparison to Ref. \cite{Davidson2017}, we include an extra imaginary unit $i$  factor into the definition of the Poisson brackets in Eqs. \ref{eq:equation2} and \ref{eq: EOM}.} with the structure constants given by the corresponding quantum commutation relations:
\begin{equation}
[\hat E^{i}_j,  \hat E^{k}_l]= \hat E^i_l \delta_{kj}-\hat E^k_j \delta_{il}\; \rightarrow\;
\{\rho_{ij},\rho_{kl}\} = \rho_{il} \delta_{kj}-\rho_{kj} \delta_{il}, \label{eq:equation2}
\end{equation}
where $\rho_{ij}=\rho_{ji}^\ast$.
We note that the operators $\hat E^i_j$ form a representation of a $su(N)$ algebra,  where $N$ is the number of sites.  The operators are mapped to functions using the Wigner-Weyl quantization~\cite{Polkovnikov2010}.  In particular, the Hamiltonian $\hat H$ is mapped to its Weyl symbol $H_W$:
\begin{equation}
H_{W} =  J\sum_{\left\langle ij\right\rangle }\left(\rho_{i j}+c.c.\right)+\sum_{i}(\Delta_{i}\rho_{i i}+V\rho_{ii}^2)+ V\sum_{\langle ij\rangle}\rho_{i i}\rho_{j j}.
\label{equation: Weyl of H}
\end{equation}
and the initial density matrix $\hat \rho^0$ is mapped to the Wigner function  
$\mathcal{W}(\{\rho^0_{ij}\})$,  which plays the role of the initial probability distribution of the phase space variables (here, the following denotation is used $\rho_{kl}^0\equiv\rho_{kl}(t=0)$). The origin of $V\rho_{ii}^2$ in $H_W$ is explained in Appendix \ref{appendix-hamiltonian}. The dynamics of the phase space variables within the fTWA is described by the classical Hamiltonian equations of motion:
\begin{equation}
i\frac{d\rho_{kl}}{d t}  = \{\rho_{kl}, H_W\}= \sum_{m}  \left({\partial H_W\over \partial \rho_{lm} } \rho_{km}-{\partial H_W\over \partial \rho_{mk}} \rho_{ml}\right)
\label{eq: EOM}
\end{equation}
To find an observable at the time, $t$, we need to evolve the phase space variables in time starting from initial conditions drawn from the Wigner function,  compute the Weyl symbol of the corresponding operator and then average over the initial conditions.  In this paper we are focusing on the expectation values of the number operators,  such that this prescription gives
\begin{equation}
\left\langle \hat{n}_{i}(t)\right\rangle \approx\int\left(\rho_{ii}(t)+\frac{1}{2}\right)W(\{\rho^0_{kl}\})D\rho^0_{kl},\label{eq: average fTWA},
\end{equation}
where $D\rho^0_{kl}$ stands for integration over all independent phase space variables. In our simulations we use open boundary conditions and start from the initial states  which are the product of single site states with 0 or 1 fermions. The latter allows us to approximate the initial Wigner function as a factorisable over different pairs of sites Gaussian distribution
\begin{equation}
\mathcal{W}(\{\rho_{kl}\})=\prod_{kl}\frac{1}{2\pi\sigma_{kl}^2} \exp{\left(\frac{(\rho_{kl}-\mu_{kl})(\rho_{kl}^*-\mu_{kl}) }{2\sigma_{kl}^2}\right)},
\end{equation}
where $\mu_{kl}$ and $\sigma_{kl}$ are fixed by the expectation values and the fluctuations of the operators $\hat E^i_j$~\cite{Davidson2017}, i.e.
\begin{equation}
Tr(\hat\rho\hat{E}^k_l) = \int \rho_{kl} \mathcal{W}(\{\rho_{mn}\})D\rho_{mn},
\end{equation}
\begin{equation}
\frac{1}{2}Tr\left(\hat\rho(\hat{E}^i_j\hat{E}^k_l+\hat{E}^k_l\hat{E}^i_j)\right) = \int\rho_{ij}\rho_{kl} \mathcal{W}(\{\rho_{mn}\})D\rho_{mn}.
\end{equation}

Let us note that the complexity of the fTWA scales polynomially with the system size $L$ as the dimensionality of the phase space, $L^2$, is much less than the dimensionality of the quantum Hilbert space $2^L$.  In the interacting systems, fTWA is guaranteed to be accurate only at early times \cite{Polkovnikov2010, Davidson2017},  while in noninteracting systems fTWA is exact at all times. This method also becomes exact for fermions with infinite range interactions and, in particular, it can accurately describe dynamics of systems with long-range interactions ~\cite{PhysRevA.102.033338}.  It can also be made asymptotically exact by increasing the number of fermion flavors~\cite{2205.06620}.  It is also expected that accuracy of fTWA increases with the dimensionality of the system.

\section{fTWA as a bound for relaxation dynamics \label{sec: Benchmark-fTWA-against}}  \label{bound}

Before we analyze 2D systems in this section we benchmark the applicability of fTWA in 1D systems
by comparing it to exact diagonalization (ED) and Lanczos method~\cite{lantime1,lantime2}.  We consider quenches from an initial charge density wave (CDW) product state at a half filling: 
\begin{equation}
|\psi(t=0)\rangle = |0\rangle|1\rangle|0\rangle|1\rangle\dots,
\end{equation}
where $|0\rangle$ and $|1\rangle$ are empty and occupied states which alternate between neighboring sites.  Such states are accessible experimentally, e.g.,  they were realized in ultracold atom experiments
\citep{mblschreiber,PhysRevLett.116.140401,smith2016}.  They are easily represented
by the approximate Wigner function in fTWA \cite{Davidson2017, S.M.Davidson.thesis, PhysRevB.99.134301,PhysRevA.102.033338, 2007.05063,2205.06620}.  In simulations we analyze the imbalance function related to the on-site densities
in the following way
\begin{equation}
I(t)=\frac{N_{o}(t)-N_{e}(t)}{N_{o}(t)+N_{e}(t)}
\end{equation}
where 
\begin{equation}
N_{o}(t)=\sum_{i\in\text{initially occupied sites}}\left\langle \hat{n}_{i}(t)\right\rangle ,
\end{equation}
\begin{equation}
N_{e}(t)=\sum_{i\in\text{initially empty sites}}\left\langle \hat{n}_{i}(t)\right\rangle .
\end{equation}
This imbalance was widely used both in ultracold atom experiments\citep{mblschreiber,Choi2016,PhysRevLett.116.140401,bordia2017_1,Lschen2017} and in numerical simulations~\cite{sierant_lewenstein_20} as an indicator of thermalization. In~Fig.~\ref{fig: time fTWA vs ED} we show the time dependence of the imbalance $I(t)$ computed within the fTWA and the Lanczos methods starting from the CDW state in a 1D system of size $L=20$.  We set interaction strength $V=J$. We see that at short and intermediate times the fTWA accurately describes the imbalance correctly predicting the initial transient dynamics followed by a crossover to a slow relaxation at strong disorder. At weaker disorder $W/J\lesssim 2$, the fTWA nearly agrees with the exact dynamics at all times.  However at $W/J>2$, we see that the fTWA predicts faster decay of the disorder-averaged imbalance over a long time.  Interestingly as the disorder keeps increasing the fTWA starts improving again successively approaching the imbalance plateau (compare the results for $W/J=6$ and $W/J=10$ in Fig. \ref{fig: time fTWA vs ED}).  This behavior is consistent with previous observations in disordered spin systems \cite{PhysRevA.96.033604, WURTZ2018341} and in the long-range Hubbard model~\cite{PhysRevA.102.033338}.

\begin{figure*}[t]
\includegraphics[scale=0.33]{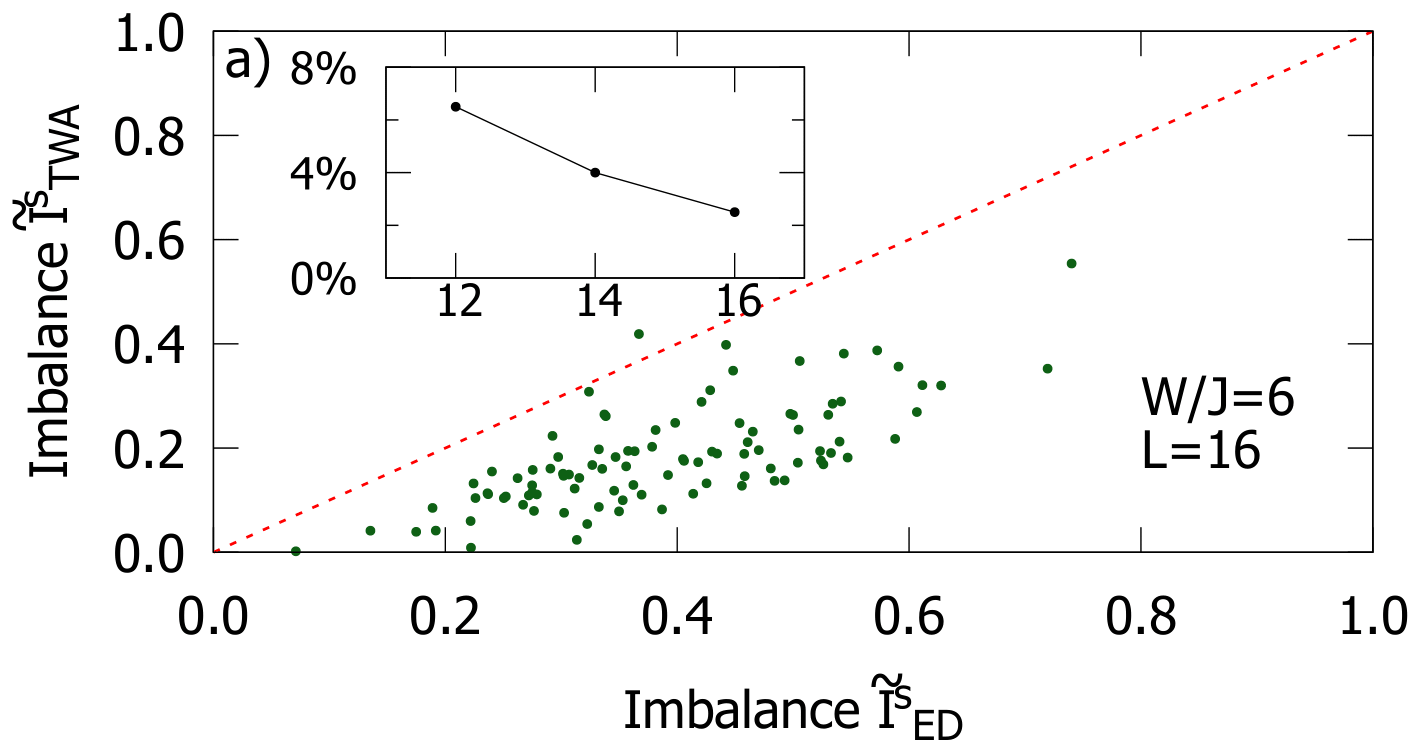}\includegraphics[scale=0.33]{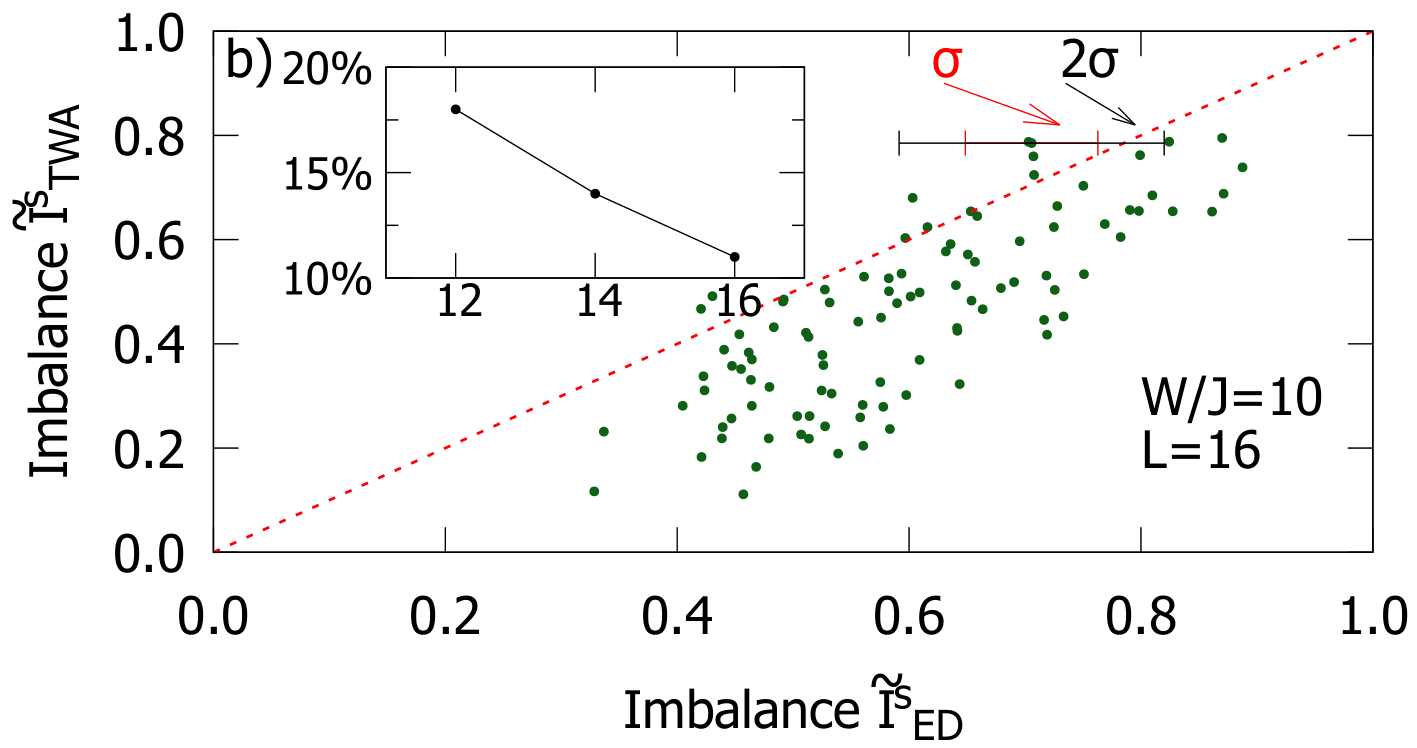}\includegraphics[scale=0.33]{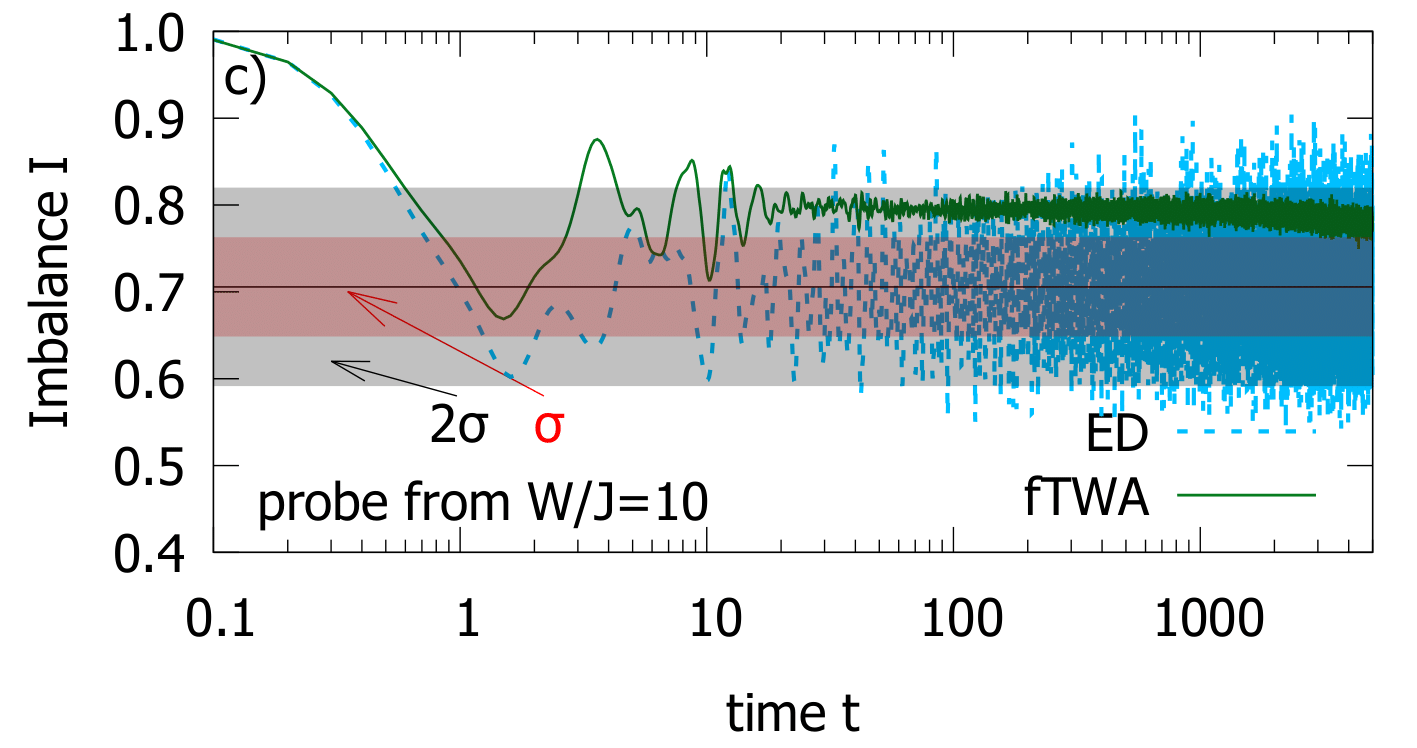}
\caption{\textcolor{black}{ Location of $(\tilde{I}_{\text{fTWA}}^{s},\ \tilde{I}_{\text{ED}}^{s})$
points for $100$ disorder realizations (Fig. (a) and (b)). Fig. (a) correspond to $W/J=6$ and Fig. (b) to $W/J=10$. $\tilde{I}_{\text{fTWA/ED}}^{s}$
are calculated from averaging imbalances in time window $t\in(1000,\,5000)$, see Eq. \ref{sigma}. }
The initial state is CDW with 16 lattice sites and due to long-time dynamics
exact diagonalization method was used. Fig. (c) represents one ($\sigma$) and two standard  deviations ($2\sigma$) of ED data with respect to fTWA dynamics for the most upper point in Fig. (b). The insets in Fig. (a) and (b) show percentage of points satisfying $\tilde{I}_{\text{fTWA}}^{s} >\tilde{I}_{\text{ED}}^{s} $ condition  (vertical axis) when the system size (horizontal axis) is varied from 12 to 16 lattice sites (in calculations 200 disorder realizations were used). For fTWA 500 trajectories were used and the interaction strength was set to $V/J=1$ with $J=0.5$. \label{fig: error}}
\end{figure*}

\begin{figure*}[t]
\includegraphics[scale=0.5]{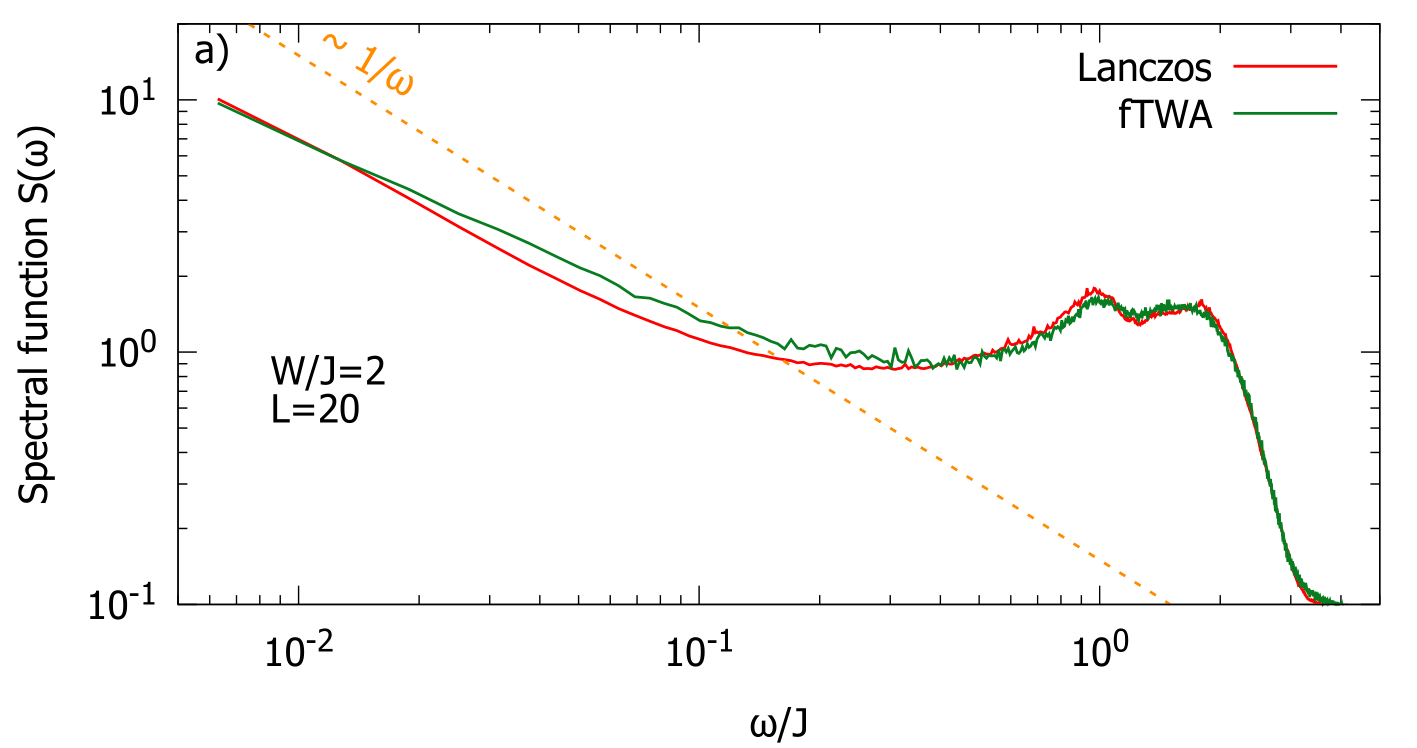}\includegraphics[scale=0.5]{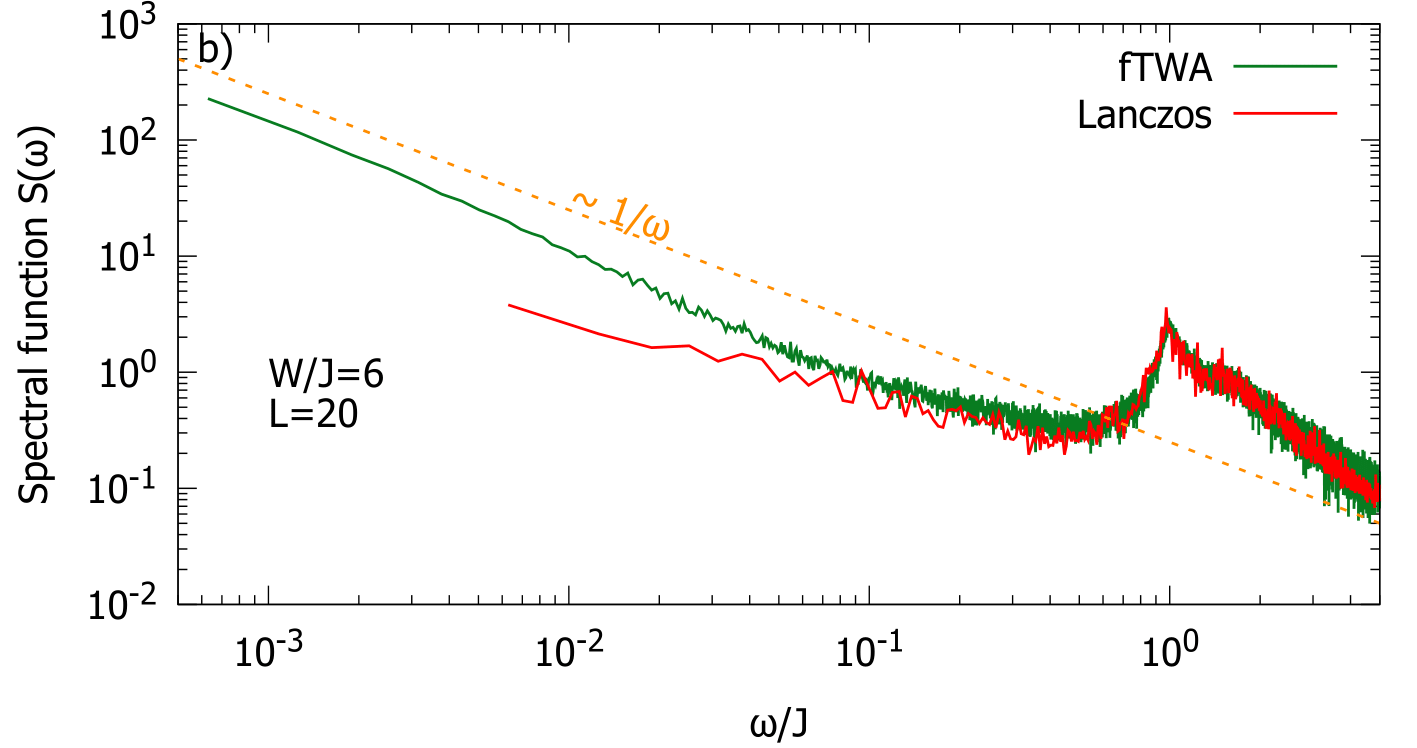}
\includegraphics[scale=0.5]{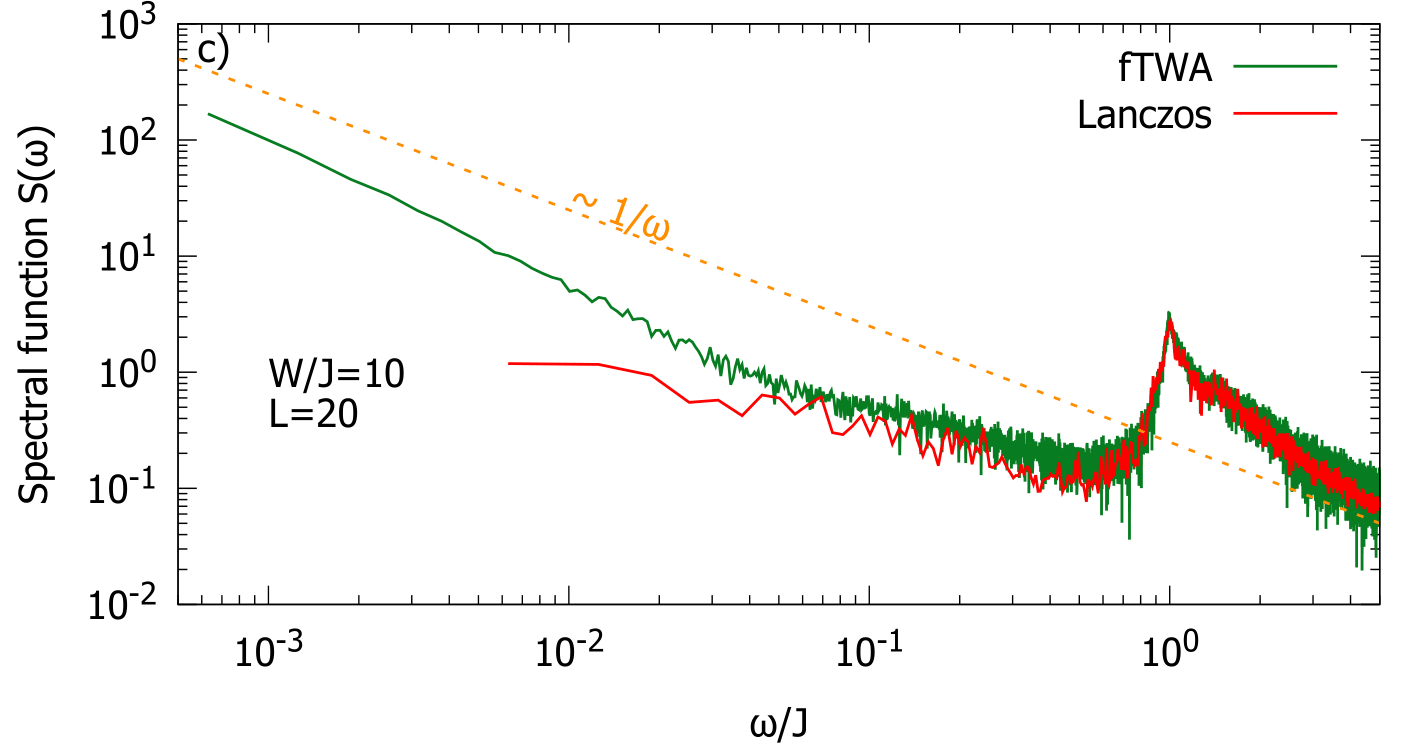}\includegraphics[scale=0.5]{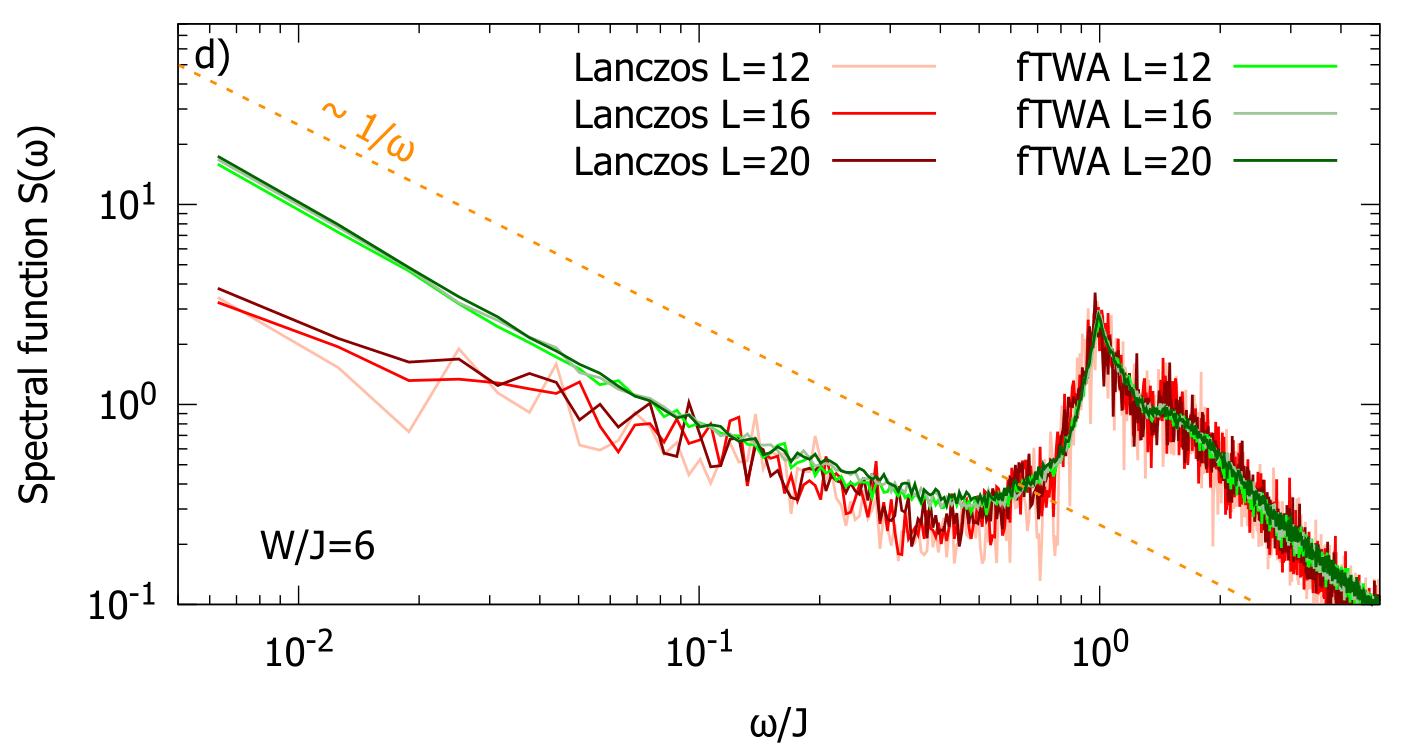}

\caption{(a-c) Spectral function $S(\omega)$ in 1D system with $L=20$ lattice sites. Disorder strength is set to (a) $W/J=2$, (b) $W/J=6$, (c) $W/J=10$. Results for $\text{fTWA}$ and Lanczos method (a-c) are shown as a green and red line, respectively. In (d), lattice sizes L=12, 16, 20 are plotted. The fTWA parameters are as follows: (a) and (d) 500 trajectories, (b) and (c) 400 trajectories due to long computation time requirements. In (a), (b) and (d) averaging over 200 disorders is taken while in (c) number of disorder realizations is increased to 1000 in order to avoid noisy Lanczos data at strong disorder strength $W/J=10$. \label{fig: 1 over f in 1D}}
\end{figure*}

\begin{figure*}[t]
\includegraphics[scale=0.5]{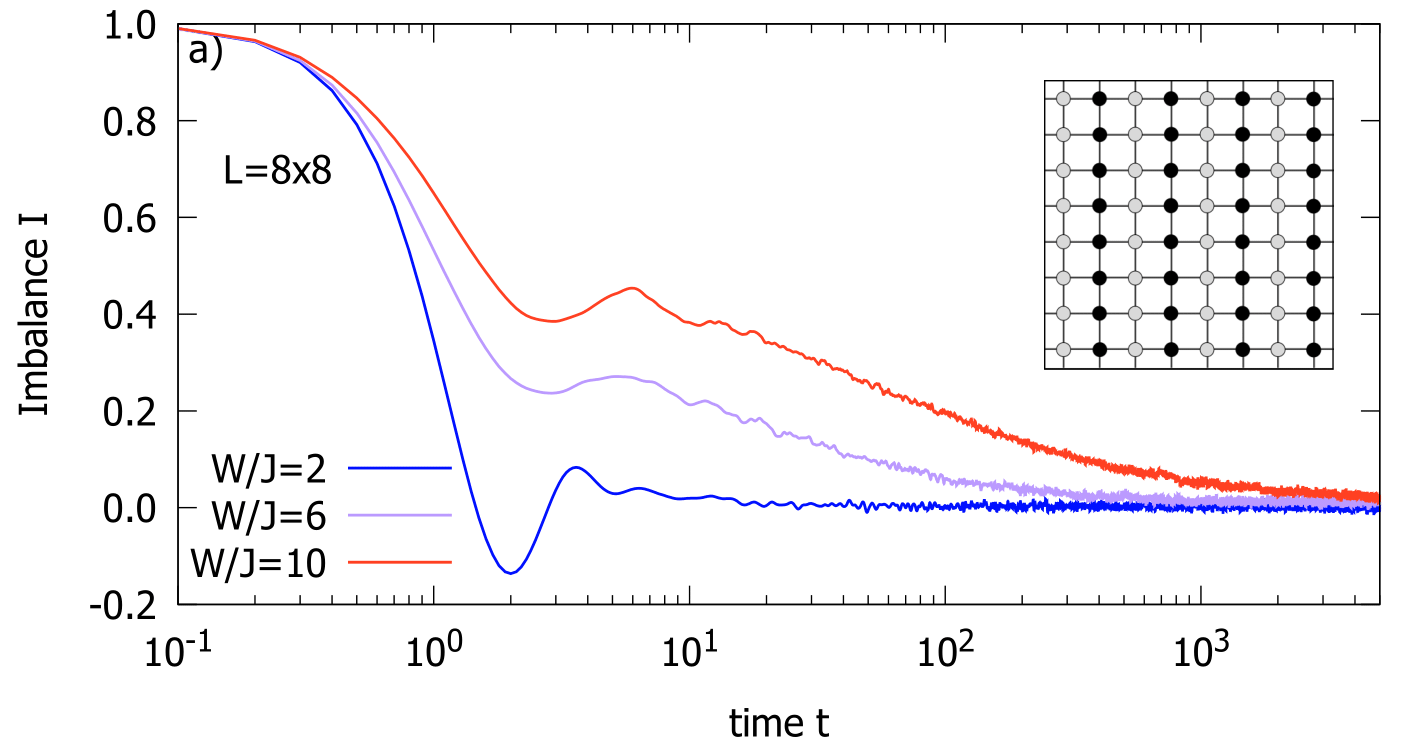}\includegraphics[scale=0.5]{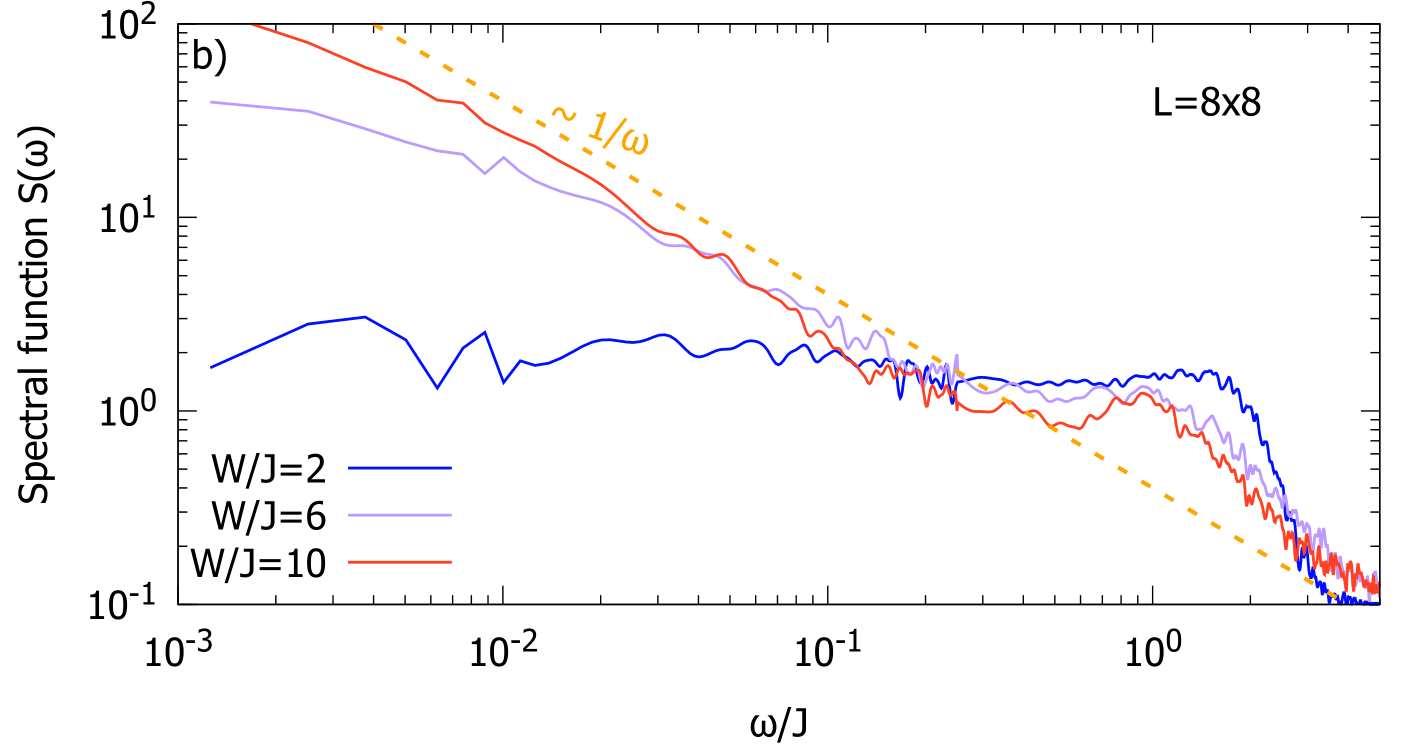}

\caption{Time dependence of imbalance (a) and spectral
function $S(\omega)$ (b) in 2D system with the size
$8\times8$. The inset in Fig. (a) represents stripes CDW initial condition (black circles represent occupied sites and white empty). Results represent data for $\text{fTWA}$
with disorder strength $W/J=2,\,6,\,10$ where average over 30 disorder
realizations were used. In fTWA there are 25 trajectories simulated
for each disorder realization. The rest of the parameters are $V/J=1$, $J=0.5$.
\label{fig: 1 over f in 2D}}
\end{figure*}

From recent literature  \cite{PhysRevA.96.033604, WURTZ2018341,PhysRevA.102.033338}, it can be concluded that in comparison to the exact numerics, nonlinearities
presented in the semiclassical description are responsible for the faster disappearance
of the memory effects encoded in the initial state of disordered systems.
Here we show that this condition obtained previously for disorder averages holds for almost
every single disorder realization. For long-time simulations we compare time averages  of fTWA and ED imbalances in the time window $t\in(1000, 5000)$:
\begin{equation}
\tilde{I}_{\text{fTWA/ED}}^{s} = \frac{1}{\Delta t} \int_{t_0}^{t_0+\Delta t} I^{s} _{\text{fTWA/ED}}(t) dt, \label{sigma}
\end{equation}
where $t_0 = 1000$, $\Delta t = 4000$ and $s$ index denotes that imbalance is calculated for single disorder realization. For larger disorder strengths points $(\tilde{I}_{\text{ED}}^{s} , \tilde{I}_{\text{fTWA}}^{s} )$ are plotted in Fig. \ref{fig: error} a and b. We observe that majority of points satisfy $\tilde{I}_{\text{fTWA}}^{s} \leqslant\tilde{I}_{\text{ED}}^{s} $ which suggest that fTWA dynamics can be regarded as an upper bound for relaxation rates. While we observe some violations of the proposed bound, we note a steady decrease of the number of disorder realizations with system size that do so, see insets in Figs. \ref{fig: error} a and b. Consequently, we expect a negligible effect of these rare realizations for the much larger systems studied in the next section.

We also check that the statistical uncertainty of  $\tilde{I}_{\text{ED}}^{s}$ for the most upper point in Fig. \ref{fig: error} b, which satisfies $\tilde{I}_{\text{fTWA}}^{s} >\tilde{I}_{\text{ED}}^{s} $, coincides with  $\tilde{I}_{\text{fTWA}}^{s} \approx\tilde{I}_{\text{ED}}^{s} $ condition within two standard deviations (2$\sigma$), see Fig. \ref{fig: error} c. Definition of $\sigma$ is given by
\begin{equation}
\sigma^2= \frac{1}{\Delta t} \int_{t_0}^{t_0+\Delta t} (I_{\text{ED}}^{s}(t) - \tilde I_{\text{ED}}^{s})^2 dt. \label{2sigma}
\end{equation}

\section{Spectral function in 1D and 2D lattices \label{sec: 1overf-noise}}  \label{spectral}

To achieve further insight into the relaxation dynamics, it is convenient to analyze the spectral function defined as the Fourier transform of the imbalance function $I(t)$
\begin{equation}
S(\omega)=\int_{-\infty}^{\infty}I(t)e^{-i\omega t} dt=2\text{Re}\left[\int_{0}^{\infty}I(t)e^{-i\omega t}dt\right].
\end{equation}
First, we focus on the 1D system. Using data presented
in Fig. \ref{fig: time fTWA vs ED}, $S(\omega)$ is calculated for
exact and fTWA imbalance, see Fig. \ref{fig: 1 over f in 1D}. From obtained
data, fTWA for the weak disorder strength ($W/J=2$) is
almost exact. For larger values of $W/J$, fTWA reflects ED results
quantitatively down to $\omega$ of order $\mathcal O(10^{-1})$. Interestingly, for larger values of disorder strength, fTWA
predicts almost $1/\omega$ behavior which was also observed in other works~\cite{mierzejewski2016, serbyn2017, sels2020,vidmar2021}. Such non-trivial behavior comes from logarithmic-in-time
decay of the imbalance and is also partially visible in the propagation
within the Lanczos method. These results suggest that fTWA upper bound
for relaxation rates is of logarithmic type.

\begin{figure*}[t]
\includegraphics[scale=0.5]{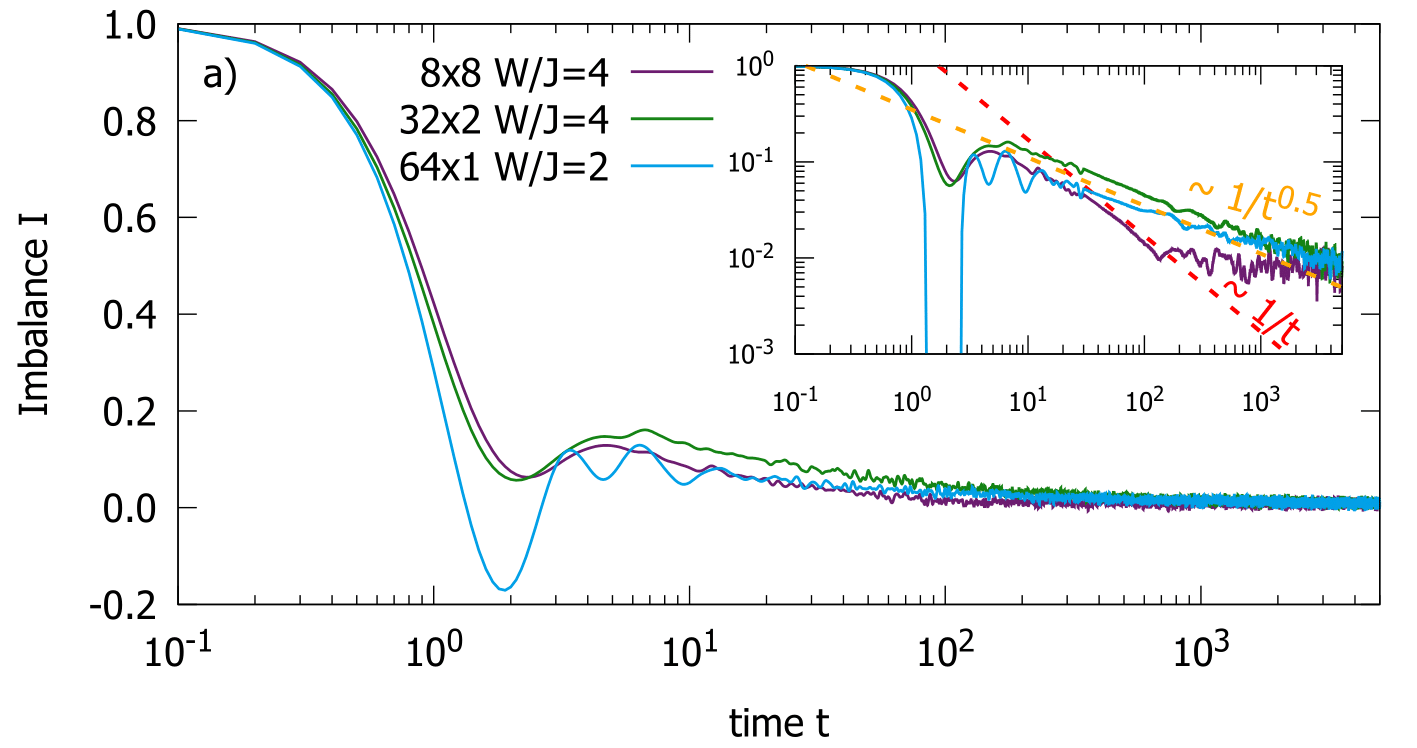}\includegraphics[scale=0.5]{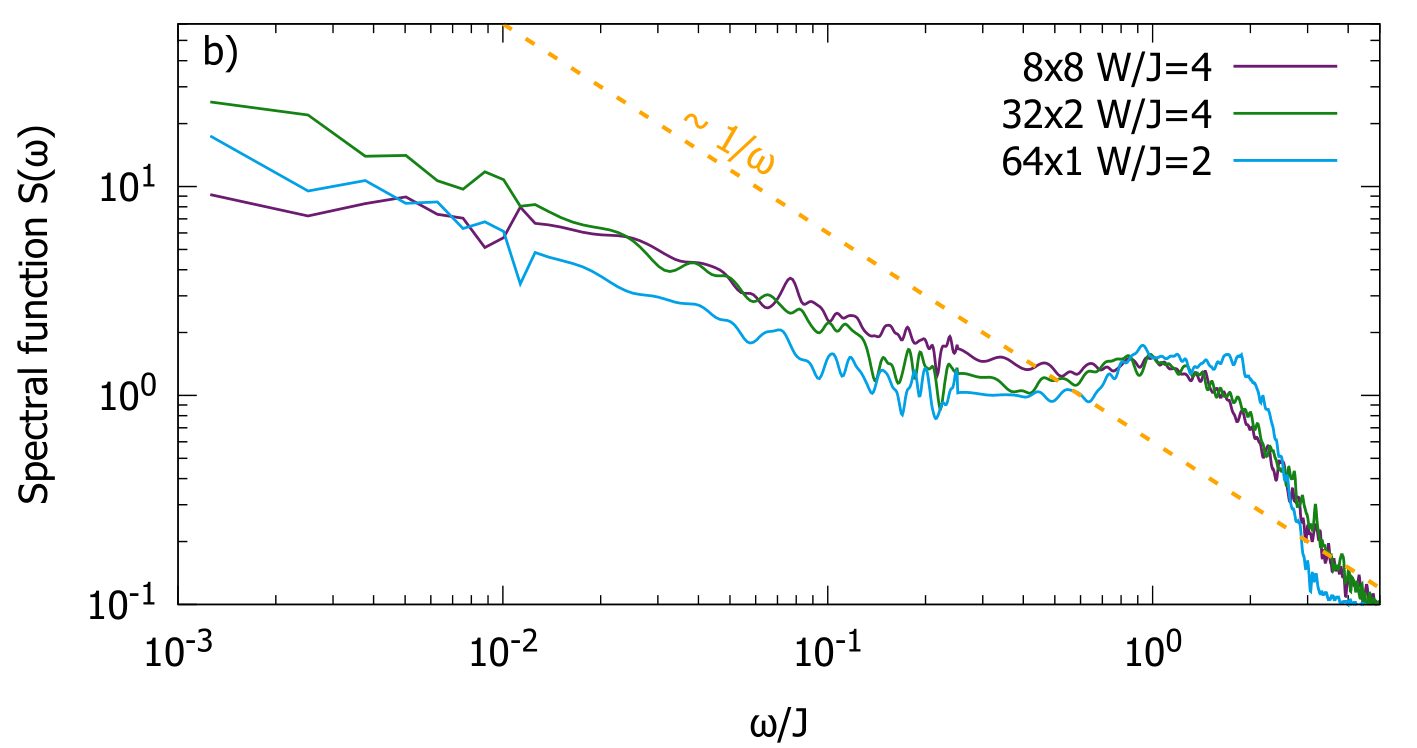}
\includegraphics[scale=0.5]{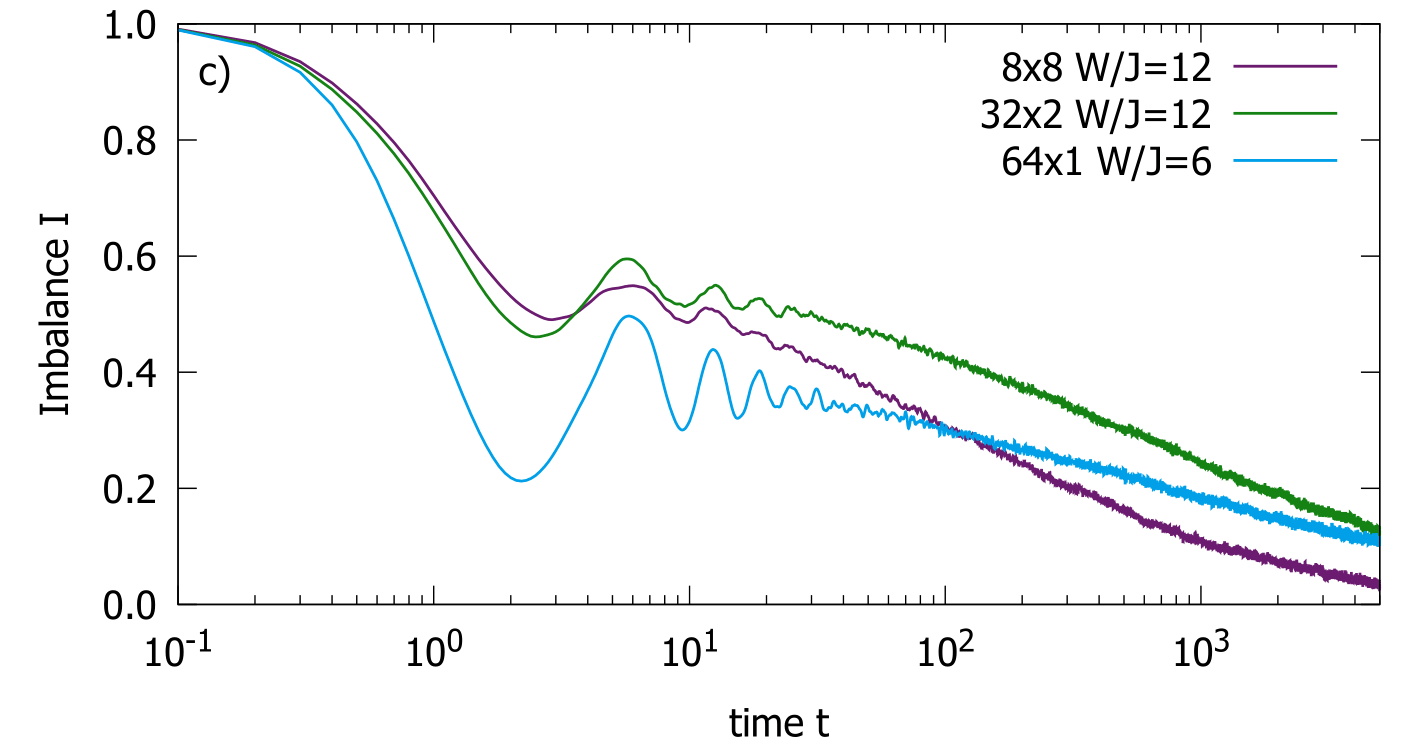}\includegraphics[scale=0.5]{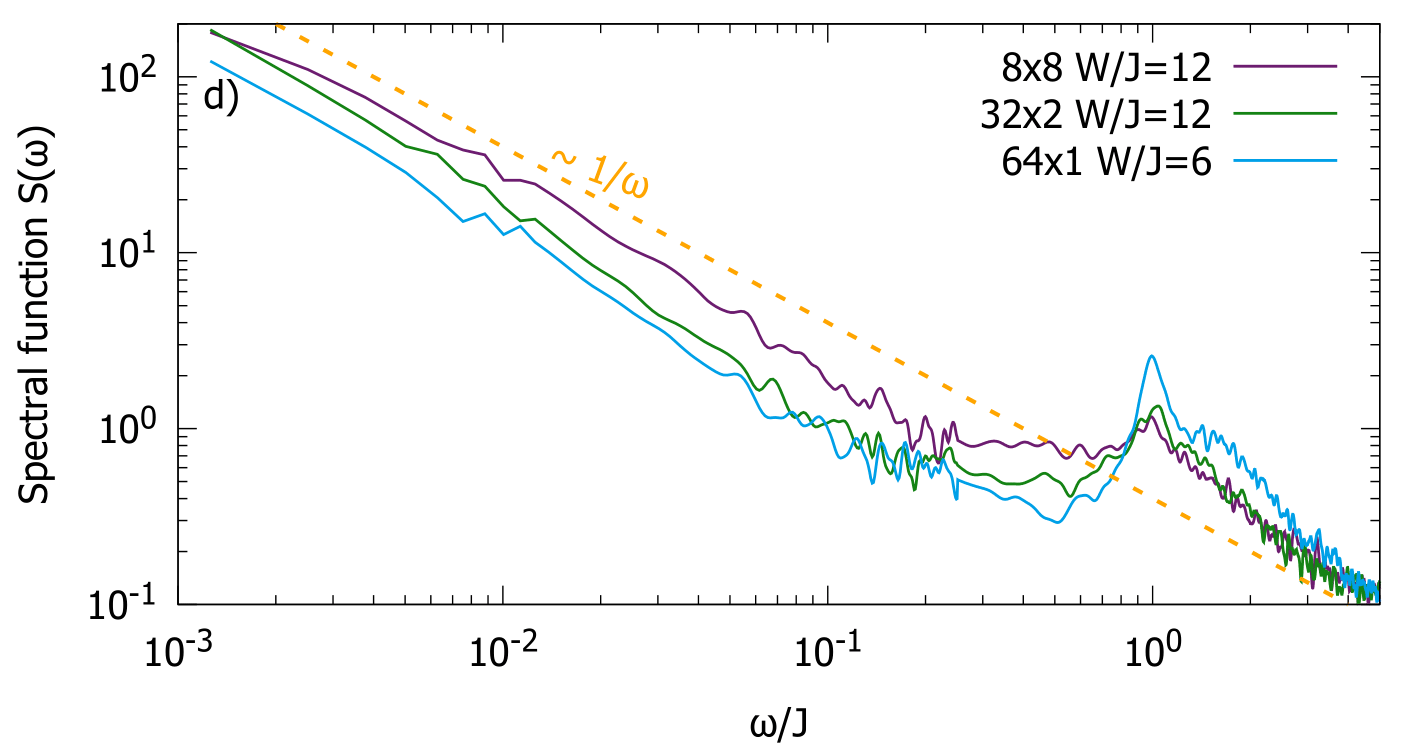}
\caption{Time dependence of imbalance ((a) and (c)) and spectral
function $S(\omega)$ ((b) and (d)) for different lattice geometry $64\times1$
(finite 1D lattice), $8\times8$ (finite 2D
lattice) and $32\times2$ (crossover region between 1D and 2D). 
Data for (Fig. (a) and (b)) represent disorder strength $W/J=2$ for 1D, $W/J=4$ for 2D and ladder-type. 
Results for (Fig. (c) and (d)) represent disorder strength $W/J=6$ for 1D, $W/J=12$ for 2D and ladder-type. 
For better comparison of 1D and 2D data, the disorder strength is doubled for 2D case.
Each line is averaged over 30 disorder realization. In fTWA there are 25 trajectories for each disorder realization. Moreover, in the inset of Fig. (a), the same data as in (a) are presented but with the additional log scale on the vertical axis.
Rest of parameters are $V/J=1$, $J=0.5$. \label{fig: 1 over f crossover}}
\end{figure*}

\begin{figure*}[t]
\includegraphics[scale=0.33]{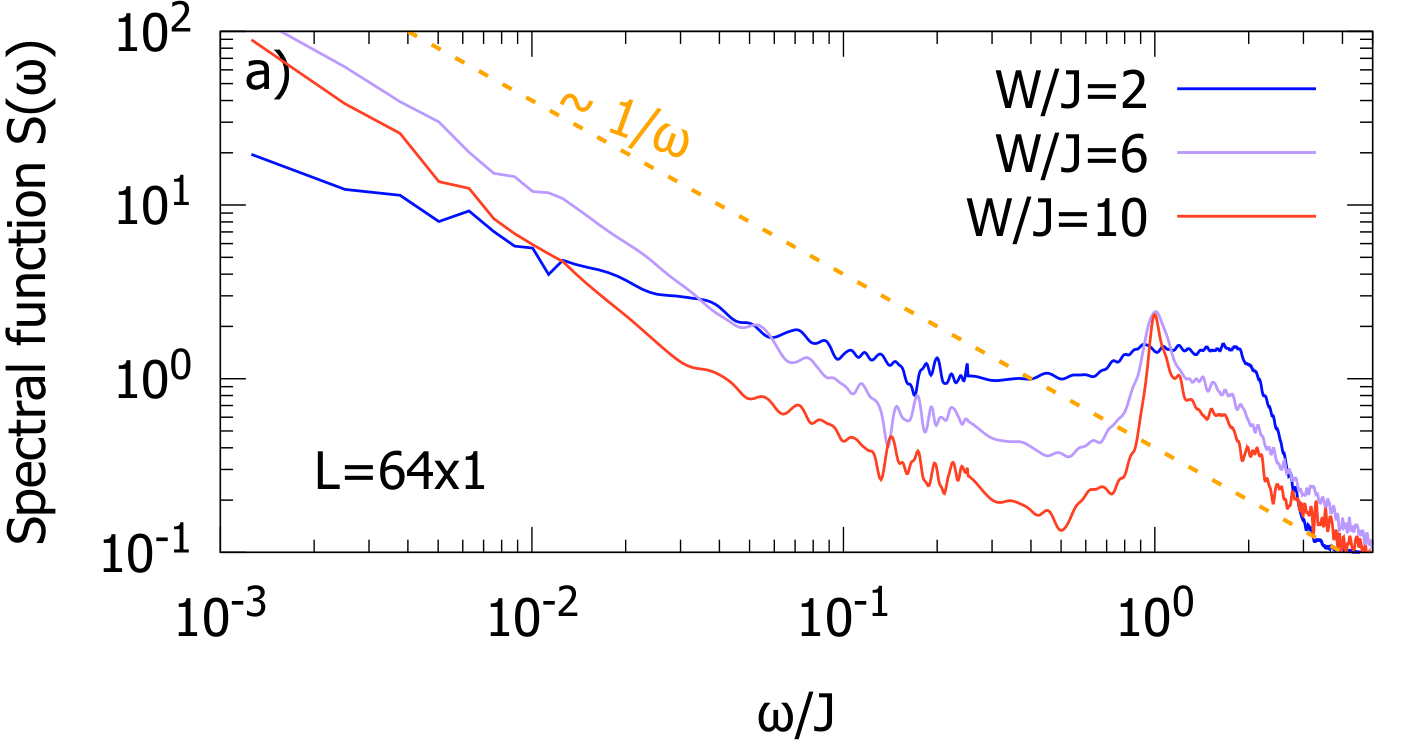}\includegraphics[scale=0.33]{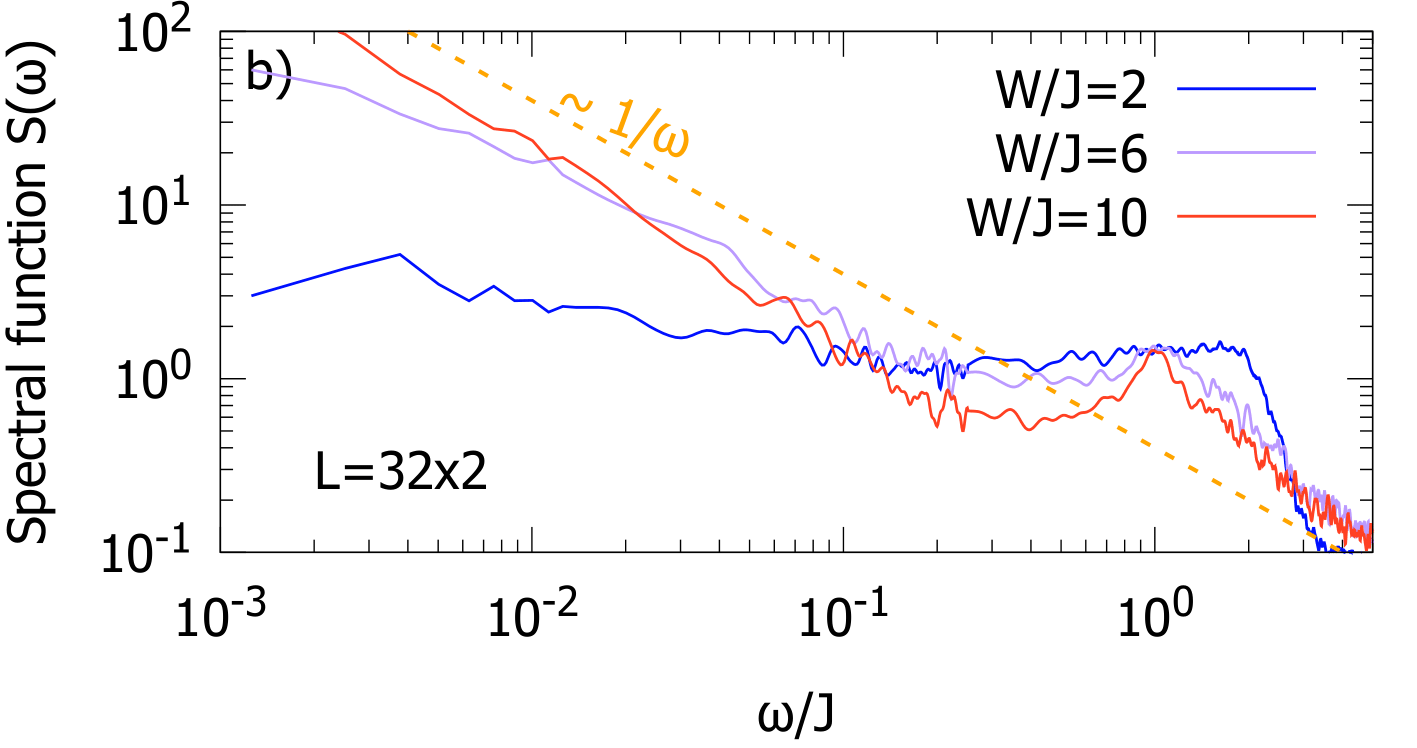}\includegraphics[scale=0.33]{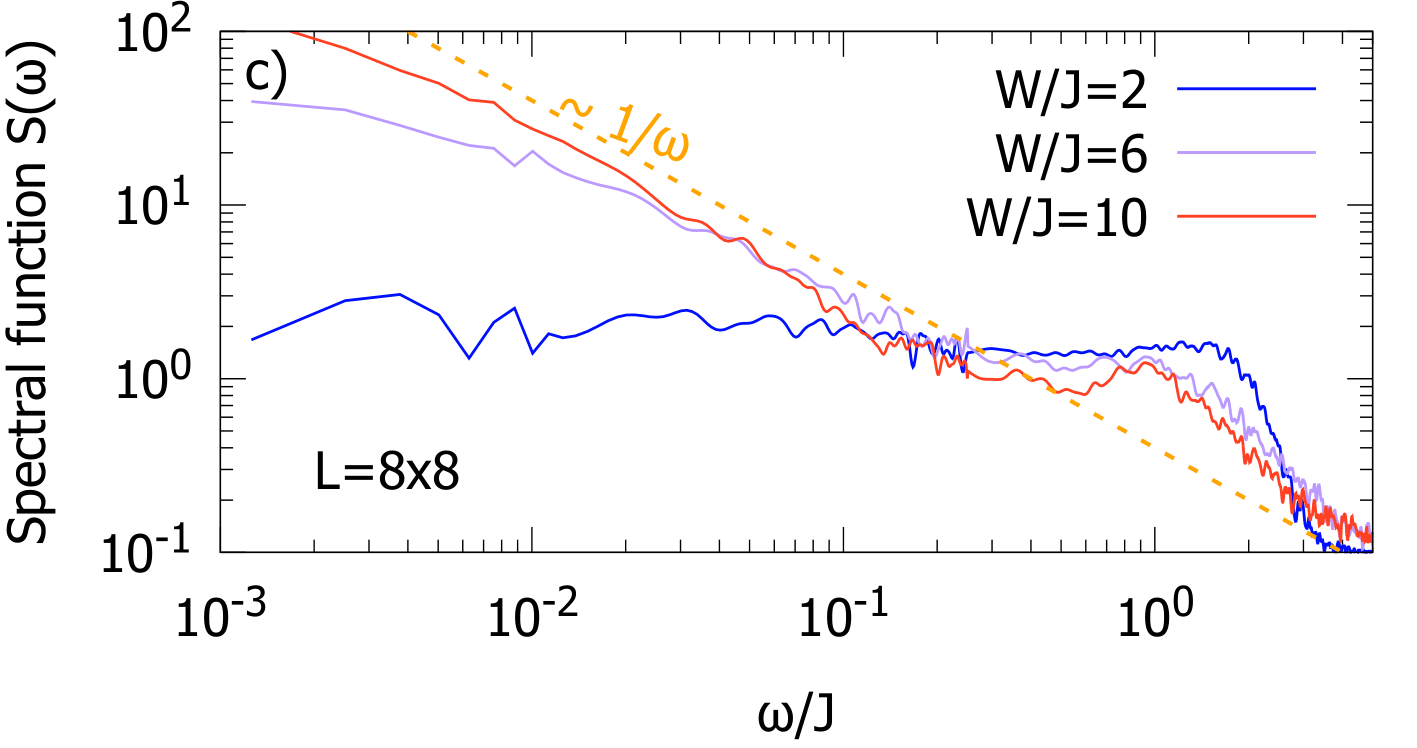}
\caption{Spectral function $S(\omega)$ of imbalance for
different lattice shapes $64\times1$ ((a) finite 1D lattice),
$32\times2$ ((b) crossover region between 1D and 2D lattice) 
and $8\times8$ ((c) finite 2D lattice). Results represent
data for $\text{fTWA}$ with disorder strength which
varies from $W/J=2$ to $W/J=10$. Each line is obtained from averaging
over 30 disorder realizations. In $\text{fTWA}_{\text{LOC}}$ 25 trajectories
were simulated for each disorder realization. The rest of the parameters are
$V/J=1$, $J=0.5$. \label{fig: 1 over f crossover ver 2}}
\end{figure*}

\begin{figure}[t]
\includegraphics[scale=0.47]{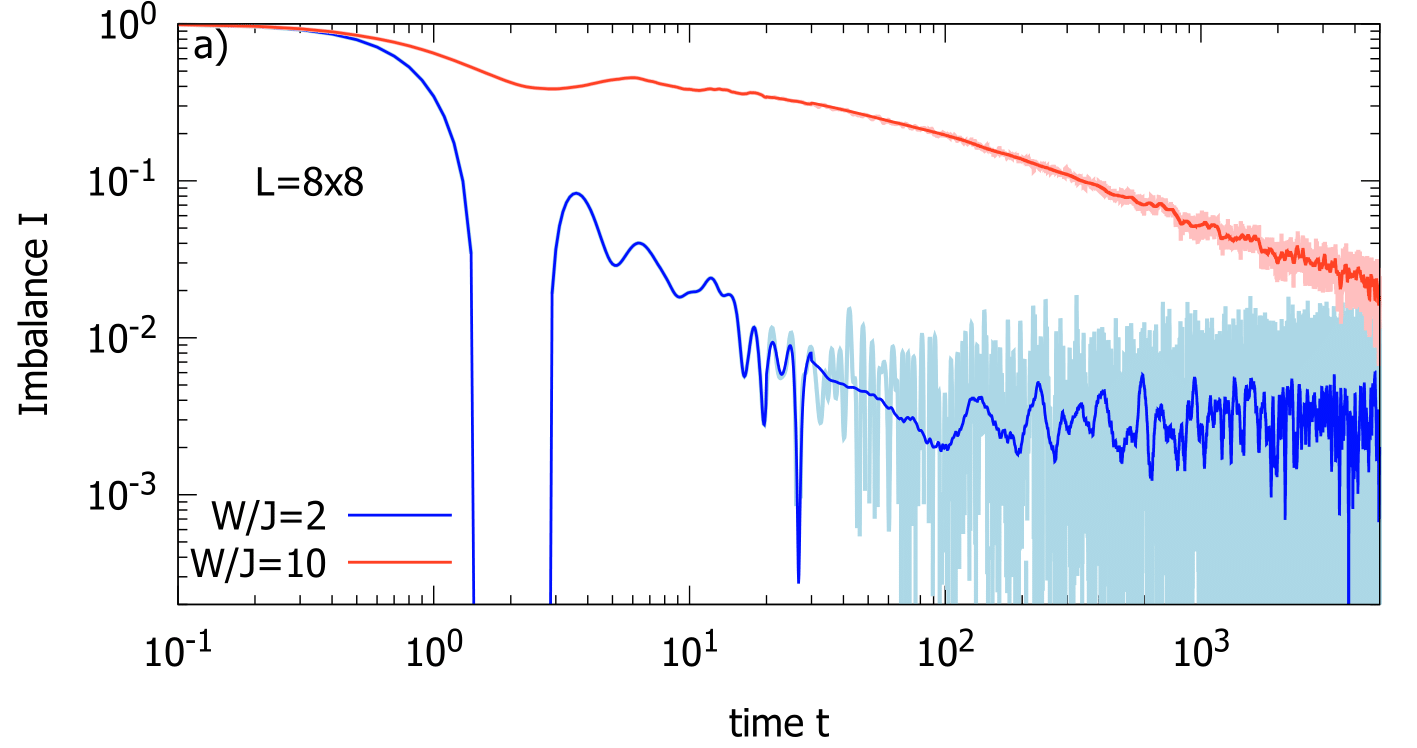}
\includegraphics[scale=0.47]{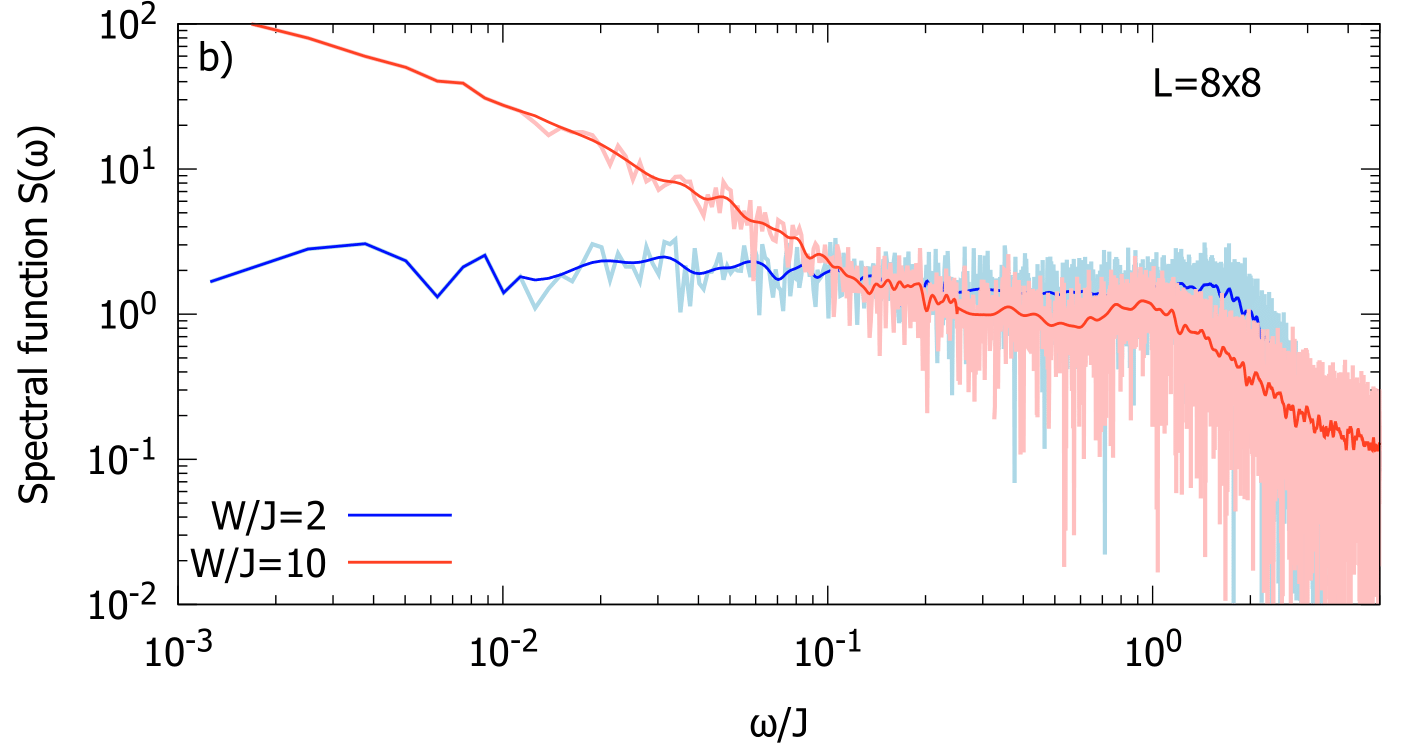}
\caption{Example of data for imbalance $I$ and spectral function $S(\omega)$  with and without filtering of the noise. (a) correspond to data in the inset of Fig. \ref{fig: 1 over f crossover} a and (b) correspond to the data from Fig. \ref{fig: 1 over f crossover ver 2} c.} \label{fig: noise-filter}
\end{figure}

\begin{figure*}[t]
\includegraphics[scale=0.49]{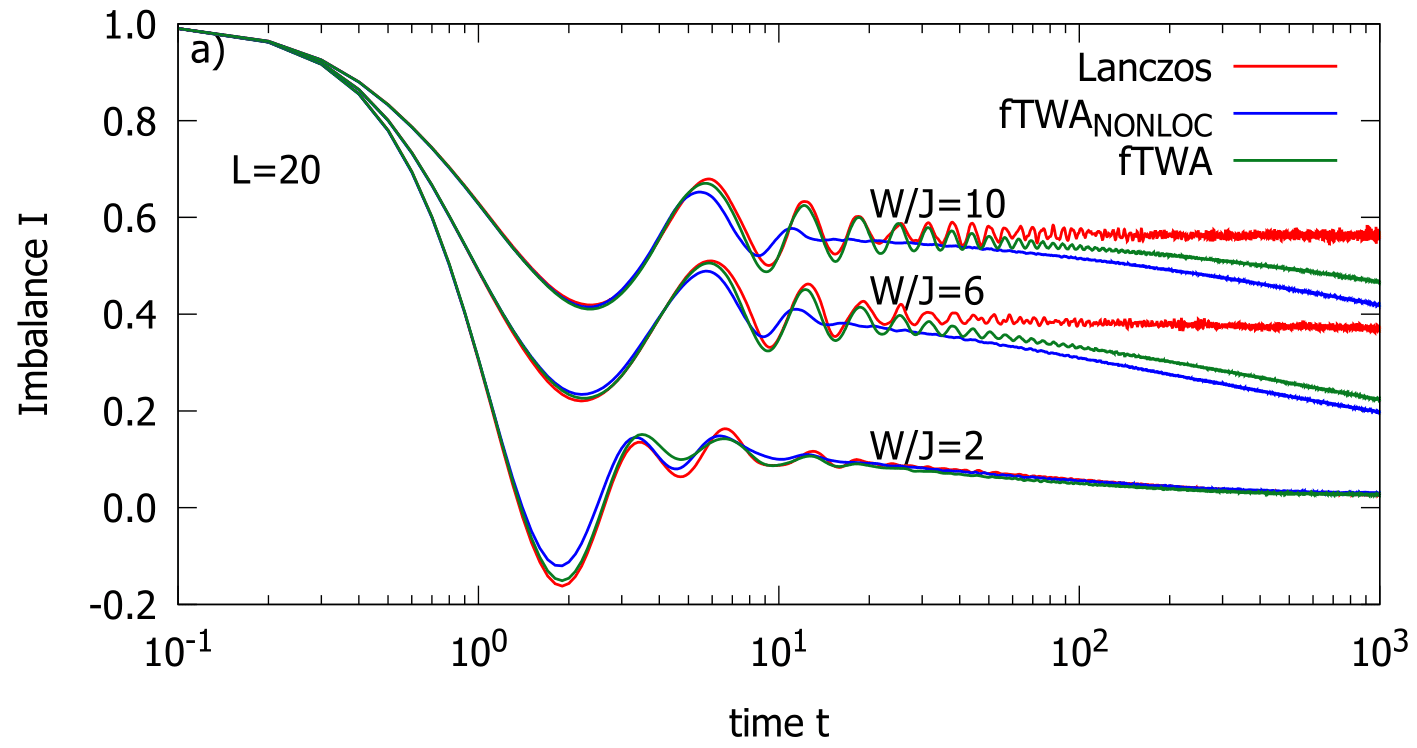} \includegraphics[scale=0.49]{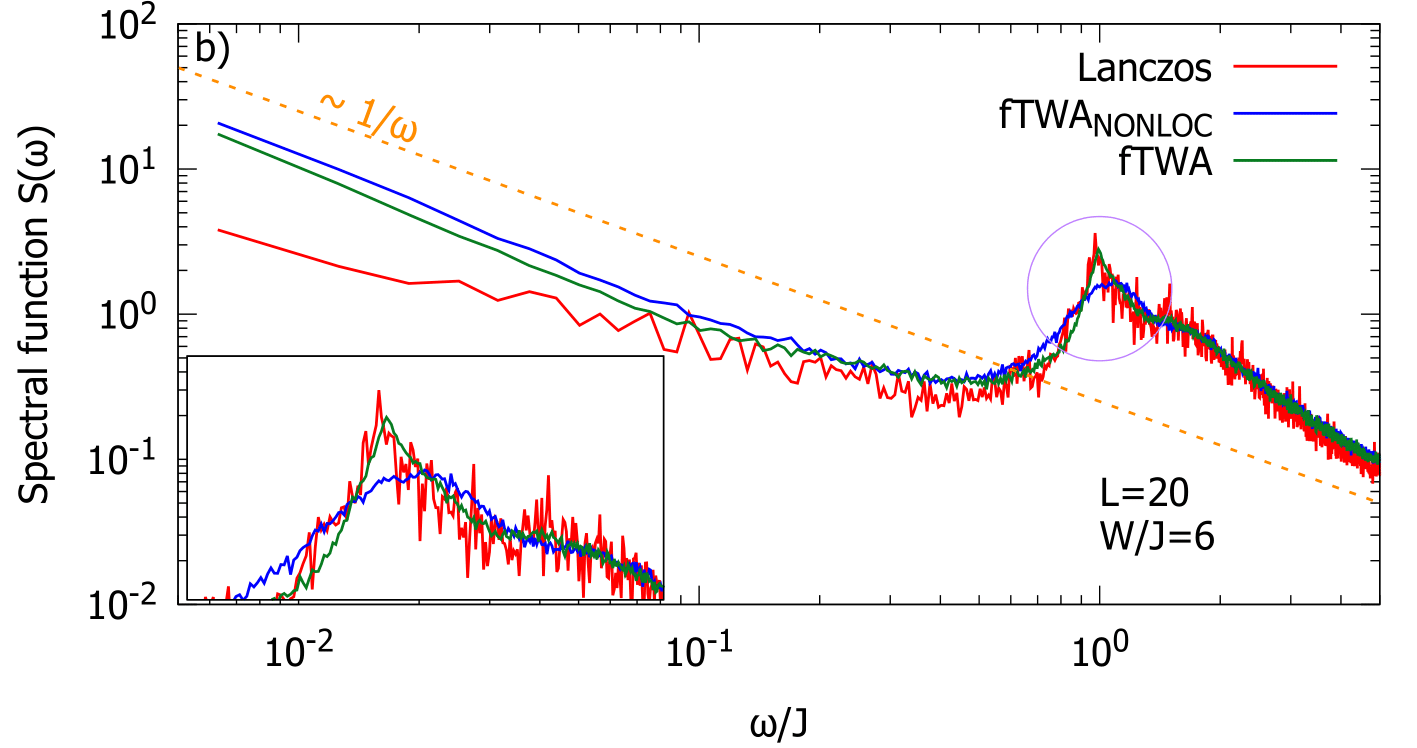}
\caption{Comparison of $\text{fTWA}$ and fTWA$_{\text{NONLOC}}$ for time-dependent imbalance $I(t)$ (a) and its spectral function $S(\omega)$ (b). There are 500 trajectories in both fTWA and fTWA$_{\text{NONLOC}}$ (green and blue line, respectively). Inset in (b) shows a sharp peak around $\omega/J=1$ frequency. The rest of the parameters on (a) and (b) correspond to Fig. \ref{fig: time fTWA vs ED} and Fig \ref{fig: 1 over f in 1D}b, respectively.  \label{fig: nonloc - time}}
\end{figure*}

\begin{figure*}[t]
\includegraphics[scale=0.5]{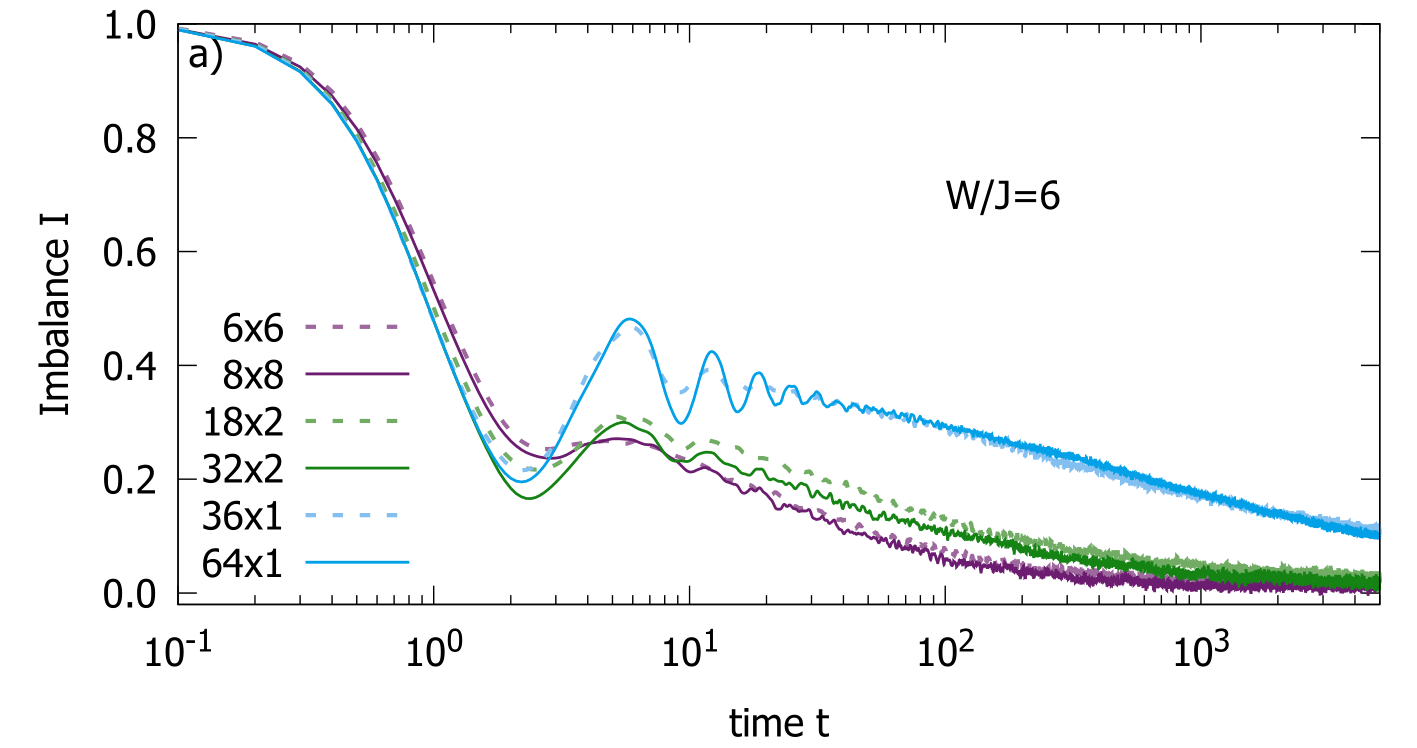}\includegraphics[scale=0.5]{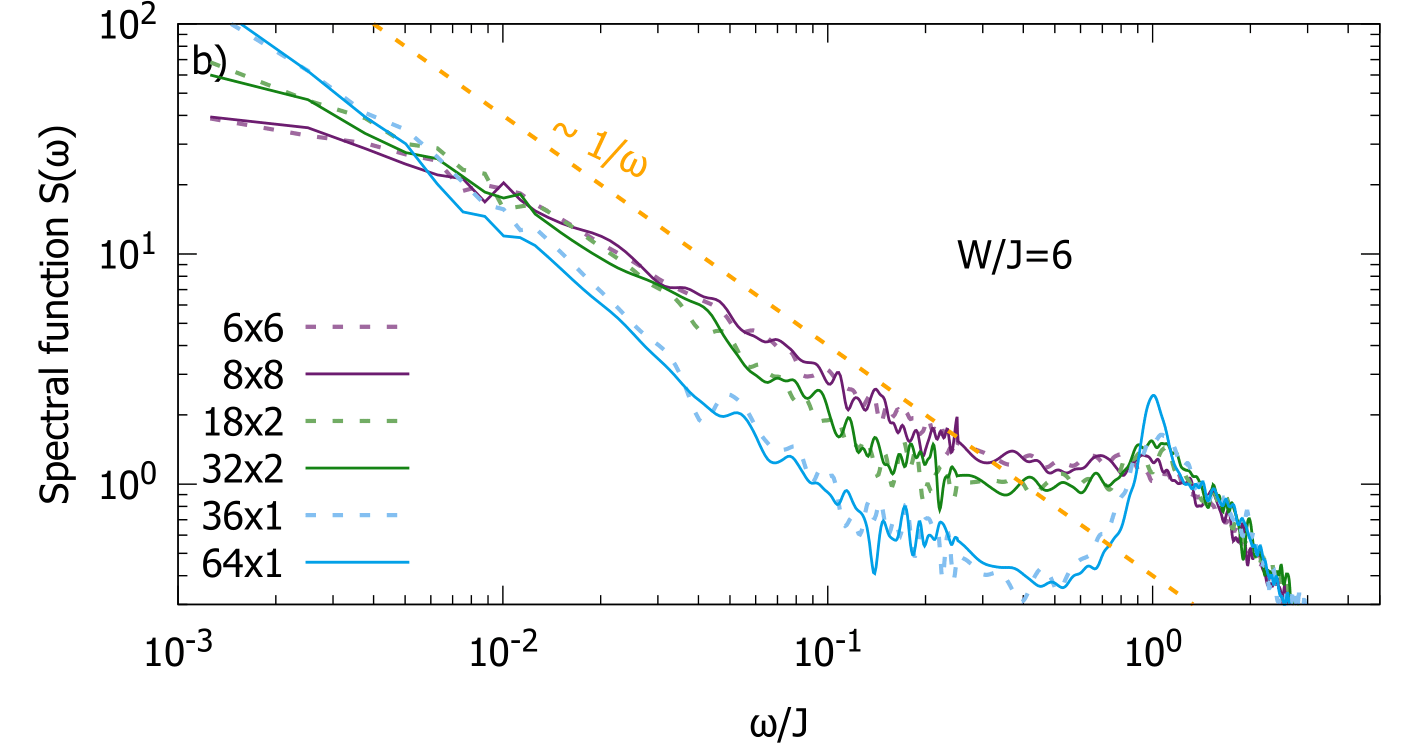}
\caption{Finite-size effects in fTWA for time-dependent imbalance (a) and spectral
function $S(\omega)$ (b) for different lattice topologies with 64 lattice sites and 36 lattice sites. Presented data represents lattices with shapes: $6\times6$, $18\times2$ and $36\times1$ where $W/J=6$ disorder strength was used. Each simulated curve is averaged over 75 disorder realizations. In fTWA there are 25 trajectories for
each disorder realization. The rest of the parameters are the same as in Fig. \ref{fig: 1 over f crossover}. \label{fig: finite-size}.}
\end{figure*}

Moreover, analyzing finite size effects of $S(\omega)$ in Fig.~\ref{fig: 1 over f in 1D}~d,  we observe that in Lanczos simulations spectral function weakly drifts to $1/\omega$ while the change in fTWA is negligible.  For a more accurate comparison,  lattice sites at boundaries were removed symmetrically when system sizes were decreased from $L=20$ to $L=12$ without changing the disorder distribution of the remaining sites.

As the main result of this work, we focus on the 2D systems in which fTWA
is capable of simulating larger system sizes. We consider the initial
product state of densities in the form of stripes (see inset in Fig.
\ref{fig: 1 over f in 2D}~a). Such stripe-like structures are directly
accessible in experiments \citep{PhysRevLett.116.140401,bordia2017_1}.
In this work we simulate numerically $8\times8$ lattice sites in
the long time limit, see Fig. \ref{fig: 1 over f in 2D}. As in the
1D system, we also observe logarithmic-in-time decay
of imbalances which is also reflected in its spectral functions as
$1/\omega$ dependence (Fig. \ref{fig: 1 over f in 2D} b). Due to the higher value of the coordination number in a 2D
lattice we do not observe a sharp resonant feature around $\omega/J\approx1$.

To compare the crossover region between the 1D and 2D system
we simulated $64\times1$, $32\times2$, $8\times8$ lattice sizes (see also finite size effects analysis in the Appendix \ref{appendix}).
Data obtained for imbalances together with their spectral functions
are plotted in Fig. \ref{fig: 1 over f crossover}. To more efficiently compare data for different lattices, the disorder strength in 2D is set to two times larger value than in 1D. We observe that for the weaker disorder strengths (Fig.~\ref{fig: 1 over f crossover}~a)  decay of imbalance at long times follows diffusive behavior, i.e., $I \sim 1/t^{0.5}$ in 1D and  $I \sim 1/t$ in 2D. Therefore $1/\omega$ behavior is  naturally not achieved in the spectral function analysis presented in Fig. \ref{fig: 1 over f crossover} b. However, for the larger strengths of disorder, we observe that $1/\omega$ behavior emerges and is immune to the shape of the lattice (see, Fig. \ref{fig: 1 over f crossover} d). This confirms the universal $1/\omega$ behavior for interacting, strongly disordered systems. 

Interestingly, obtained data also show that it is enough to consider ladder-type lattice
to observe almost two-dimensional behavior of spectral function (compare Fig.
\ref{fig: 1 over f crossover ver 2} b and c). However, proper scaling of disorder strength with lattice dimension shows that the role of dimensionality is limited, see  Fig. \ref{fig: 1 over f crossover} b and d. 

In this section all presented data for $S(\omega)$ starting from Fig. \ref{fig: 1 over f in 2D} and data in the inset of Fig. \ref{fig: 1 over f crossover} have been passed through a Kaiser filter to remove some of the noise coming mostly from sampling of the initial Wigner function (for comparison of data with and without a filtered noise see Fig. \ref{fig: noise-filter}).

\section{Summary} \label{summary}
In this work we analyze the slow dynamics of spinless interacting fermions on one and two-dimensional lattices with disorder.  Using fTWA and exact simulations we show that fTWA method gives an upper bound on the relaxations rates at single disorder realization. We exploit this method to analyze fermionic dynamics for up to 64 lattice sites at half filling and for times of order $\mathcal{O}(10^3)$,  obtaining bounds on the quantum dynamics of the system. 

Moreover, we demonstrate that fTWA exhibits $1/\omega$ behavior of the spectral function  in one and two-dimension suggesting the universality of such behavior in strongly disordered systems. The $1/\omega$ feature is a footprint of logarithmic-in-time imbalance decay which was previously observed also in one-dimensional disorder systems~\cite{mierzejewski2016, serbyn2017, sels2020, vidmar2021}.
Analyzing the spectral functions  in 1D systems we observe that upon increasing the system size, results from the Lanczos method  drift to the fTWA results while the size-dependence in fTWA is negligible. The origin of this unexpected feature remain an open problem. It deserves further investigations also for other quantum system.

\section{Acknowledgments} \label{acknowledgments}
A.P. acknowledges support from NSF under Grant DMR- 2103658, the AFOSR under Grant  FA9550-21-1-0342.  D.S.  was partially supported by AFOSR: Grant FA9550-21-1-0236.  Flatiron Institute is a division of the Simons Foundation. A.S.S. acknowledges the funding from the Polish Ministry of Science and Higher Education through a ''Mobilno\'{s}\'{c} Plus'' program nr 1651/MOB/V/2017/0. 
 \L.I and M.M. acknowledge  support by the National Science Centre, Poland via project 2020/37/B/ST3/00020.
Numerical studies in this work have been carried out using resources provided by the Wroclaw Centre for Networking and Supercomputing  \footnote{http://wcss.pl}, Grant No. 551.

\section{Appendix: Phase space representation of spinless fermionic Hamiltonian} \label{appendix-hamiltonian}
We check that adding to the Hamiltonian $\hat{H}$ effective local
interaction term $V\sum_{i}(\hat{n}_{i}-1/2)^{2}$ (Eq. \ref{eq: hamiltonian}),
leads to the significant improvement of fTWA at early times up to
order $\mathcal{O}(10)$. At later times improvement of fTWA is also
visible. We present these results in Fig. \ref{fig: nonloc - time} a in which results without local interaction term are denoted by $\text{fTWA}{}_{\text{NONLOC}}$. The explanation of this behavior
is the following. In Ref. \cite{PhysRevA.102.033338} it was shown that for the long-range
interacting model, fTWA dynamics can be significantly improved by including
local interaction term between the same fermions species. This is
because the semiclassical dynamics becomes exact in the long-range
limit, only if this term is explicitly included in the equations of
motion. Formally it means that the term $V\sum_{i}(\hat{n}_{i}-1/2)^{2}$, which
is irrelevant in the exact dynamics (because it is proportional to the
total number of particles), have to be implemented in the phase space
description as $V\sum_{i}\rho_{ii}^{2}$. Such a term
introduces nonlinearity in the equations of motion needed for recovery
of exact long-range behavior within fTWA.

Moreover, one can also give an alternative explanation of $V\sum_{i}\rho_{ii}^{2}$ term presence in the semiclassical representation. Namely, such a term naturally appears in the su(N) invariant Hubbard model and it has been recently studied in the large-N flavor limit within fTWA \cite{2007.05063}.

In the end, it is also worth stressing that $\text{fTWA}$ perfectly recovers imbalance
oscillations at initial times, which are seen as sharp peak around $\omega/J=1$  \citep{PhysRevB.93.235151}
and which do not appear in the standard $\text{fTWA}{}_{\text{NONLOC}}$ description (see, Fig. \ref{fig: nonloc - time} a and b).

\section{Appendix: Finite-size effects} \label{appendix}
In Sec. \ref{bound} and \ref{spectral} we analyze imbalance function and its Fourier transform for the system at half filling with 64 lattice sites. In order to analyze finite-size effects we compare dimensional crossover  with simulations for 32 lattice sites in Fig. \ref{fig: finite-size}. We conclude that for disorder strength $W/J=6$, finite-size effects have a small impact on the analyzed dynamics in this manuscript. We expect that the finite-size effect can be more important for weaker disorder strength. However main results of our work concern strong disorder so we omit this analysis.

\bibliographystyle{apsrev4-1}
\bibliography{library.bib,ref_mbl.bib}

\begin{thebibliography}{92}%
\makeatletter
\providecommand \@ifxundefined [1]{%
 \@ifx{#1\undefined}
}%
\providecommand \@ifnum [1]{%
 \ifnum #1\expandafter \@firstoftwo
 \else \expandafter \@secondoftwo
 \fi
}%
\providecommand \@ifx [1]{%
 \ifx #1\expandafter \@firstoftwo
 \else \expandafter \@secondoftwo
 \fi
}%
\providecommand \natexlab [1]{#1}%
\providecommand \enquote  [1]{``#1''}%
\providecommand \bibnamefont  [1]{#1}%
\providecommand \bibfnamefont [1]{#1}%
\providecommand \citenamefont [1]{#1}%
\providecommand \href@noop [0]{\@secondoftwo}%
\providecommand \href [0]{\begingroup \@sanitize@url \@href}%
\providecommand \@href[1]{\@@startlink{#1}\@@href}%
\providecommand \@@href[1]{\endgroup#1\@@endlink}%
\providecommand \@sanitize@url [0]{\catcode `\\12\catcode `\$12\catcode
  `\&12\catcode `\#12\catcode `\^12\catcode `\_12\catcode `\%12\relax}%
\providecommand \@@startlink[1]{}%
\providecommand \@@endlink[0]{}%
\providecommand \url  [0]{\begingroup\@sanitize@url \@url }%
\providecommand \@url [1]{\endgroup\@href {#1}{\urlprefix }}%
\providecommand \urlprefix  [0]{URL }%
\providecommand \Eprint [0]{\href }%
\providecommand \doibase [0]{http://dx.doi.org/}%
\providecommand \selectlanguage [0]{\@gobble}%
\providecommand \bibinfo  [0]{\@secondoftwo}%
\providecommand \bibfield  [0]{\@secondoftwo}%
\providecommand \translation [1]{[#1]}%
\providecommand \BibitemOpen [0]{}%
\providecommand \bibitemStop [0]{}%
\providecommand \bibitemNoStop [0]{.\EOS\space}%
\providecommand \EOS [0]{\spacefactor3000\relax}%
\providecommand \BibitemShut  [1]{\csname bibitem#1\endcsname}%
\let\auto@bib@innerbib\@empty
%</preamble>
\bibitem [{\citenamefont {Basko}\ \emph {et~al.}(2006)\citenamefont {Basko},
  \citenamefont {Aleiner},\ and\ \citenamefont {Altshuler}}]{basko06}%
  \BibitemOpen
  \bibfield  {author} {\bibinfo {author} {\bibfnamefont {D.}~\bibnamefont
  {Basko}}, \bibinfo {author} {\bibfnamefont {I.}~\bibnamefont {Aleiner}}, \
  and\ \bibinfo {author} {\bibfnamefont {B.}~\bibnamefont {Altshuler}},\ }\href
  {\doibase 10.1016/j.aop.2005.11.014} {\bibfield  {journal} {\bibinfo
  {journal} {Ann. Phys.}\ }\textbf {\bibinfo {volume} {321}},\ \bibinfo {pages}
  {1126} (\bibinfo {year} {2006})}\BibitemShut {NoStop}%
\bibitem [{\citenamefont {Oganesyan}\ and\ \citenamefont
  {Huse}(2007)}]{oganesyan07}%
  \BibitemOpen
  \bibfield  {author} {\bibinfo {author} {\bibfnamefont {V.}~\bibnamefont
  {Oganesyan}}\ and\ \bibinfo {author} {\bibfnamefont {D.~A.}\ \bibnamefont
  {Huse}},\ }\href {\doibase 10.1103/PhysRevB.75.155111} {\bibfield  {journal}
  {\bibinfo  {journal} {Phys. Rev. B}\ }\textbf {\bibinfo {volume} {75}},\
  \bibinfo {pages} {155111} (\bibinfo {year} {2007})}\BibitemShut {NoStop}%
\bibitem [{\citenamefont {Monthus}\ and\ \citenamefont
  {Garel}(2010)}]{monthus10}%
  \BibitemOpen
  \bibfield  {author} {\bibinfo {author} {\bibfnamefont {C.}~\bibnamefont
  {Monthus}}\ and\ \bibinfo {author} {\bibfnamefont {T.}~\bibnamefont
  {Garel}},\ }\href {\doibase 10.1103/PhysRevB.81.134202} {\bibfield  {journal}
  {\bibinfo  {journal} {Phys. Rev. B}\ }\textbf {\bibinfo {volume} {81}},\
  \bibinfo {pages} {134202} (\bibinfo {year} {2010})}\BibitemShut {NoStop}%
\bibitem [{\citenamefont {Luitz}\ \emph {et~al.}(2015)\citenamefont {Luitz},
  \citenamefont {Laflorencie},\ and\ \citenamefont {Alet}}]{luitz15}%
  \BibitemOpen
  \bibfield  {author} {\bibinfo {author} {\bibfnamefont {D.~J.}\ \bibnamefont
  {Luitz}}, \bibinfo {author} {\bibfnamefont {N.}~\bibnamefont {Laflorencie}},
  \ and\ \bibinfo {author} {\bibfnamefont {F.}~\bibnamefont {Alet}},\ }\href
  {\doibase 10.1103/PhysRevB.91.081103} {\bibfield  {journal} {\bibinfo
  {journal} {Phys. Rev. B}\ }\textbf {\bibinfo {volume} {91}},\ \bibinfo
  {pages} {081103(R)} (\bibinfo {year} {2015})}\BibitemShut {NoStop}%
\bibitem [{\citenamefont {Andraschko}\ \emph {et~al.}(2014)\citenamefont
  {Andraschko}, \citenamefont {Enss},\ and\ \citenamefont {Sirker}}]{ZZZ5_4}%
  \BibitemOpen
  \bibfield  {author} {\bibinfo {author} {\bibfnamefont {F.}~\bibnamefont
  {Andraschko}}, \bibinfo {author} {\bibfnamefont {T.}~\bibnamefont {Enss}}, \
  and\ \bibinfo {author} {\bibfnamefont {J.}~\bibnamefont {Sirker}},\ }\href
  {\doibase 10.1103/PhysRevLett.113.217201} {\bibfield  {journal} {\bibinfo
  {journal} {Phys. Rev. Lett.}\ }\textbf {\bibinfo {volume} {113}},\ \bibinfo
  {pages} {217201} (\bibinfo {year} {2014})}\BibitemShut {NoStop}%
\bibitem [{\citenamefont {Ponte}\ \emph {et~al.}(2015)\citenamefont {Ponte},
  \citenamefont {Papi\ifmmode~\acute{c}\else \'{c}\fi{}}, \citenamefont
  {Huveneers},\ and\ \citenamefont {Abanin}}]{Ponte2015}%
  \BibitemOpen
  \bibfield  {author} {\bibinfo {author} {\bibfnamefont {P.}~\bibnamefont
  {Ponte}}, \bibinfo {author} {\bibfnamefont {Z.}~\bibnamefont
  {Papi\ifmmode~\acute{c}\else \'{c}\fi{}}}, \bibinfo {author} {\bibfnamefont
  {F.}~\bibnamefont {Huveneers}}, \ and\ \bibinfo {author} {\bibfnamefont
  {D.~A.}\ \bibnamefont {Abanin}},\ }\href {\doibase
  10.1103/PhysRevLett.114.140401} {\bibfield  {journal} {\bibinfo  {journal}
  {Phys. Rev. Lett.}\ }\textbf {\bibinfo {volume} {114}},\ \bibinfo {pages}
  {140401} (\bibinfo {year} {2015})}\BibitemShut {NoStop}%
\bibitem [{\citenamefont {Lazarides}\ \emph {et~al.}(2015)\citenamefont
  {Lazarides}, \citenamefont {Das},\ and\ \citenamefont
  {Moessner}}]{lazarides15}%
  \BibitemOpen
  \bibfield  {author} {\bibinfo {author} {\bibfnamefont {A.}~\bibnamefont
  {Lazarides}}, \bibinfo {author} {\bibfnamefont {A.}~\bibnamefont {Das}}, \
  and\ \bibinfo {author} {\bibfnamefont {R.}~\bibnamefont {Moessner}},\ }\href
  {\doibase 10.1103/PhysRevLett.115.030402} {\bibfield  {journal} {\bibinfo
  {journal} {Phys. Rev. Lett.}\ }\textbf {\bibinfo {volume} {115}},\ \bibinfo
  {pages} {030402} (\bibinfo {year} {2015})}\BibitemShut {NoStop}%
\bibitem [{\citenamefont {Vasseur}\ \emph {et~al.}(2015)\citenamefont
  {Vasseur}, \citenamefont {Parameswaran},\ and\ \citenamefont
  {Moore}}]{vasseur15a}%
  \BibitemOpen
  \bibfield  {author} {\bibinfo {author} {\bibfnamefont {R.}~\bibnamefont
  {Vasseur}}, \bibinfo {author} {\bibfnamefont {S.~A.}\ \bibnamefont
  {Parameswaran}}, \ and\ \bibinfo {author} {\bibfnamefont {J.~E.}\
  \bibnamefont {Moore}},\ }\href {\doibase 10.1103/PhysRevB.91.140202}
  {\bibfield  {journal} {\bibinfo  {journal} {Phys. Rev. B}\ }\textbf {\bibinfo
  {volume} {91}},\ \bibinfo {pages} {140202(R)} (\bibinfo {year}
  {2015})}\BibitemShut {NoStop}%
\bibitem [{\citenamefont {Serbyn}\ \emph {et~al.}(2014)\citenamefont {Serbyn},
  \citenamefont {Papi\'{c}},\ and\ \citenamefont {Abanin}}]{serbyn2014a}%
  \BibitemOpen
  \bibfield  {author} {\bibinfo {author} {\bibfnamefont {M.}~\bibnamefont
  {Serbyn}}, \bibinfo {author} {\bibfnamefont {Z.}~\bibnamefont {Papi\'{c}}}, \
  and\ \bibinfo {author} {\bibfnamefont {D.~A.}\ \bibnamefont {Abanin}},\
  }\href {\doibase 10.1103/PhysRevB.90.174302} {\bibfield  {journal} {\bibinfo
  {journal} {Phys. Rev. B}\ }\textbf {\bibinfo {volume} {90}},\ \bibinfo
  {pages} {174302} (\bibinfo {year} {2014})}\BibitemShut {NoStop}%
\bibitem [{\citenamefont {Pekker}\ \emph {et~al.}(2014)\citenamefont {Pekker},
  \citenamefont {Refael}, \citenamefont {Altman}, \citenamefont {Demler},\ and\
  \citenamefont {Oganesyan}}]{pekker2014}%
  \BibitemOpen
  \bibfield  {author} {\bibinfo {author} {\bibfnamefont {D.}~\bibnamefont
  {Pekker}}, \bibinfo {author} {\bibfnamefont {G.}~\bibnamefont {Refael}},
  \bibinfo {author} {\bibfnamefont {E.}~\bibnamefont {Altman}}, \bibinfo
  {author} {\bibfnamefont {E.}~\bibnamefont {Demler}}, \ and\ \bibinfo {author}
  {\bibfnamefont {V.}~\bibnamefont {Oganesyan}},\ }\href {\doibase
  10.1103/PhysRevX.4.011052} {\bibfield  {journal} {\bibinfo  {journal} {Phys.
  Rev. X}\ }\textbf {\bibinfo {volume} {4}},\ \bibinfo {pages} {011052}
  (\bibinfo {year} {2014})}\BibitemShut {NoStop}%
\bibitem [{\citenamefont {Torres-Herrera}\ and\ \citenamefont
  {Santos}(2015)}]{torres15}%
  \BibitemOpen
  \bibfield  {author} {\bibinfo {author} {\bibfnamefont {E.~J.}\ \bibnamefont
  {Torres-Herrera}}\ and\ \bibinfo {author} {\bibfnamefont {L.~F.}\
  \bibnamefont {Santos}},\ }\href {\doibase 10.1103/PhysRevB.92.014208}
  {\bibfield  {journal} {\bibinfo  {journal} {Phys. Rev. B}\ }\textbf {\bibinfo
  {volume} {92}},\ \bibinfo {pages} {014208} (\bibinfo {year}
  {2015})}\BibitemShut {NoStop}%
\bibitem [{\citenamefont {T\'avora}\ \emph {et~al.}(2016)\citenamefont
  {T\'avora}, \citenamefont {Torres-Herrera},\ and\ \citenamefont
  {Santos}}]{torres16}%
  \BibitemOpen
  \bibfield  {author} {\bibinfo {author} {\bibfnamefont {M.}~\bibnamefont
  {T\'avora}}, \bibinfo {author} {\bibfnamefont {E.~J.}\ \bibnamefont
  {Torres-Herrera}}, \ and\ \bibinfo {author} {\bibfnamefont {L.~F.}\
  \bibnamefont {Santos}},\ }\href {\doibase 10.1103/PhysRevA.94.041603}
  {\bibfield  {journal} {\bibinfo  {journal} {Phys. Rev. A}\ }\textbf {\bibinfo
  {volume} {94}},\ \bibinfo {pages} {041603(R)} (\bibinfo {year}
  {2016})}\BibitemShut {NoStop}%
\bibitem [{\citenamefont {Laumann}\ \emph {et~al.}(2014)\citenamefont
  {Laumann}, \citenamefont {Pal},\ and\ \citenamefont
  {Scardicchio}}]{laumann2015}%
  \BibitemOpen
  \bibfield  {author} {\bibinfo {author} {\bibfnamefont {C.~R.}\ \bibnamefont
  {Laumann}}, \bibinfo {author} {\bibfnamefont {A.}~\bibnamefont {Pal}}, \ and\
  \bibinfo {author} {\bibfnamefont {A.}~\bibnamefont {Scardicchio}},\ }\href
  {\doibase 10.1103/PhysRevLett.113.200405} {\bibfield  {journal} {\bibinfo
  {journal} {Phys. Rev. Lett.}\ }\textbf {\bibinfo {volume} {113}},\ \bibinfo
  {pages} {200405} (\bibinfo {year} {2014})}\BibitemShut {NoStop}%
\bibitem [{\citenamefont {Huse}\ \emph {et~al.}(2014)\citenamefont {Huse},
  \citenamefont {Nandkishore},\ and\ \citenamefont {Oganesyan}}]{huse14}%
  \BibitemOpen
  \bibfield  {author} {\bibinfo {author} {\bibfnamefont {D.~A.}\ \bibnamefont
  {Huse}}, \bibinfo {author} {\bibfnamefont {R.}~\bibnamefont {Nandkishore}}, \
  and\ \bibinfo {author} {\bibfnamefont {V.}~\bibnamefont {Oganesyan}},\ }\href
  {\doibase 10.1103/PhysRevB.90.174202} {\bibfield  {journal} {\bibinfo
  {journal} {Phys. Rev. B}\ }\textbf {\bibinfo {volume} {90}},\ \bibinfo
  {pages} {174202} (\bibinfo {year} {2014})}\BibitemShut {NoStop}%
\bibitem [{\citenamefont {Gopalakrishnan}\ \emph {et~al.}(2017)\citenamefont
  {Gopalakrishnan}, \citenamefont {Islam},\ and\ \citenamefont
  {Knap}}]{gopal17}%
  \BibitemOpen
  \bibfield  {author} {\bibinfo {author} {\bibfnamefont {S.}~\bibnamefont
  {Gopalakrishnan}}, \bibinfo {author} {\bibfnamefont {K.~R.}\ \bibnamefont
  {Islam}}, \ and\ \bibinfo {author} {\bibfnamefont {M.}~\bibnamefont {Knap}},\
  }\href {\doibase 10.1103/PhysRevLett.119.046601} {\bibfield  {journal}
  {\bibinfo  {journal} {Phys. Rev. Lett.}\ }\textbf {\bibinfo {volume} {119}},\
  \bibinfo {pages} {046601} (\bibinfo {year} {2017})}\BibitemShut {NoStop}%
\bibitem [{\citenamefont {Hauschild}\ \emph {et~al.}(2016)\citenamefont
  {Hauschild}, \citenamefont {Heidrich-Meisner},\ and\ \citenamefont
  {Pollmann}}]{Hauschild_2016}%
  \BibitemOpen
  \bibfield  {author} {\bibinfo {author} {\bibfnamefont {J.}~\bibnamefont
  {Hauschild}}, \bibinfo {author} {\bibfnamefont {F.}~\bibnamefont
  {Heidrich-Meisner}}, \ and\ \bibinfo {author} {\bibfnamefont
  {F.}~\bibnamefont {Pollmann}},\ }\href {\doibase 10.1103/physrevb.94.161109}
  {\bibfield  {journal} {\bibinfo  {journal} {Physical Review B}\ }\textbf
  {\bibinfo {volume} {94}},\ \bibinfo {pages} {161109(R)} (\bibinfo {year}
  {2016})}\BibitemShut {NoStop}%
\bibitem [{\citenamefont {Herbrych}\ \emph {et~al.}(2013)\citenamefont
  {Herbrych}, \citenamefont {Kokalj},\ and\ \citenamefont
  {Prelov\ifmmode~\check{s}\else \v{s}\fi{}ek}}]{herbrych13}%
  \BibitemOpen
  \bibfield  {author} {\bibinfo {author} {\bibfnamefont {J.}~\bibnamefont
  {Herbrych}}, \bibinfo {author} {\bibfnamefont {J.}~\bibnamefont {Kokalj}}, \
  and\ \bibinfo {author} {\bibfnamefont {P.}~\bibnamefont
  {Prelov\ifmmode~\check{s}\else \v{s}\fi{}ek}},\ }\href {\doibase
  10.1103/PhysRevLett.111.147203} {\bibfield  {journal} {\bibinfo  {journal}
  {Phys. Rev. Lett.}\ }\textbf {\bibinfo {volume} {111}},\ \bibinfo {pages}
  {147203} (\bibinfo {year} {2013})}\BibitemShut {NoStop}%
\bibitem [{\citenamefont {Imbrie}(2016)}]{imbrie16}%
  \BibitemOpen
  \bibfield  {author} {\bibinfo {author} {\bibfnamefont {J.~Z.}\ \bibnamefont
  {Imbrie}},\ }\href {\doibase 10.1103/PhysRevLett.117.027201} {\bibfield
  {journal} {\bibinfo  {journal} {Phys. Rev. Lett.}\ }\textbf {\bibinfo
  {volume} {117}},\ \bibinfo {pages} {027201} (\bibinfo {year}
  {2016})}\BibitemShut {NoStop}%
\bibitem [{\citenamefont {Steinigeweg}\ \emph {et~al.}(2016)\citenamefont
  {Steinigeweg}, \citenamefont {Herbrych}, \citenamefont {Pollmann},\ and\
  \citenamefont {Brenig}}]{steinigeweg16}%
  \BibitemOpen
  \bibfield  {author} {\bibinfo {author} {\bibfnamefont {R.}~\bibnamefont
  {Steinigeweg}}, \bibinfo {author} {\bibfnamefont {J.}~\bibnamefont
  {Herbrych}}, \bibinfo {author} {\bibfnamefont {F.}~\bibnamefont {Pollmann}},
  \ and\ \bibinfo {author} {\bibfnamefont {W.}~\bibnamefont {Brenig}},\ }\href
  {\doibase 10.1103/PhysRevB.94.180401} {\bibfield  {journal} {\bibinfo
  {journal} {Phys. Rev. B}\ }\textbf {\bibinfo {volume} {94}},\ \bibinfo
  {pages} {180401(R)} (\bibinfo {year} {2016})}\BibitemShut {NoStop}%
\bibitem [{\citenamefont {Herbrych}\ and\ \citenamefont
  {Kokalj}(2017)}]{Herbrych17}%
  \BibitemOpen
  \bibfield  {author} {\bibinfo {author} {\bibfnamefont {J.}~\bibnamefont
  {Herbrych}}\ and\ \bibinfo {author} {\bibfnamefont {J.}~\bibnamefont
  {Kokalj}},\ }\href {\doibase 10.1103/PhysRevB.95.125129} {\bibfield
  {journal} {\bibinfo  {journal} {Phys. Rev. B}\ }\textbf {\bibinfo {volume}
  {95}},\ \bibinfo {pages} {125129} (\bibinfo {year} {2017})}\BibitemShut
  {NoStop}%
\bibitem [{\citenamefont {Panda}\ \emph {et~al.}(2020)\citenamefont {Panda},
  \citenamefont {Scardicchio}, \citenamefont {Schulz}, \citenamefont {Taylor},\
  and\ \citenamefont {{\v{Z}}nidari{\v{c}}}}]{Panda2020}%
  \BibitemOpen
  \bibfield  {author} {\bibinfo {author} {\bibfnamefont {R.~K.}\ \bibnamefont
  {Panda}}, \bibinfo {author} {\bibfnamefont {A.}~\bibnamefont {Scardicchio}},
  \bibinfo {author} {\bibfnamefont {M.}~\bibnamefont {Schulz}}, \bibinfo
  {author} {\bibfnamefont {S.~R.}\ \bibnamefont {Taylor}}, \ and\ \bibinfo
  {author} {\bibfnamefont {M.}~\bibnamefont {{\v{Z}}nidari{\v{c}}}},\ }\href
  {\doibase 10.1209/0295-5075/128/67003} {\bibfield  {journal} {\bibinfo
  {journal} {{EPL} (Europhysics Letters)}\ }\textbf {\bibinfo {volume} {128}},\
  \bibinfo {pages} {67003} (\bibinfo {year} {2020})}\BibitemShut {NoStop}%
\bibitem [{\citenamefont {Sierant}\ \emph
  {et~al.}(2020{\natexlab{a}})\citenamefont {Sierant}, \citenamefont
  {Delande},\ and\ \citenamefont {Zakrzewski}}]{Sierant2020}%
  \BibitemOpen
  \bibfield  {author} {\bibinfo {author} {\bibfnamefont {P.}~\bibnamefont
  {Sierant}}, \bibinfo {author} {\bibfnamefont {D.}~\bibnamefont {Delande}}, \
  and\ \bibinfo {author} {\bibfnamefont {J.}~\bibnamefont {Zakrzewski}},\
  }\href {\doibase 10.1103/PhysRevLett.124.186601} {\bibfield  {journal}
  {\bibinfo  {journal} {Phys. Rev. Lett.}\ }\textbf {\bibinfo {volume} {124}},\
  \bibinfo {pages} {186601} (\bibinfo {year} {2020}{\natexlab{a}})}\BibitemShut
  {NoStop}%
\bibitem [{\citenamefont {Sierant}\ \emph
  {et~al.}(2020{\natexlab{b}})\citenamefont {Sierant}, \citenamefont
  {Lewenstein},\ and\ \citenamefont {Zakrzewski}}]{sierant_lewenstein_20}%
  \BibitemOpen
  \bibfield  {author} {\bibinfo {author} {\bibfnamefont {P.}~\bibnamefont
  {Sierant}}, \bibinfo {author} {\bibfnamefont {M.}~\bibnamefont {Lewenstein}},
  \ and\ \bibinfo {author} {\bibfnamefont {J.}~\bibnamefont {Zakrzewski}},\
  }\href {\doibase 10.1103/PhysRevLett.125.156601} {\bibfield  {journal}
  {\bibinfo  {journal} {Phys. Rev. Lett.}\ }\textbf {\bibinfo {volume} {125}},\
  \bibinfo {pages} {156601} (\bibinfo {year} {2020}{\natexlab{b}})}\BibitemShut
  {NoStop}%
\bibitem [{\citenamefont {Morningstar}\ \emph {et~al.}(2022)\citenamefont
  {Morningstar}, \citenamefont {Colmenarez}, \citenamefont {Khemani},
  \citenamefont {Luitz},\ and\ \citenamefont {Huse}}]{Morningstar2022}%
  \BibitemOpen
  \bibfield  {author} {\bibinfo {author} {\bibfnamefont {A.}~\bibnamefont
  {Morningstar}}, \bibinfo {author} {\bibfnamefont {L.}~\bibnamefont
  {Colmenarez}}, \bibinfo {author} {\bibfnamefont {V.}~\bibnamefont {Khemani}},
  \bibinfo {author} {\bibfnamefont {D.~J.}\ \bibnamefont {Luitz}}, \ and\
  \bibinfo {author} {\bibfnamefont {D.~A.}\ \bibnamefont {Huse}},\ }\href
  {\doibase 10.1103/PhysRevB.105.174205} {\bibfield  {journal} {\bibinfo
  {journal} {Phys. Rev. B}\ }\textbf {\bibinfo {volume} {105}},\ \bibinfo
  {pages} {174205} (\bibinfo {year} {2022})}\BibitemShut {NoStop}%
\bibitem [{\citenamefont {Abanin}\ \emph {et~al.}(2021)\citenamefont {Abanin},
  \citenamefont {Bardarson}, \citenamefont {{De Tomasi}}, \citenamefont
  {Gopalakrishnan}, \citenamefont {Khemani}, \citenamefont {Parameswaran},
  \citenamefont {Pollmann}, \citenamefont {Potter}, \citenamefont {Serbyn},\
  and\ \citenamefont {Vasseur}}]{abanin_bardarson_21}%
  \BibitemOpen
  \bibfield  {author} {\bibinfo {author} {\bibfnamefont {D.}~\bibnamefont
  {Abanin}}, \bibinfo {author} {\bibfnamefont {J.}~\bibnamefont {Bardarson}},
  \bibinfo {author} {\bibfnamefont {G.}~\bibnamefont {{De Tomasi}}}, \bibinfo
  {author} {\bibfnamefont {S.}~\bibnamefont {Gopalakrishnan}}, \bibinfo
  {author} {\bibfnamefont {V.}~\bibnamefont {Khemani}}, \bibinfo {author}
  {\bibfnamefont {S.}~\bibnamefont {Parameswaran}}, \bibinfo {author}
  {\bibfnamefont {F.}~\bibnamefont {Pollmann}}, \bibinfo {author}
  {\bibfnamefont {A.}~\bibnamefont {Potter}}, \bibinfo {author} {\bibfnamefont
  {M.}~\bibnamefont {Serbyn}}, \ and\ \bibinfo {author} {\bibfnamefont
  {R.}~\bibnamefont {Vasseur}},\ }\href {\doibase
  https://doi.org/10.1016/j.aop.2021.168415} {\bibfield  {journal} {\bibinfo
  {journal} {Annals of Physics}\ }\textbf {\bibinfo {volume} {427}},\ \bibinfo
  {pages} {168415} (\bibinfo {year} {2021})}\BibitemShut {NoStop}%
\bibitem [{\citenamefont {\v{S}untajs}\ \emph
  {et~al.}(2020{\natexlab{a}})\citenamefont {\v{S}untajs}, \citenamefont
  {Bon\v{c}a}, \citenamefont {Prosen},\ and\ \citenamefont
  {Vidmar}}]{suntajs_bonca_20a}%
  \BibitemOpen
  \bibfield  {author} {\bibinfo {author} {\bibfnamefont {J.}~\bibnamefont
  {\v{S}untajs}}, \bibinfo {author} {\bibfnamefont {J.}~\bibnamefont
  {Bon\v{c}a}}, \bibinfo {author} {\bibfnamefont {T.}~\bibnamefont {Prosen}}, \
  and\ \bibinfo {author} {\bibfnamefont {L.}~\bibnamefont {Vidmar}},\ }\href
  {\doibase 10.1103/PhysRevE.102.062144} {\bibfield  {journal} {\bibinfo
  {journal} {Phys. Rev. E}\ }\textbf {\bibinfo {volume} {102}},\ \bibinfo
  {pages} {062144} (\bibinfo {year} {2020}{\natexlab{a}})}\BibitemShut
  {NoStop}%
\bibitem [{\citenamefont {\v{S}untajs}\ \emph
  {et~al.}(2020{\natexlab{b}})\citenamefont {\v{S}untajs}, \citenamefont
  {Bon\v{c}a}, \citenamefont {Prosen},\ and\ \citenamefont
  {Vidmar}}]{suntajs_bonca_20}%
  \BibitemOpen
  \bibfield  {author} {\bibinfo {author} {\bibfnamefont {J.}~\bibnamefont
  {\v{S}untajs}}, \bibinfo {author} {\bibfnamefont {J.}~\bibnamefont
  {Bon\v{c}a}}, \bibinfo {author} {\bibfnamefont {T.}~\bibnamefont {Prosen}}, \
  and\ \bibinfo {author} {\bibfnamefont {L.}~\bibnamefont {Vidmar}},\ }\href
  {\doibase 10.1103/PhysRevB.102.064207} {\bibfield  {journal} {\bibinfo
  {journal} {Phys. Rev. B}\ }\textbf {\bibinfo {volume} {102}},\ \bibinfo
  {pages} {064207} (\bibinfo {year} {2020}{\natexlab{b}})}\BibitemShut
  {NoStop}%
\bibitem [{\citenamefont {Sels}\ and\ \citenamefont
  {Polkovnikov}(2021{\natexlab{a}})}]{sels2020}%
  \BibitemOpen
  \bibfield  {author} {\bibinfo {author} {\bibfnamefont {D.}~\bibnamefont
  {Sels}}\ and\ \bibinfo {author} {\bibfnamefont {A.}~\bibnamefont
  {Polkovnikov}},\ }\href {\doibase 10.1103/PhysRevE.104.054105} {\bibfield
  {journal} {\bibinfo  {journal} {Phys. Rev. E}\ }\textbf {\bibinfo {volume}
  {104}},\ \bibinfo {pages} {054105} (\bibinfo {year}
  {2021}{\natexlab{a}})}\BibitemShut {NoStop}%
\bibitem [{\citenamefont {Sels}(2022)}]{Sels_2022}%
  \BibitemOpen
  \bibfield  {author} {\bibinfo {author} {\bibfnamefont {D.}~\bibnamefont
  {Sels}},\ }\href {\doibase 10.1103/PhysRevB.106.L020202} {\bibfield
  {journal} {\bibinfo  {journal} {Phys. Rev. B}\ }\textbf {\bibinfo {volume}
  {106}},\ \bibinfo {pages} {L020202} (\bibinfo {year} {2022})}\BibitemShut
  {NoStop}%
\bibitem [{\citenamefont {Sels}\ and\ \citenamefont
  {Polkovnikov}(2021{\natexlab{b}})}]{Sels_dilute_2021}%
  \BibitemOpen
  \bibfield  {author} {\bibinfo {author} {\bibfnamefont {D.}~\bibnamefont
  {Sels}}\ and\ \bibinfo {author} {\bibfnamefont {A.}~\bibnamefont
  {Polkovnikov}},\ }\href {\doibase 10.48550/ARXIV.2105.09348} {\enquote
  {\bibinfo {title} {Thermalization of dilute impurities in one dimensional
  spin chains},}\ } (\bibinfo {year} {2021}{\natexlab{b}})\BibitemShut
  {NoStop}%
\bibitem [{\citenamefont {\v{Z}nidari\v{c}}\ \emph {et~al.}(2008)\citenamefont
  {\v{Z}nidari\v{c}}, \citenamefont {Prosen},\ and\ \citenamefont
  {Prelov\ifmmode~\check{s}\else \v{s}\fi{}ek}}]{znidaric08}%
  \BibitemOpen
  \bibfield  {author} {\bibinfo {author} {\bibfnamefont {M.}~\bibnamefont
  {\v{Z}nidari\v{c}}}, \bibinfo {author} {\bibfnamefont {T.}~\bibnamefont
  {Prosen}}, \ and\ \bibinfo {author} {\bibfnamefont {P.}~\bibnamefont
  {Prelov\ifmmode~\check{s}\else \v{s}\fi{}ek}},\ }\href {\doibase
  10.1103/PhysRevB.77.064426} {\bibfield  {journal} {\bibinfo  {journal} {Phys.
  Rev. B}\ }\textbf {\bibinfo {volume} {77}},\ \bibinfo {pages} {064426}
  (\bibinfo {year} {2008})}\BibitemShut {NoStop}%
\bibitem [{\citenamefont {Bardarson}\ \emph {et~al.}(2012)\citenamefont
  {Bardarson}, \citenamefont {Pollmann},\ and\ \citenamefont
  {Moore}}]{bardarson12}%
  \BibitemOpen
  \bibfield  {author} {\bibinfo {author} {\bibfnamefont {J.~H.}\ \bibnamefont
  {Bardarson}}, \bibinfo {author} {\bibfnamefont {F.}~\bibnamefont {Pollmann}},
  \ and\ \bibinfo {author} {\bibfnamefont {J.~E.}\ \bibnamefont {Moore}},\
  }\href {\doibase 10.1103/PhysRevLett.109.017202} {\bibfield  {journal}
  {\bibinfo  {journal} {Phys. Rev. Lett.}\ }\textbf {\bibinfo {volume} {109}},\
  \bibinfo {pages} {017202} (\bibinfo {year} {2012})}\BibitemShut {NoStop}%
\bibitem [{\citenamefont {Kj{\"{a}}ll}\ \emph {et~al.}(2014)\citenamefont
  {Kj{\"{a}}ll}, \citenamefont {Bardarson},\ and\ \citenamefont
  {Pollmann}}]{kjall14}%
  \BibitemOpen
  \bibfield  {author} {\bibinfo {author} {\bibfnamefont {J.~A.}\ \bibnamefont
  {Kj{\"{a}}ll}}, \bibinfo {author} {\bibfnamefont {J.~H.}\ \bibnamefont
  {Bardarson}}, \ and\ \bibinfo {author} {\bibfnamefont {F.}~\bibnamefont
  {Pollmann}},\ }\href {\doibase 10.1103/PhysRevLett.113.107204} {\bibfield
  {journal} {\bibinfo  {journal} {Phys. Rev. Lett.}\ }\textbf {\bibinfo
  {volume} {113}},\ \bibinfo {pages} {107204} (\bibinfo {year}
  {2014})}\BibitemShut {NoStop}%
\bibitem [{\citenamefont {Serbyn}\ \emph {et~al.}(2015)\citenamefont {Serbyn},
  \citenamefont {Papi{\'{c}}},\ and\ \citenamefont {Abanin}}]{serbyn15}%
  \BibitemOpen
  \bibfield  {author} {\bibinfo {author} {\bibfnamefont {M.}~\bibnamefont
  {Serbyn}}, \bibinfo {author} {\bibfnamefont {Z.}~\bibnamefont {Papi{\'{c}}}},
  \ and\ \bibinfo {author} {\bibfnamefont {D.~A.}\ \bibnamefont {Abanin}},\
  }\href {\doibase 10.1103/PhysRevX.5.041047} {\bibfield  {journal} {\bibinfo
  {journal} {Phys. Rev. X}\ }\textbf {\bibinfo {volume} {5}},\ \bibinfo {pages}
  {041047} (\bibinfo {year} {2015})}\BibitemShut {NoStop}%
\bibitem [{\citenamefont {Luitz}\ \emph {et~al.}(2016)\citenamefont {Luitz},
  \citenamefont {Laflorencie},\ and\ \citenamefont {Alet}}]{luitz16}%
  \BibitemOpen
  \bibfield  {author} {\bibinfo {author} {\bibfnamefont {D.~J.}\ \bibnamefont
  {Luitz}}, \bibinfo {author} {\bibfnamefont {N.}~\bibnamefont {Laflorencie}},
  \ and\ \bibinfo {author} {\bibfnamefont {F.}~\bibnamefont {Alet}},\ }\href
  {\doibase 10.1103/PhysRevB.93.060201} {\bibfield  {journal} {\bibinfo
  {journal} {Phys. Rev. B}\ }\textbf {\bibinfo {volume} {93}},\ \bibinfo
  {pages} {060201(R)} (\bibinfo {year} {2016})}\BibitemShut {NoStop}%
\bibitem [{\citenamefont {Serbyn}\ \emph {et~al.}(2013)\citenamefont {Serbyn},
  \citenamefont {Papi\'{c}},\ and\ \citenamefont {Abanin}}]{serbyn13_1}%
  \BibitemOpen
  \bibfield  {author} {\bibinfo {author} {\bibfnamefont {M.}~\bibnamefont
  {Serbyn}}, \bibinfo {author} {\bibfnamefont {Z.}~\bibnamefont {Papi\'{c}}}, \
  and\ \bibinfo {author} {\bibfnamefont {D.~A.}\ \bibnamefont {Abanin}},\
  }\href {\doibase 10.1103/PhysRevLett.110.260601} {\bibfield  {journal}
  {\bibinfo  {journal} {Phys. Rev. Lett.}\ }\textbf {\bibinfo {volume} {110}},\
  \bibinfo {pages} {260601} (\bibinfo {year} {2013})}\BibitemShut {NoStop}%
\bibitem [{\citenamefont {Bera}\ \emph {et~al.}(2015)\citenamefont {Bera},
  \citenamefont {Schomerus}, \citenamefont {Heidrich-Meisner},\ and\
  \citenamefont {Bardarson}}]{bera15}%
  \BibitemOpen
  \bibfield  {author} {\bibinfo {author} {\bibfnamefont {S.}~\bibnamefont
  {Bera}}, \bibinfo {author} {\bibfnamefont {H.}~\bibnamefont {Schomerus}},
  \bibinfo {author} {\bibfnamefont {F.}~\bibnamefont {Heidrich-Meisner}}, \
  and\ \bibinfo {author} {\bibfnamefont {J.~H.}\ \bibnamefont {Bardarson}},\
  }\href {\doibase 10.1103/PhysRevLett.115.046603} {\bibfield  {journal}
  {\bibinfo  {journal} {Phys. Rev. Lett.}\ }\textbf {\bibinfo {volume} {115}},\
  \bibinfo {pages} {046603} (\bibinfo {year} {2015})}\BibitemShut {NoStop}%
\bibitem [{\citenamefont {Altman}\ and\ \citenamefont {Vosk}(2015)}]{altman15}%
  \BibitemOpen
  \bibfield  {author} {\bibinfo {author} {\bibfnamefont {E.}~\bibnamefont
  {Altman}}\ and\ \bibinfo {author} {\bibfnamefont {R.}~\bibnamefont {Vosk}},\
  }\href {\doibase 10.1146/annurev-conmatphys-031214-014701} {\bibfield
  {journal} {\bibinfo  {journal} {Annu. Rev. Condens. Matter Phys.}\ }\textbf
  {\bibinfo {volume} {6}},\ \bibinfo {pages} {383} (\bibinfo {year}
  {2015})}\BibitemShut {NoStop}%
\bibitem [{\citenamefont {Agarwal}\ \emph {et~al.}(2015)\citenamefont
  {Agarwal}, \citenamefont {Gopalakrishnan}, \citenamefont {Knap},
  \citenamefont {M\"uller},\ and\ \citenamefont {Demler}}]{agarwal15}%
  \BibitemOpen
  \bibfield  {author} {\bibinfo {author} {\bibfnamefont {K.}~\bibnamefont
  {Agarwal}}, \bibinfo {author} {\bibfnamefont {S.}~\bibnamefont
  {Gopalakrishnan}}, \bibinfo {author} {\bibfnamefont {M.}~\bibnamefont
  {Knap}}, \bibinfo {author} {\bibfnamefont {M.}~\bibnamefont {M\"uller}}, \
  and\ \bibinfo {author} {\bibfnamefont {E.}~\bibnamefont {Demler}},\ }\href
  {\doibase 10.1103/PhysRevLett.114.160401} {\bibfield  {journal} {\bibinfo
  {journal} {Phys. Rev. Lett.}\ }\textbf {\bibinfo {volume} {114}},\ \bibinfo
  {pages} {160401} (\bibinfo {year} {2015})}\BibitemShut {NoStop}%
\bibitem [{\citenamefont {Gopalakrishnan}\ \emph {et~al.}(2015)\citenamefont
  {Gopalakrishnan}, \citenamefont {M\"uller}, \citenamefont {Khemani},
  \citenamefont {Knap}, \citenamefont {Demler},\ and\ \citenamefont
  {Huse}}]{gopal15}%
  \BibitemOpen
  \bibfield  {author} {\bibinfo {author} {\bibfnamefont {S.}~\bibnamefont
  {Gopalakrishnan}}, \bibinfo {author} {\bibfnamefont {M.}~\bibnamefont
  {M\"uller}}, \bibinfo {author} {\bibfnamefont {V.}~\bibnamefont {Khemani}},
  \bibinfo {author} {\bibfnamefont {M.}~\bibnamefont {Knap}}, \bibinfo {author}
  {\bibfnamefont {E.}~\bibnamefont {Demler}}, \ and\ \bibinfo {author}
  {\bibfnamefont {D.~A.}\ \bibnamefont {Huse}},\ }\href {\doibase
  10.1103/PhysRevB.92.104202} {\bibfield  {journal} {\bibinfo  {journal} {Phys.
  Rev. B}\ }\textbf {\bibinfo {volume} {92}},\ \bibinfo {pages} {104202}
  (\bibinfo {year} {2015})}\BibitemShut {NoStop}%
\bibitem [{\citenamefont {{\v Znidari\v c}}\ \emph {et~al.}(2016)\citenamefont
  {{\v Znidari\v c}}, \citenamefont {Scardicchio},\ and\ \citenamefont
  {Varma}}]{znidaric16}%
  \BibitemOpen
  \bibfield  {author} {\bibinfo {author} {\bibfnamefont {M.}~\bibnamefont {{\v
  Znidari\v c}}}, \bibinfo {author} {\bibfnamefont {A.}~\bibnamefont
  {Scardicchio}}, \ and\ \bibinfo {author} {\bibfnamefont {V.~K.}\ \bibnamefont
  {Varma}},\ }\href {http://dx.doi.org/10.1103/PhysRevLett.117.040601}
  {\bibfield  {journal} {\bibinfo  {journal} {Phys. Rev. Lett.}\ }\textbf
  {\bibinfo {volume} {117}},\ \bibinfo {pages} {040601} (\bibinfo {year}
  {2016})}\BibitemShut {NoStop}%
\bibitem [{\citenamefont {Mierzejewski}\ \emph {et~al.}(2016)\citenamefont
  {Mierzejewski}, \citenamefont {Herbrych},\ and\ \citenamefont
  {Prelov\ifmmode~\check{s}\else \v{s}\fi{}ek}}]{mierzejewski2016}%
  \BibitemOpen
  \bibfield  {author} {\bibinfo {author} {\bibfnamefont {M.}~\bibnamefont
  {Mierzejewski}}, \bibinfo {author} {\bibfnamefont {J.}~\bibnamefont
  {Herbrych}}, \ and\ \bibinfo {author} {\bibfnamefont {P.}~\bibnamefont
  {Prelov\ifmmode~\check{s}\else \v{s}\fi{}ek}},\ }\href {\doibase
  10.1103/PhysRevB.94.224207} {\bibfield  {journal} {\bibinfo  {journal} {Phys.
  Rev. B}\ }\textbf {\bibinfo {volume} {94}},\ \bibinfo {pages} {224207}
  (\bibinfo {year} {2016})}\BibitemShut {NoStop}%
\bibitem [{\citenamefont {Bar~Lev}\ and\ \citenamefont
  {Reichman}(2014)}]{lev14}%
  \BibitemOpen
  \bibfield  {author} {\bibinfo {author} {\bibfnamefont {Y.}~\bibnamefont
  {Bar~Lev}}\ and\ \bibinfo {author} {\bibfnamefont {D.~R.}\ \bibnamefont
  {Reichman}},\ }\href {\doibase 10.1103/PhysRevB.89.220201} {\bibfield
  {journal} {\bibinfo  {journal} {Phys. Rev. B}\ }\textbf {\bibinfo {volume}
  {89}},\ \bibinfo {pages} {220201(R)} (\bibinfo {year} {2014})}\BibitemShut
  {NoStop}%
\bibitem [{\citenamefont {Bar~Lev}\ \emph {et~al.}(2015)\citenamefont
  {Bar~Lev}, \citenamefont {Cohen},\ and\ \citenamefont {Reichman}}]{lev15}%
  \BibitemOpen
  \bibfield  {author} {\bibinfo {author} {\bibfnamefont {Y.}~\bibnamefont
  {Bar~Lev}}, \bibinfo {author} {\bibfnamefont {G.}~\bibnamefont {Cohen}}, \
  and\ \bibinfo {author} {\bibfnamefont {D.~R.}\ \bibnamefont {Reichman}},\
  }\href {\doibase 10.1103/PhysRevLett.114.100601} {\bibfield  {journal}
  {\bibinfo  {journal} {Phys. Rev. Lett.}\ }\textbf {\bibinfo {volume} {114}},\
  \bibinfo {pages} {100601} (\bibinfo {year} {2015})}\BibitemShut {NoStop}%
\bibitem [{\citenamefont {Bari\ifmmode \check{s}\else
  \v{s}\fi{}i\ifmmode~\acute{c}\else \'{c}\fi{}}\ \emph
  {et~al.}(2016)\citenamefont {Bari\ifmmode \check{s}\else
  \v{s}\fi{}i\ifmmode~\acute{c}\else \'{c}\fi{}}, \citenamefont {Kokalj},
  \citenamefont {Balog},\ and\ \citenamefont {Prelov\ifmmode~\check{s}\else
  \v{s}\fi{}ek}}]{barisic16}%
  \BibitemOpen
  \bibfield  {author} {\bibinfo {author} {\bibfnamefont {O.~S.}\ \bibnamefont
  {Bari\ifmmode \check{s}\else \v{s}\fi{}i\ifmmode~\acute{c}\else \'{c}\fi{}}},
  \bibinfo {author} {\bibfnamefont {J.}~\bibnamefont {Kokalj}}, \bibinfo
  {author} {\bibfnamefont {I.}~\bibnamefont {Balog}}, \ and\ \bibinfo {author}
  {\bibfnamefont {P.}~\bibnamefont {Prelov\ifmmode~\check{s}\else
  \v{s}\fi{}ek}},\ }\href {\doibase 10.1103/PhysRevB.94.045126} {\bibfield
  {journal} {\bibinfo  {journal} {Phys. Rev. B}\ }\textbf {\bibinfo {volume}
  {94}},\ \bibinfo {pages} {045126} (\bibinfo {year} {2016})}\BibitemShut
  {NoStop}%
\bibitem [{\citenamefont {Bon\ifmmode~\check{c}\else \v{c}\fi{}a}\ and\
  \citenamefont {Mierzejewski}(2017)}]{bonca17}%
  \BibitemOpen
  \bibfield  {author} {\bibinfo {author} {\bibfnamefont {J.}~\bibnamefont
  {Bon\ifmmode~\check{c}\else \v{c}\fi{}a}}\ and\ \bibinfo {author}
  {\bibfnamefont {M.}~\bibnamefont {Mierzejewski}},\ }\href {\doibase
  10.1103/PhysRevB.95.214201} {\bibfield  {journal} {\bibinfo  {journal} {Phys.
  Rev. B}\ }\textbf {\bibinfo {volume} {95}},\ \bibinfo {pages} {214201}
  (\bibinfo {year} {2017})}\BibitemShut {NoStop}%
\bibitem [{\citenamefont {Bordia}\ \emph {et~al.}(2017)\citenamefont {Bordia},
  \citenamefont {L\"uschen}, \citenamefont {Scherg}, \citenamefont
  {Gopalakrishnan}, \citenamefont {Knap}, \citenamefont {Schneider},\ and\
  \citenamefont {Bloch}}]{bordia2017_1}%
  \BibitemOpen
  \bibfield  {author} {\bibinfo {author} {\bibfnamefont {P.}~\bibnamefont
  {Bordia}}, \bibinfo {author} {\bibfnamefont {H.}~\bibnamefont {L\"uschen}},
  \bibinfo {author} {\bibfnamefont {S.}~\bibnamefont {Scherg}}, \bibinfo
  {author} {\bibfnamefont {S.}~\bibnamefont {Gopalakrishnan}}, \bibinfo
  {author} {\bibfnamefont {M.}~\bibnamefont {Knap}}, \bibinfo {author}
  {\bibfnamefont {U.}~\bibnamefont {Schneider}}, \ and\ \bibinfo {author}
  {\bibfnamefont {I.}~\bibnamefont {Bloch}},\ }\href {\doibase
  10.1103/PhysRevX.7.041047} {\bibfield  {journal} {\bibinfo  {journal} {Phys.
  Rev. X}\ }\textbf {\bibinfo {volume} {7}},\ \bibinfo {pages} {041047}
  (\bibinfo {year} {2017})}\BibitemShut {NoStop}%
\bibitem [{\citenamefont {Sierant}\ \emph {et~al.}(2017)\citenamefont
  {Sierant}, \citenamefont {Delande},\ and\ \citenamefont
  {Zakrzewski}}]{zakrzewski16}%
  \BibitemOpen
  \bibfield  {author} {\bibinfo {author} {\bibfnamefont {P.}~\bibnamefont
  {Sierant}}, \bibinfo {author} {\bibfnamefont {D.}~\bibnamefont {Delande}}, \
  and\ \bibinfo {author} {\bibfnamefont {J.}~\bibnamefont {Zakrzewski}},\
  }\href {\doibase 10.1103/PhysRevA.95.021601} {\bibfield  {journal} {\bibinfo
  {journal} {Phys. Rev. A}\ }\textbf {\bibinfo {volume} {95}},\ \bibinfo
  {pages} {021601(R)} (\bibinfo {year} {2017})}\BibitemShut {NoStop}%
\bibitem [{\citenamefont {Protopopov}\ and\ \citenamefont
  {Abanin}(2019)}]{protopopov2018}%
  \BibitemOpen
  \bibfield  {author} {\bibinfo {author} {\bibfnamefont {I.~V.}\ \bibnamefont
  {Protopopov}}\ and\ \bibinfo {author} {\bibfnamefont {D.~A.}\ \bibnamefont
  {Abanin}},\ }\href {\doibase 10.1103/PhysRevB.99.115111} {\bibfield
  {journal} {\bibinfo  {journal} {Phys. Rev. B}\ }\textbf {\bibinfo {volume}
  {99}},\ \bibinfo {pages} {115111} (\bibinfo {year} {2019})}\BibitemShut
  {NoStop}%
\bibitem [{\citenamefont {Schecter}\ \emph {et~al.}(2018)\citenamefont
  {Schecter}, \citenamefont {Iadecola},\ and\ \citenamefont
  {Das~Sarma}}]{sankar2018}%
  \BibitemOpen
  \bibfield  {author} {\bibinfo {author} {\bibfnamefont {M.}~\bibnamefont
  {Schecter}}, \bibinfo {author} {\bibfnamefont {T.}~\bibnamefont {Iadecola}},
  \ and\ \bibinfo {author} {\bibfnamefont {S.}~\bibnamefont {Das~Sarma}},\
  }\href {\doibase 10.1103/PhysRevB.98.174201} {\bibfield  {journal} {\bibinfo
  {journal} {Phys. Rev. B}\ }\textbf {\bibinfo {volume} {98}},\ \bibinfo
  {pages} {174201} (\bibinfo {year} {2018})}\BibitemShut {NoStop}%
\bibitem [{\citenamefont {Zakrzewski}\ and\ \citenamefont
  {Delande}(2018)}]{zakrzewski2018}%
  \BibitemOpen
  \bibfield  {author} {\bibinfo {author} {\bibfnamefont {J.}~\bibnamefont
  {Zakrzewski}}\ and\ \bibinfo {author} {\bibfnamefont {D.}~\bibnamefont
  {Delande}},\ }\href {\doibase 10.1103/PhysRevB.98.014203} {\bibfield
  {journal} {\bibinfo  {journal} {Phys. Rev. B}\ }\textbf {\bibinfo {volume}
  {98}},\ \bibinfo {pages} {014203} (\bibinfo {year} {2018})}\BibitemShut
  {NoStop}%
\bibitem [{\citenamefont {Chandran}\ \emph {et~al.}(2014)\citenamefont
  {Chandran}, \citenamefont {Khemani}, \citenamefont {Laumann},\ and\
  \citenamefont {Sondhi}}]{Chandran2014}%
  \BibitemOpen
  \bibfield  {author} {\bibinfo {author} {\bibfnamefont {A.}~\bibnamefont
  {Chandran}}, \bibinfo {author} {\bibfnamefont {V.}~\bibnamefont {Khemani}},
  \bibinfo {author} {\bibfnamefont {C.~R.}\ \bibnamefont {Laumann}}, \ and\
  \bibinfo {author} {\bibfnamefont {S.~L.}\ \bibnamefont {Sondhi}},\ }\href
  {http://journals.aps.org/prb/abstract/10.1103/PhysRevB.89.144201} {\bibfield
  {journal} {\bibinfo  {journal} {Phys. Rev. B}\ }\textbf {\bibinfo {volume}
  {89}},\ \bibinfo {pages} {144201} (\bibinfo {year} {2014})}\BibitemShut
  {NoStop}%
\bibitem [{\citenamefont {Potter}\ and\ \citenamefont
  {Vasseur}(2016)}]{potter16}%
  \BibitemOpen
  \bibfield  {author} {\bibinfo {author} {\bibfnamefont {A.~C.}\ \bibnamefont
  {Potter}}\ and\ \bibinfo {author} {\bibfnamefont {R.}~\bibnamefont
  {Vasseur}},\ }\href {\doibase 10.1103/PhysRevB.94.224206} {\bibfield
  {journal} {\bibinfo  {journal} {Phys. Rev. B}\ }\textbf {\bibinfo {volume}
  {94}},\ \bibinfo {pages} {224206} (\bibinfo {year} {2016})}\BibitemShut
  {NoStop}%
\bibitem [{\citenamefont {Prelov\ifmmode~\check{s}\else \v{s}\fi{}ek}\ \emph
  {et~al.}(2016)\citenamefont {Prelov\ifmmode~\check{s}\else \v{s}\fi{}ek},
  \citenamefont {Bari\ifmmode \check{s}\else \v{s}\fi{}i\ifmmode~\acute{c}\else
  \'{c}\fi{}},\ and\ \citenamefont {\ifmmode \check{Z}\else
  \v{Z}\fi{}nidari\ifmmode~\check{c}\else \v{c}\fi{}}}]{prelovsek16}%
  \BibitemOpen
  \bibfield  {author} {\bibinfo {author} {\bibfnamefont {P.}~\bibnamefont
  {Prelov\ifmmode~\check{s}\else \v{s}\fi{}ek}}, \bibinfo {author}
  {\bibfnamefont {O.~S.}\ \bibnamefont {Bari\ifmmode \check{s}\else
  \v{s}\fi{}i\ifmmode~\acute{c}\else \'{c}\fi{}}}, \ and\ \bibinfo {author}
  {\bibfnamefont {M.}~\bibnamefont {\ifmmode \check{Z}\else
  \v{Z}\fi{}nidari\ifmmode~\check{c}\else \v{c}\fi{}}},\ }\href {\doibase
  10.1103/PhysRevB.94.241104} {\bibfield  {journal} {\bibinfo  {journal} {Phys.
  Rev. B}\ }\textbf {\bibinfo {volume} {94}},\ \bibinfo {pages} {241104(R)}
  (\bibinfo {year} {2016})}\BibitemShut {NoStop}%
\bibitem [{\citenamefont {Protopopov}\ \emph {et~al.}(2017)\citenamefont
  {Protopopov}, \citenamefont {Ho},\ and\ \citenamefont {Abanin}}]{proto2017}%
  \BibitemOpen
  \bibfield  {author} {\bibinfo {author} {\bibfnamefont {I.~V.}\ \bibnamefont
  {Protopopov}}, \bibinfo {author} {\bibfnamefont {W.~W.}\ \bibnamefont {Ho}},
  \ and\ \bibinfo {author} {\bibfnamefont {D.~A.}\ \bibnamefont {Abanin}},\
  }\href {\doibase 10.1103/PhysRevB.96.041122} {\bibfield  {journal} {\bibinfo
  {journal} {Phys. Rev. B}\ }\textbf {\bibinfo {volume} {96}},\ \bibinfo
  {pages} {041122(R)} (\bibinfo {year} {2017})}\BibitemShut {NoStop}%
\bibitem [{\citenamefont {Friedman}\ \emph {et~al.}(2018)\citenamefont
  {Friedman}, \citenamefont {Vasseur}, \citenamefont {Potter},\ and\
  \citenamefont {Parameswaran}}]{friedman2017}%
  \BibitemOpen
  \bibfield  {author} {\bibinfo {author} {\bibfnamefont {A.~J.}\ \bibnamefont
  {Friedman}}, \bibinfo {author} {\bibfnamefont {R.}~\bibnamefont {Vasseur}},
  \bibinfo {author} {\bibfnamefont {A.~C.}\ \bibnamefont {Potter}}, \ and\
  \bibinfo {author} {\bibfnamefont {S.~A.}\ \bibnamefont {Parameswaran}},\
  }\href {\doibase 10.1103/PhysRevB.98.064203} {\bibfield  {journal} {\bibinfo
  {journal} {Phys. Rev. B}\ }\textbf {\bibinfo {volume} {98}},\ \bibinfo
  {pages} {064203} (\bibinfo {year} {2018})}\BibitemShut {NoStop}%
\bibitem [{\citenamefont {Lev}\ and\ \citenamefont
  {Reichman}(2016)}]{barlev2016}%
  \BibitemOpen
  \bibfield  {author} {\bibinfo {author} {\bibfnamefont {Y.~B.}\ \bibnamefont
  {Lev}}\ and\ \bibinfo {author} {\bibfnamefont {D.~R.}\ \bibnamefont
  {Reichman}},\ }\href {\doibase 10.1209/0295-5075/113/46001} {\bibfield
  {journal} {\bibinfo  {journal} {{EPL} (Europhysics Letters)}\ }\textbf
  {\bibinfo {volume} {113}},\ \bibinfo {pages} {46001} (\bibinfo {year}
  {2016})}\BibitemShut {NoStop}%
\bibitem [{\citenamefont {Li}\ \emph {et~al.}(2017)\citenamefont {Li},
  \citenamefont {Deng}, \citenamefont {Wu},\ and\ \citenamefont
  {Das~Sarma}}]{lisarma17}%
  \BibitemOpen
  \bibfield  {author} {\bibinfo {author} {\bibfnamefont {X.}~\bibnamefont
  {Li}}, \bibinfo {author} {\bibfnamefont {D.-L.}\ \bibnamefont {Deng}},
  \bibinfo {author} {\bibfnamefont {Y.-L.}\ \bibnamefont {Wu}}, \ and\ \bibinfo
  {author} {\bibfnamefont {S.}~\bibnamefont {Das~Sarma}},\ }\href {\doibase
  10.1103/PhysRevB.95.020201} {\bibfield  {journal} {\bibinfo  {journal} {Phys.
  Rev. B}\ }\textbf {\bibinfo {volume} {95}},\ \bibinfo {pages} {020201(R)}
  (\bibinfo {year} {2017})}\BibitemShut {NoStop}%
\bibitem [{\citenamefont {Luitz}\ and\ \citenamefont
  {Bar~Lev}(2016{\natexlab{a}})}]{luitz2016prl}%
  \BibitemOpen
  \bibfield  {author} {\bibinfo {author} {\bibfnamefont {D.~J.}\ \bibnamefont
  {Luitz}}\ and\ \bibinfo {author} {\bibfnamefont {Y.}~\bibnamefont
  {Bar~Lev}},\ }\href {\doibase 10.1103/PhysRevLett.117.170404} {\bibfield
  {journal} {\bibinfo  {journal} {Phys. Rev. Lett.}\ }\textbf {\bibinfo
  {volume} {117}},\ \bibinfo {pages} {170404} (\bibinfo {year}
  {2016}{\natexlab{a}})}\BibitemShut {NoStop}%
\bibitem [{\citenamefont {Luitz}\ and\ \citenamefont
  {Bar~Lev}(2016{\natexlab{b}})}]{luitz116}%
  \BibitemOpen
  \bibfield  {author} {\bibinfo {author} {\bibfnamefont {D.~J.}\ \bibnamefont
  {Luitz}}\ and\ \bibinfo {author} {\bibfnamefont {Y.}~\bibnamefont
  {Bar~Lev}},\ }\href {\doibase 10.1002/andp.201600350} {\bibfield  {journal}
  {\bibinfo  {journal} {Annalen der Physik}\ }\textbf {\bibinfo {volume}
  {529}},\ \bibinfo {pages} {1600350} (\bibinfo {year}
  {2016}{\natexlab{b}})}\BibitemShut {NoStop}%
\bibitem [{\citenamefont {Kozarzewski}\ \emph {et~al.}(2018)\citenamefont
  {Kozarzewski}, \citenamefont {Prelov\ifmmode~\check{s}\else \v{s}\fi{}ek},\
  and\ \citenamefont {Mierzejewski}}]{kozarzewski18}%
  \BibitemOpen
  \bibfield  {author} {\bibinfo {author} {\bibfnamefont {M.}~\bibnamefont
  {Kozarzewski}}, \bibinfo {author} {\bibfnamefont {P.}~\bibnamefont
  {Prelov\ifmmode~\check{s}\else \v{s}\fi{}ek}}, \ and\ \bibinfo {author}
  {\bibfnamefont {M.}~\bibnamefont {Mierzejewski}},\ }\href {\doibase
  10.1103/PhysRevLett.120.246602} {\bibfield  {journal} {\bibinfo  {journal}
  {Phys. Rev. Lett.}\ }\textbf {\bibinfo {volume} {120}},\ \bibinfo {pages}
  {246602} (\bibinfo {year} {2018})}\BibitemShut {NoStop}%
\bibitem [{\citenamefont {Prelov\ifmmode~\check{s}\else \v{s}\fi{}ek}\ and\
  \citenamefont {Herbrych}(2017)}]{prelovsek217}%
  \BibitemOpen
  \bibfield  {author} {\bibinfo {author} {\bibfnamefont {P.}~\bibnamefont
  {Prelov\ifmmode~\check{s}\else \v{s}\fi{}ek}}\ and\ \bibinfo {author}
  {\bibfnamefont {J.}~\bibnamefont {Herbrych}},\ }\href {\doibase
  10.1103/PhysRevB.96.035130} {\bibfield  {journal} {\bibinfo  {journal} {Phys.
  Rev. B}\ }\textbf {\bibinfo {volume} {96}},\ \bibinfo {pages} {035130}
  (\bibinfo {year} {2017})}\BibitemShut {NoStop}%
\bibitem [{\citenamefont {Lev}\ \emph {et~al.}(2017)\citenamefont {Lev},
  \citenamefont {Kennes}, \citenamefont {Kl{\"o}ckner}, \citenamefont
  {Reichman},\ and\ \citenamefont {Karrasch}}]{new_karrasch}%
  \BibitemOpen
  \bibfield  {author} {\bibinfo {author} {\bibfnamefont {Y.~B.}\ \bibnamefont
  {Lev}}, \bibinfo {author} {\bibfnamefont {D.~M.}\ \bibnamefont {Kennes}},
  \bibinfo {author} {\bibfnamefont {C.}~\bibnamefont {Kl{\"o}ckner}}, \bibinfo
  {author} {\bibfnamefont {D.~R.}\ \bibnamefont {Reichman}}, \ and\ \bibinfo
  {author} {\bibfnamefont {C.}~\bibnamefont {Karrasch}},\ }\href
  {http://stacks.iop.org/0295-5075/119/i=3/a=37003} {\bibfield  {journal}
  {\bibinfo  {journal} {EPL (Europhysics Letters)}\ }\textbf {\bibinfo {volume}
  {119}},\ \bibinfo {pages} {37003} (\bibinfo {year} {2017})}\BibitemShut
  {NoStop}%
\bibitem [{\citenamefont {Prelov\ifmmode~\check{s}\else \v{s}\fi{}ek}\ \emph
  {et~al.}(2018)\citenamefont {Prelov\ifmmode~\check{s}\else \v{s}\fi{}ek},
  \citenamefont {Bon\ifmmode~\check{c}\else \v{c}\fi{}a},\ and\ \citenamefont
  {Mierzejewski}}]{prelovsek2018a}%
  \BibitemOpen
  \bibfield  {author} {\bibinfo {author} {\bibfnamefont {P.}~\bibnamefont
  {Prelov\ifmmode~\check{s}\else \v{s}\fi{}ek}}, \bibinfo {author}
  {\bibfnamefont {J.}~\bibnamefont {Bon\ifmmode~\check{c}\else \v{c}\fi{}a}}, \
  and\ \bibinfo {author} {\bibfnamefont {M.}~\bibnamefont {Mierzejewski}},\
  }\href {\doibase 10.1103/PhysRevB.98.125119} {\bibfield  {journal} {\bibinfo
  {journal} {Phys. Rev. B}\ }\textbf {\bibinfo {volume} {98}},\ \bibinfo
  {pages} {125119} (\bibinfo {year} {2018})}\BibitemShut {NoStop}%
\bibitem [{\citenamefont {Agarwal}\ \emph {et~al.}(2016)\citenamefont
  {Agarwal}, \citenamefont {Altman}, \citenamefont {Demler}, \citenamefont
  {Gopalakrishnan}, \citenamefont {Huse},\ and\ \citenamefont
  {Knap}}]{agarwal16}%
  \BibitemOpen
  \bibfield  {author} {\bibinfo {author} {\bibfnamefont {K.}~\bibnamefont
  {Agarwal}}, \bibinfo {author} {\bibfnamefont {E.}~\bibnamefont {Altman}},
  \bibinfo {author} {\bibfnamefont {E.}~\bibnamefont {Demler}}, \bibinfo
  {author} {\bibfnamefont {S.}~\bibnamefont {Gopalakrishnan}}, \bibinfo
  {author} {\bibfnamefont {D.~A.}\ \bibnamefont {Huse}}, \ and\ \bibinfo
  {author} {\bibfnamefont {M.}~\bibnamefont {Knap}},\ }\href {\doibase
  10.1002/andp.201600326} {\bibfield  {journal} {\bibinfo  {journal} {Annalen
  der Physik}\ }\textbf {\bibinfo {volume} {529}},\ \bibinfo {pages} {1600326}
  (\bibinfo {year} {2016})}\BibitemShut {NoStop}%
\bibitem [{\citenamefont {L\"uschen}\ \emph
  {et~al.}(2017{\natexlab{a}})\citenamefont {L\"uschen}, \citenamefont
  {Bordia}, \citenamefont {Scherg}, \citenamefont {Alet}, \citenamefont
  {Altman}, \citenamefont {Schneider},\ and\ \citenamefont
  {Bloch}}]{luschen17}%
  \BibitemOpen
  \bibfield  {author} {\bibinfo {author} {\bibfnamefont {H.~P.}\ \bibnamefont
  {L\"uschen}}, \bibinfo {author} {\bibfnamefont {P.}~\bibnamefont {Bordia}},
  \bibinfo {author} {\bibfnamefont {S.}~\bibnamefont {Scherg}}, \bibinfo
  {author} {\bibfnamefont {F.}~\bibnamefont {Alet}}, \bibinfo {author}
  {\bibfnamefont {E.}~\bibnamefont {Altman}}, \bibinfo {author} {\bibfnamefont
  {U.}~\bibnamefont {Schneider}}, \ and\ \bibinfo {author} {\bibfnamefont
  {I.}~\bibnamefont {Bloch}},\ }\href {\doibase 10.1103/PhysRevLett.119.260401}
  {\bibfield  {journal} {\bibinfo  {journal} {Phys. Rev. Lett.}\ }\textbf
  {\bibinfo {volume} {119}},\ \bibinfo {pages} {260401} (\bibinfo {year}
  {2017}{\natexlab{a}})}\BibitemShut {NoStop}%
\bibitem [{\citenamefont {y.~Choi}\ \emph {et~al.}(2016)\citenamefont
  {y.~Choi}, \citenamefont {Hild}, \citenamefont {Zeiher}, \citenamefont
  {Schauss}, \citenamefont {Rubio-Abadal}, \citenamefont {Yefsah},
  \citenamefont {Khemani}, \citenamefont {Huse}, \citenamefont {Bloch},\ and\
  \citenamefont {Gross}}]{Choi2016}%
  \BibitemOpen
  \bibfield  {author} {\bibinfo {author} {\bibfnamefont {J.}~\bibnamefont
  {y.~Choi}}, \bibinfo {author} {\bibfnamefont {S.}~\bibnamefont {Hild}},
  \bibinfo {author} {\bibfnamefont {J.}~\bibnamefont {Zeiher}}, \bibinfo
  {author} {\bibfnamefont {P.}~\bibnamefont {Schauss}}, \bibinfo {author}
  {\bibfnamefont {A.}~\bibnamefont {Rubio-Abadal}}, \bibinfo {author}
  {\bibfnamefont {T.}~\bibnamefont {Yefsah}}, \bibinfo {author} {\bibfnamefont
  {V.}~\bibnamefont {Khemani}}, \bibinfo {author} {\bibfnamefont {D.~A.}\
  \bibnamefont {Huse}}, \bibinfo {author} {\bibfnamefont {I.}~\bibnamefont
  {Bloch}}, \ and\ \bibinfo {author} {\bibfnamefont {C.}~\bibnamefont
  {Gross}},\ }\href {\doibase 10.1126/science.aaf8834} {\bibfield  {journal}
  {\bibinfo  {journal} {Science}\ }\textbf {\bibinfo {volume} {352}},\ \bibinfo
  {pages} {1547} (\bibinfo {year} {2016})}\BibitemShut {NoStop}%
\bibitem [{\citenamefont {Chiaro}\ \emph {et~al.}(2022)\citenamefont {Chiaro},
  \citenamefont {Neill}, \citenamefont {Bohrdt}, \citenamefont {Filippone},
  \citenamefont {Arute}, \citenamefont {Arya}, \citenamefont {Babbush},
  \citenamefont {Bacon}, \citenamefont {Bardin}, \citenamefont {Barends},
  \citenamefont {Boixo}, \citenamefont {Buell}, \citenamefont {Burkett},
  \citenamefont {Chen}, \citenamefont {Chen}, \citenamefont {Collins},
  \citenamefont {Dunsworth}, \citenamefont {Farhi}, \citenamefont {Fowler},
  \citenamefont {Foxen}, \citenamefont {Gidney}, \citenamefont {Giustina},
  \citenamefont {Harrigan}, \citenamefont {Huang}, \citenamefont {Isakov},
  \citenamefont {Jeffrey}, \citenamefont {Jiang}, \citenamefont {Kafri},
  \citenamefont {Kechedzhi}, \citenamefont {Kelly}, \citenamefont {Klimov},
  \citenamefont {Korotkov}, \citenamefont {Kostritsa}, \citenamefont
  {Landhuis}, \citenamefont {Lucero}, \citenamefont {McClean}, \citenamefont
  {Mi}, \citenamefont {Megrant}, \citenamefont {Mohseni}, \citenamefont
  {Mutus}, \citenamefont {McEwen}, \citenamefont {Naaman}, \citenamefont
  {Neeley}, \citenamefont {Niu}, \citenamefont {Petukhov}, \citenamefont
  {Quintana}, \citenamefont {Rubin}, \citenamefont {Sank}, \citenamefont
  {Satzinger}, \citenamefont {White}, \citenamefont {Yao}, \citenamefont {Yeh},
  \citenamefont {Zalcman}, \citenamefont {Smelyanskiy}, \citenamefont {Neven},
  \citenamefont {Gopalakrishnan}, \citenamefont {Abanin}, \citenamefont {Knap},
  \citenamefont {Martinis},\ and\ \citenamefont {Roushan}}]{1910.06024}%
  \BibitemOpen
  \bibfield  {author} {\bibinfo {author} {\bibfnamefont {B.}~\bibnamefont
  {Chiaro}}, \bibinfo {author} {\bibfnamefont {C.}~\bibnamefont {Neill}},
  \bibinfo {author} {\bibfnamefont {A.}~\bibnamefont {Bohrdt}}, \bibinfo
  {author} {\bibfnamefont {M.}~\bibnamefont {Filippone}}, \bibinfo {author}
  {\bibfnamefont {F.}~\bibnamefont {Arute}}, \bibinfo {author} {\bibfnamefont
  {K.}~\bibnamefont {Arya}}, \bibinfo {author} {\bibfnamefont {R.}~\bibnamefont
  {Babbush}}, \bibinfo {author} {\bibfnamefont {D.}~\bibnamefont {Bacon}},
  \bibinfo {author} {\bibfnamefont {J.}~\bibnamefont {Bardin}}, \bibinfo
  {author} {\bibfnamefont {R.}~\bibnamefont {Barends}}, \bibinfo {author}
  {\bibfnamefont {S.}~\bibnamefont {Boixo}}, \bibinfo {author} {\bibfnamefont
  {D.}~\bibnamefont {Buell}}, \bibinfo {author} {\bibfnamefont
  {B.}~\bibnamefont {Burkett}}, \bibinfo {author} {\bibfnamefont
  {Y.}~\bibnamefont {Chen}}, \bibinfo {author} {\bibfnamefont {Z.}~\bibnamefont
  {Chen}}, \bibinfo {author} {\bibfnamefont {R.}~\bibnamefont {Collins}},
  \bibinfo {author} {\bibfnamefont {A.}~\bibnamefont {Dunsworth}}, \bibinfo
  {author} {\bibfnamefont {E.}~\bibnamefont {Farhi}}, \bibinfo {author}
  {\bibfnamefont {A.}~\bibnamefont {Fowler}}, \bibinfo {author} {\bibfnamefont
  {B.}~\bibnamefont {Foxen}}, \bibinfo {author} {\bibfnamefont
  {C.}~\bibnamefont {Gidney}}, \bibinfo {author} {\bibfnamefont
  {M.}~\bibnamefont {Giustina}}, \bibinfo {author} {\bibfnamefont
  {M.}~\bibnamefont {Harrigan}}, \bibinfo {author} {\bibfnamefont
  {T.}~\bibnamefont {Huang}}, \bibinfo {author} {\bibfnamefont
  {S.}~\bibnamefont {Isakov}}, \bibinfo {author} {\bibfnamefont
  {E.}~\bibnamefont {Jeffrey}}, \bibinfo {author} {\bibfnamefont
  {Z.}~\bibnamefont {Jiang}}, \bibinfo {author} {\bibfnamefont
  {D.}~\bibnamefont {Kafri}}, \bibinfo {author} {\bibfnamefont
  {K.}~\bibnamefont {Kechedzhi}}, \bibinfo {author} {\bibfnamefont
  {J.}~\bibnamefont {Kelly}}, \bibinfo {author} {\bibfnamefont
  {P.}~\bibnamefont {Klimov}}, \bibinfo {author} {\bibfnamefont
  {A.}~\bibnamefont {Korotkov}}, \bibinfo {author} {\bibfnamefont
  {F.}~\bibnamefont {Kostritsa}}, \bibinfo {author} {\bibfnamefont
  {D.}~\bibnamefont {Landhuis}}, \bibinfo {author} {\bibfnamefont
  {E.}~\bibnamefont {Lucero}}, \bibinfo {author} {\bibfnamefont
  {J.}~\bibnamefont {McClean}}, \bibinfo {author} {\bibfnamefont
  {X.}~\bibnamefont {Mi}}, \bibinfo {author} {\bibfnamefont {A.}~\bibnamefont
  {Megrant}}, \bibinfo {author} {\bibfnamefont {M.}~\bibnamefont {Mohseni}},
  \bibinfo {author} {\bibfnamefont {J.}~\bibnamefont {Mutus}}, \bibinfo
  {author} {\bibfnamefont {M.}~\bibnamefont {McEwen}}, \bibinfo {author}
  {\bibfnamefont {O.}~\bibnamefont {Naaman}}, \bibinfo {author} {\bibfnamefont
  {M.}~\bibnamefont {Neeley}}, \bibinfo {author} {\bibfnamefont
  {M.}~\bibnamefont {Niu}}, \bibinfo {author} {\bibfnamefont {A.}~\bibnamefont
  {Petukhov}}, \bibinfo {author} {\bibfnamefont {C.}~\bibnamefont {Quintana}},
  \bibinfo {author} {\bibfnamefont {N.}~\bibnamefont {Rubin}}, \bibinfo
  {author} {\bibfnamefont {D.}~\bibnamefont {Sank}}, \bibinfo {author}
  {\bibfnamefont {K.}~\bibnamefont {Satzinger}}, \bibinfo {author}
  {\bibfnamefont {T.}~\bibnamefont {White}}, \bibinfo {author} {\bibfnamefont
  {Z.}~\bibnamefont {Yao}}, \bibinfo {author} {\bibfnamefont {P.}~\bibnamefont
  {Yeh}}, \bibinfo {author} {\bibfnamefont {A.}~\bibnamefont {Zalcman}},
  \bibinfo {author} {\bibfnamefont {V.}~\bibnamefont {Smelyanskiy}}, \bibinfo
  {author} {\bibfnamefont {H.}~\bibnamefont {Neven}}, \bibinfo {author}
  {\bibfnamefont {S.}~\bibnamefont {Gopalakrishnan}}, \bibinfo {author}
  {\bibfnamefont {D.}~\bibnamefont {Abanin}}, \bibinfo {author} {\bibfnamefont
  {M.}~\bibnamefont {Knap}}, \bibinfo {author} {\bibfnamefont {J.}~\bibnamefont
  {Martinis}}, \ and\ \bibinfo {author} {\bibfnamefont {P.}~\bibnamefont
  {Roushan}},\ }\href {\doibase 10.1103/PhysRevResearch.4.013148} {\bibfield
  {journal} {\bibinfo  {journal} {Phys. Rev. Research}\ }\textbf {\bibinfo
  {volume} {4}},\ \bibinfo {pages} {013148} (\bibinfo {year}
  {2022})}\BibitemShut {NoStop}%
\bibitem [{\citenamefont {Kondov}\ \emph {et~al.}(2015)\citenamefont {Kondov},
  \citenamefont {McGehee}, \citenamefont {Xu},\ and\ \citenamefont
  {DeMarco}}]{kondov15}%
  \BibitemOpen
  \bibfield  {author} {\bibinfo {author} {\bibfnamefont {S.~S.}\ \bibnamefont
  {Kondov}}, \bibinfo {author} {\bibfnamefont {W.~R.}\ \bibnamefont {McGehee}},
  \bibinfo {author} {\bibfnamefont {W.}~\bibnamefont {Xu}}, \ and\ \bibinfo
  {author} {\bibfnamefont {B.}~\bibnamefont {DeMarco}},\ }\href {\doibase
  10.1103/PhysRevLett.114.083002} {\bibfield  {journal} {\bibinfo  {journal}
  {Phys. Rev. Lett.}\ }\textbf {\bibinfo {volume} {114}},\ \bibinfo {pages}
  {083002} (\bibinfo {year} {2015})}\BibitemShut {NoStop}%
\bibitem [{\citenamefont {Mierzejewski}\ \emph {et~al.}(2020)\citenamefont
  {Mierzejewski}, \citenamefont {\ifmmode~\acute{S}\else \'{S}\fi{}roda},
  \citenamefont {Herbrych},\ and\ \citenamefont {Prelov\ifmmode~\check{s}\else
  \v{s}\fi{}ek}}]{mierzejewski2020}%
  \BibitemOpen
  \bibfield  {author} {\bibinfo {author} {\bibfnamefont {M.}~\bibnamefont
  {Mierzejewski}}, \bibinfo {author} {\bibfnamefont {M.}~\bibnamefont
  {\ifmmode~\acute{S}\else \'{S}\fi{}roda}}, \bibinfo {author} {\bibfnamefont
  {J.}~\bibnamefont {Herbrych}}, \ and\ \bibinfo {author} {\bibfnamefont
  {P.}~\bibnamefont {Prelov\ifmmode~\check{s}\else \v{s}\fi{}ek}},\ }\href
  {\doibase 10.1103/PhysRevB.102.161111} {\bibfield  {journal} {\bibinfo
  {journal} {Phys. Rev. B}\ }\textbf {\bibinfo {volume} {102}},\ \bibinfo
  {pages} {161111(R)} (\bibinfo {year} {2020})}\BibitemShut {NoStop}%
\bibitem [{\citenamefont {{\v S}trkalj}\ \emph {et~al.}(2022)\citenamefont {{\v
  S}trkalj}, \citenamefont {Doggen},\ and\ \citenamefont
  {Castelnovo}}]{strkalj2022}%
  \BibitemOpen
  \bibfield  {author} {\bibinfo {author} {\bibfnamefont {A.}~\bibnamefont {{\v
  S}trkalj}}, \bibinfo {author} {\bibfnamefont {E.~V.~H.}\ \bibnamefont
  {Doggen}}, \ and\ \bibinfo {author} {\bibfnamefont {C.}~\bibnamefont
  {Castelnovo}},\ }\href {\doibase arXiv:2204.05198} {\  (\bibinfo {year}
  {2022}),\ arXiv:2204.05198}\BibitemShut {NoStop}%
\bibitem [{\citenamefont {Davidson}\ \emph {et~al.}(2017)\citenamefont
  {Davidson}, \citenamefont {Sels},\ and\ \citenamefont
  {Polkovnikov}}]{Davidson2017}%
  \BibitemOpen
  \bibfield  {author} {\bibinfo {author} {\bibfnamefont {S.~M.}\ \bibnamefont
  {Davidson}}, \bibinfo {author} {\bibfnamefont {D.}~\bibnamefont {Sels}}, \
  and\ \bibinfo {author} {\bibfnamefont {A.}~\bibnamefont {Polkovnikov}},\
  }\href {\doibase 10.1016/j.aop.2017.07.003} {\bibfield  {journal} {\bibinfo
  {journal} {Annals of Physics}\ }\textbf {\bibinfo {volume} {384}},\ \bibinfo
  {pages} {128} (\bibinfo {year} {2017})}\BibitemShut {NoStop}%
\bibitem [{\citenamefont {Davidson}(2017)}]{S.M.Davidson.thesis}%
  \BibitemOpen
  \bibfield  {author} {\bibinfo {author} {\bibfnamefont {S.~M.}\ \bibnamefont
  {Davidson}},\ }\href@noop {} {\bibfield  {journal} {\bibinfo  {journal}
  {Ph.D. Thesis}\ } (\bibinfo {year} {Boston University, Boston
  2017})}\BibitemShut {NoStop}%
\bibitem [{\citenamefont {Schmitt}\ \emph {et~al.}(2019)\citenamefont
  {Schmitt}, \citenamefont {Sels}, \citenamefont {Kehrein},\ and\ \citenamefont
  {Polkovnikov}}]{PhysRevB.99.134301}%
  \BibitemOpen
  \bibfield  {author} {\bibinfo {author} {\bibfnamefont {M.}~\bibnamefont
  {Schmitt}}, \bibinfo {author} {\bibfnamefont {D.}~\bibnamefont {Sels}},
  \bibinfo {author} {\bibfnamefont {S.}~\bibnamefont {Kehrein}}, \ and\
  \bibinfo {author} {\bibfnamefont {A.}~\bibnamefont {Polkovnikov}},\ }\href
  {\doibase 10.1103/PhysRevB.99.134301} {\bibfield  {journal} {\bibinfo
  {journal} {Phys. Rev. B}\ }\textbf {\bibinfo {volume} {99}},\ \bibinfo
  {pages} {134301} (\bibinfo {year} {2019})}\BibitemShut {NoStop}%
\bibitem [{\citenamefont {Sajna}\ and\ \citenamefont
  {Polkovnikov}(2020)}]{PhysRevA.102.033338}%
  \BibitemOpen
  \bibfield  {author} {\bibinfo {author} {\bibfnamefont {A.~S.}\ \bibnamefont
  {Sajna}}\ and\ \bibinfo {author} {\bibfnamefont {A.}~\bibnamefont
  {Polkovnikov}},\ }\href {\doibase 10.1103/PhysRevA.102.033338} {\bibfield
  {journal} {\bibinfo  {journal} {Phys. Rev. A}\ }\textbf {\bibinfo {volume}
  {102}},\ \bibinfo {pages} {033338} (\bibinfo {year} {2020})}\BibitemShut
  {NoStop}%
\bibitem [{\citenamefont {Osterkorn}\ and\ \citenamefont
  {Kehrein}(2020)}]{2007.05063}%
  \BibitemOpen
  \bibfield  {author} {\bibinfo {author} {\bibfnamefont {A.}~\bibnamefont
  {Osterkorn}}\ and\ \bibinfo {author} {\bibfnamefont {S.}~\bibnamefont
  {Kehrein}},\ }\href@noop {} {\  (\bibinfo {year} {2020})},\ \Eprint
  {http://arxiv.org/abs/arXiv:2007.05063} {arXiv:2007.05063} \BibitemShut
  {NoStop}%
\bibitem [{\citenamefont {Osterkorn}\ and\ \citenamefont
  {Kehrein}(2022)}]{2205.06620}%
  \BibitemOpen
  \bibfield  {author} {\bibinfo {author} {\bibfnamefont {A.}~\bibnamefont
  {Osterkorn}}\ and\ \bibinfo {author} {\bibfnamefont {S.}~\bibnamefont
  {Kehrein}},\ }\href@noop {} {\  (\bibinfo {year} {2022})},\ \Eprint
  {http://arxiv.org/abs/arXiv:2205.06620} {arXiv:2205.06620} \BibitemShut
  {NoStop}%
\bibitem [{\citenamefont {Serbyn}\ \emph {et~al.}(2017)\citenamefont {Serbyn},
  \citenamefont {Papi\ifmmode~\acute{c}\else \'{c}\fi{}},\ and\ \citenamefont
  {Abanin}}]{serbyn2017}%
  \BibitemOpen
  \bibfield  {author} {\bibinfo {author} {\bibfnamefont {M.}~\bibnamefont
  {Serbyn}}, \bibinfo {author} {\bibfnamefont {Z.}~\bibnamefont
  {Papi\ifmmode~\acute{c}\else \'{c}\fi{}}}, \ and\ \bibinfo {author}
  {\bibfnamefont {D.~A.}\ \bibnamefont {Abanin}},\ }\href {\doibase
  10.1103/PhysRevB.96.104201} {\bibfield  {journal} {\bibinfo  {journal} {Phys.
  Rev. B}\ }\textbf {\bibinfo {volume} {96}},\ \bibinfo {pages} {104201}
  (\bibinfo {year} {2017})}\BibitemShut {NoStop}%
\bibitem [{\citenamefont {Vidmar}\ \emph {et~al.}(2021)\citenamefont {Vidmar},
  \citenamefont {Krajewski}, \citenamefont {Bon\ifmmode~\check{c}\else
  \v{c}\fi{}a},\ and\ \citenamefont {Mierzejewski}}]{vidmar2021}%
  \BibitemOpen
  \bibfield  {author} {\bibinfo {author} {\bibfnamefont {L.}~\bibnamefont
  {Vidmar}}, \bibinfo {author} {\bibfnamefont {B.}~\bibnamefont {Krajewski}},
  \bibinfo {author} {\bibfnamefont {J.}~\bibnamefont
  {Bon\ifmmode~\check{c}\else \v{c}\fi{}a}}, \ and\ \bibinfo {author}
  {\bibfnamefont {M.}~\bibnamefont {Mierzejewski}},\ }\href {\doibase
  10.1103/PhysRevLett.127.230603} {\bibfield  {journal} {\bibinfo  {journal}
  {Phys. Rev. Lett.}\ }\textbf {\bibinfo {volume} {127}},\ \bibinfo {pages}
  {230603} (\bibinfo {year} {2021})}\BibitemShut {NoStop}%
\bibitem [{\citenamefont {Ward}\ and\ \citenamefont
  {Greenwood}(2007)}]{Ward:2007}%
  \BibitemOpen
  \bibfield  {author} {\bibinfo {author} {\bibfnamefont {L.~M.}\ \bibnamefont
  {Ward}}\ and\ \bibinfo {author} {\bibfnamefont {P.~E.}\ \bibnamefont
  {Greenwood}},\ }\href {\doibase 10.4249/scholarpedia.1537} {\bibfield
  {journal} {\bibinfo  {journal} {Scholarpedia}\ }\textbf {\bibinfo {volume}
  {2}},\ \bibinfo {pages} {1537} (\bibinfo {year} {2007})}\BibitemShut
  {NoStop}%
\bibitem [{Note1()}]{Note1}%
  \BibitemOpen
  \bibinfo {note} {In comparison to Ref. \cite {Davidson2017}, we include an
  extra imaginary unit $i$ factor into the definition of the Poisson brackets
  in Eqs. \ref {eq:equation2} and \ref {eq: EOM}.}\BibitemShut {Stop}%
\bibitem [{\citenamefont {Polkovnikov}(2010)}]{Polkovnikov2010}%
  \BibitemOpen
  \bibfield  {author} {\bibinfo {author} {\bibfnamefont {A.}~\bibnamefont
  {Polkovnikov}},\ }\href {\doibase 10.1016/j.aop.2010.02.006} {\bibfield
  {journal} {\bibinfo  {journal} {Annals of Physics}\ }\textbf {\bibinfo
  {volume} {325}},\ \bibinfo {pages} {1790} (\bibinfo {year}
  {2010})}\BibitemShut {NoStop}%
\bibitem [{\citenamefont {Park}\ and\ \citenamefont {Light}(1986)}]{lantime1}%
  \BibitemOpen
  \bibfield  {author} {\bibinfo {author} {\bibfnamefont {T.~J.}\ \bibnamefont
  {Park}}\ and\ \bibinfo {author} {\bibfnamefont {J.~C.}\ \bibnamefont
  {Light}},\ }\href {\doibase 10.1063/1.451548} {\bibfield  {journal} {\bibinfo
   {journal} {The Journal of Chemical Physics}\ }\textbf {\bibinfo {volume}
  {85}},\ \bibinfo {pages} {5870} (\bibinfo {year} {1986})}\BibitemShut
  {NoStop}%
\bibitem [{\citenamefont {Mierzejewski}\ and\ \citenamefont
  {Prelov\ifmmode~\check{s}\else \v{s}\fi{}ek}(2010)}]{lantime2}%
  \BibitemOpen
  \bibfield  {author} {\bibinfo {author} {\bibfnamefont {M.}~\bibnamefont
  {Mierzejewski}}\ and\ \bibinfo {author} {\bibfnamefont {P.}~\bibnamefont
  {Prelov\ifmmode~\check{s}\else \v{s}\fi{}ek}},\ }\href {\doibase
  10.1103/PhysRevLett.105.186405} {\bibfield  {journal} {\bibinfo  {journal}
  {Phys. Rev. Lett.}\ }\textbf {\bibinfo {volume} {105}},\ \bibinfo {pages}
  {186405} (\bibinfo {year} {2010})}\BibitemShut {NoStop}%
\bibitem [{\citenamefont {Schreiber}\ \emph {et~al.}(2015)\citenamefont
  {Schreiber}, \citenamefont {Hodgman}, \citenamefont {Bordia}, \citenamefont
  {L{\"u}schen}, \citenamefont {Fischer}, \citenamefont {Vosk}, \citenamefont
  {Altman}, \citenamefont {Schneider},\ and\ \citenamefont
  {Bloch}}]{mblschreiber}%
  \BibitemOpen
  \bibfield  {author} {\bibinfo {author} {\bibfnamefont {M.}~\bibnamefont
  {Schreiber}}, \bibinfo {author} {\bibfnamefont {S.~S.}\ \bibnamefont
  {Hodgman}}, \bibinfo {author} {\bibfnamefont {P.}~\bibnamefont {Bordia}},
  \bibinfo {author} {\bibfnamefont {H.~P.}\ \bibnamefont {L{\"u}schen}},
  \bibinfo {author} {\bibfnamefont {M.~H.}\ \bibnamefont {Fischer}}, \bibinfo
  {author} {\bibfnamefont {R.}~\bibnamefont {Vosk}}, \bibinfo {author}
  {\bibfnamefont {E.}~\bibnamefont {Altman}}, \bibinfo {author} {\bibfnamefont
  {U.}~\bibnamefont {Schneider}}, \ and\ \bibinfo {author} {\bibfnamefont
  {I.}~\bibnamefont {Bloch}},\ }\href {\doibase
  doi.org/10.1126/science.aaa7432} {\bibfield  {journal} {\bibinfo  {journal}
  {Science}\ }\textbf {\bibinfo {volume} {349}},\ \bibinfo {pages} {842}
  (\bibinfo {year} {2015})}\BibitemShut {NoStop}%
\bibitem [{\citenamefont {Bordia}\ \emph {et~al.}(2016)\citenamefont {Bordia},
  \citenamefont {L\"uschen}, \citenamefont {Hodgman}, \citenamefont
  {Schreiber}, \citenamefont {Bloch},\ and\ \citenamefont
  {Schneider}}]{PhysRevLett.116.140401}%
  \BibitemOpen
  \bibfield  {author} {\bibinfo {author} {\bibfnamefont {P.}~\bibnamefont
  {Bordia}}, \bibinfo {author} {\bibfnamefont {H.~P.}\ \bibnamefont
  {L\"uschen}}, \bibinfo {author} {\bibfnamefont {S.~S.}\ \bibnamefont
  {Hodgman}}, \bibinfo {author} {\bibfnamefont {M.}~\bibnamefont {Schreiber}},
  \bibinfo {author} {\bibfnamefont {I.}~\bibnamefont {Bloch}}, \ and\ \bibinfo
  {author} {\bibfnamefont {U.}~\bibnamefont {Schneider}},\ }\href {\doibase
  10.1103/PhysRevLett.116.140401} {\bibfield  {journal} {\bibinfo  {journal}
  {Phys. Rev. Lett.}\ }\textbf {\bibinfo {volume} {116}},\ \bibinfo {pages}
  {140401} (\bibinfo {year} {2016})}\BibitemShut {NoStop}%
\bibitem [{\citenamefont {Smith}\ \emph {et~al.}(2016)\citenamefont {Smith},
  \citenamefont {Lee}, \citenamefont {Richerme}, \citenamefont {Neyenhuis},
  \citenamefont {Hess}, \citenamefont {Hauke}, \citenamefont {Heyl},
  \citenamefont {Huse},\ and\ \citenamefont {Monroe}}]{smith2016}%
  \BibitemOpen
  \bibfield  {author} {\bibinfo {author} {\bibfnamefont {J.}~\bibnamefont
  {Smith}}, \bibinfo {author} {\bibfnamefont {A.}~\bibnamefont {Lee}}, \bibinfo
  {author} {\bibfnamefont {P.}~\bibnamefont {Richerme}}, \bibinfo {author}
  {\bibfnamefont {B.}~\bibnamefont {Neyenhuis}}, \bibinfo {author}
  {\bibfnamefont {P.~W.}\ \bibnamefont {Hess}}, \bibinfo {author}
  {\bibfnamefont {P.}~\bibnamefont {Hauke}}, \bibinfo {author} {\bibfnamefont
  {M.}~\bibnamefont {Heyl}}, \bibinfo {author} {\bibfnamefont {D.~A.}\
  \bibnamefont {Huse}}, \ and\ \bibinfo {author} {\bibfnamefont
  {C.}~\bibnamefont {Monroe}},\ }\href {\doibase 10.1038/nphys3783} {\bibfield
  {journal} {\bibinfo  {journal} {Nature Physics}\ }\textbf {\bibinfo {volume}
  {12}},\ \bibinfo {pages} {907} (\bibinfo {year} {2016})}\BibitemShut
  {NoStop}%
\bibitem [{\citenamefont {L\"uschen}\ \emph
  {et~al.}(2017{\natexlab{b}})\citenamefont {L\"uschen}, \citenamefont
  {Bordia}, \citenamefont {Hodgman}, \citenamefont {Schreiber}, \citenamefont
  {Sarkar}, \citenamefont {Daley}, \citenamefont {Fischer}, \citenamefont
  {Altman}, \citenamefont {Bloch},\ and\ \citenamefont
  {Schneider}}]{Lschen2017}%
  \BibitemOpen
  \bibfield  {author} {\bibinfo {author} {\bibfnamefont {H.~P.}\ \bibnamefont
  {L\"uschen}}, \bibinfo {author} {\bibfnamefont {P.}~\bibnamefont {Bordia}},
  \bibinfo {author} {\bibfnamefont {S.~S.}\ \bibnamefont {Hodgman}}, \bibinfo
  {author} {\bibfnamefont {M.}~\bibnamefont {Schreiber}}, \bibinfo {author}
  {\bibfnamefont {S.}~\bibnamefont {Sarkar}}, \bibinfo {author} {\bibfnamefont
  {A.~J.}\ \bibnamefont {Daley}}, \bibinfo {author} {\bibfnamefont {M.~H.}\
  \bibnamefont {Fischer}}, \bibinfo {author} {\bibfnamefont {E.}~\bibnamefont
  {Altman}}, \bibinfo {author} {\bibfnamefont {I.}~\bibnamefont {Bloch}}, \
  and\ \bibinfo {author} {\bibfnamefont {U.}~\bibnamefont {Schneider}},\ }\href
  {\doibase 10.1103/PhysRevX.7.011034} {\bibfield  {journal} {\bibinfo
  {journal} {Phys. Rev. X}\ }\textbf {\bibinfo {volume} {7}},\ \bibinfo {pages}
  {011034} (\bibinfo {year} {2017}{\natexlab{b}})}\BibitemShut {NoStop}%
\bibitem [{\citenamefont {Acevedo}\ \emph {et~al.}(2017)\citenamefont
  {Acevedo}, \citenamefont {Safavi-Naini}, \citenamefont {Schachenmayer},
  \citenamefont {Wall}, \citenamefont {Nandkishore},\ and\ \citenamefont
  {Rey}}]{PhysRevA.96.033604}%
  \BibitemOpen
  \bibfield  {author} {\bibinfo {author} {\bibfnamefont {O.~L.}\ \bibnamefont
  {Acevedo}}, \bibinfo {author} {\bibfnamefont {A.}~\bibnamefont
  {Safavi-Naini}}, \bibinfo {author} {\bibfnamefont {J.}~\bibnamefont
  {Schachenmayer}}, \bibinfo {author} {\bibfnamefont {M.~L.}\ \bibnamefont
  {Wall}}, \bibinfo {author} {\bibfnamefont {R.}~\bibnamefont {Nandkishore}}, \
  and\ \bibinfo {author} {\bibfnamefont {A.~M.}\ \bibnamefont {Rey}},\ }\href
  {\doibase 10.1103/PhysRevA.96.033604} {\bibfield  {journal} {\bibinfo
  {journal} {Phys. Rev. A}\ }\textbf {\bibinfo {volume} {96}},\ \bibinfo
  {pages} {033604} (\bibinfo {year} {2017})}\BibitemShut {NoStop}%
\bibitem [{\citenamefont {Wurtz}\ \emph {et~al.}(2018)\citenamefont {Wurtz},
  \citenamefont {Polkovnikov},\ and\ \citenamefont {Sels}}]{WURTZ2018341}%
  \BibitemOpen
  \bibfield  {author} {\bibinfo {author} {\bibfnamefont {J.}~\bibnamefont
  {Wurtz}}, \bibinfo {author} {\bibfnamefont {A.}~\bibnamefont {Polkovnikov}},
  \ and\ \bibinfo {author} {\bibfnamefont {D.}~\bibnamefont {Sels}},\ }\href
  {\doibase https://doi.org/10.1016/j.aop.2018.06.001} {\bibfield  {journal}
  {\bibinfo  {journal} {Annals of Physics}\ }\textbf {\bibinfo {volume}
  {395}},\ \bibinfo {pages} {341 } (\bibinfo {year} {2018})}\BibitemShut
  {NoStop}%
\bibitem [{Note2()}]{Note2}%
  \BibitemOpen
  \bibinfo {note} {Http://wcss.pl}\BibitemShut {NoStop}%
\bibitem [{\citenamefont {Kozarzewski}\ \emph {et~al.}(2016)\citenamefont
  {Kozarzewski}, \citenamefont {Prelov\ifmmode~\check{s}\else \v{s}\fi{}ek},\
  and\ \citenamefont {Mierzejewski}}]{PhysRevB.93.235151}%
  \BibitemOpen
  \bibfield  {author} {\bibinfo {author} {\bibfnamefont {M.}~\bibnamefont
  {Kozarzewski}}, \bibinfo {author} {\bibfnamefont {P.}~\bibnamefont
  {Prelov\ifmmode~\check{s}\else \v{s}\fi{}ek}}, \ and\ \bibinfo {author}
  {\bibfnamefont {M.}~\bibnamefont {Mierzejewski}},\ }\href {\doibase
  10.1103/PhysRevB.93.235151} {\bibfield  {journal} {\bibinfo  {journal} {Phys.
  Rev. B}\ }\textbf {\bibinfo {volume} {93}},\ \bibinfo {pages} {235151}
  (\bibinfo {year} {2016})}\BibitemShut {NoStop}%
\end{thebibliography}%

\end{document}